\documentclass[12pt]{article}
\usepackage{amsmath}
\usepackage{graphicx,psfrag,epsf} 
\usepackage{enumerate}
\usepackage{natbib}
\usepackage{url} 
\usepackage{rotating}
\usepackage{multirow}
\usepackage{amsfonts}
\usepackage{epstopdf}
\usepackage{xr}
\usepackage{multicol}
\usepackage{rotating,booktabs}
\usepackage{pdflscape}
\usepackage{array}
\usepackage{color}

\newcommand{\blind}{1}

\newtheorem{proposition}{Proposition}
\newtheorem{theorem}{Theorem}
\newtheorem{remark}{Remark}

\begin{document}

\def\spacingset#1{\renewcommand{\baselinestretch}%
{#1}\small\normalsize} \spacingset{1}


\if1\blind
{
  \title{\bf Mode testing, critical bandwidth\\ and excess mass}
  \author{Jose Ameijeiras--Alonso \hspace{.2cm}\\
    and \\
    Rosa M. Crujeiras \\
    and \\
    Alberto Rodr\'iguez--Casal\thanks{
The authors gratefully acknowledge the support of Projects MTM2016--76969--P (Spanish State Research Agency, AEI) and MTM2013--41383--P (Spanish Ministry of Economy, Industry and Competitiveness), both co--funded by the European Regional Development Fund (ERDF), IAP network from Belgian Science Policy. Work of J. Ameijeiras-Alonso has been supported by the predoctoral grant BES--2014--071006 from the Spanish Ministry of Economy, Industry and Competitiveness.
		}\\
    Department of Statistics, Mathematical Analysis and Optimization\\
    Universidade de Santiago de Compostela}
  \maketitle
} \fi

\if0\blind
{
  \bigskip
  \bigskip
  \bigskip
  \begin{center}
    {\LARGE\bf Mode testing, critical bandwidth and excess mass}
\end{center}
  \medskip
} \fi

\bigskip
\begin{abstract}
The identification of peaks or maxima in probability densities, by mode testing or bump hunting, has become an important problem in applied fields.  This task has been approached in the statistical literature from different perspectives, with the proposal of testing procedures which are based on kernel density estimators or on the quantification of excess mass. However, none of the existing proposals provides a satisfactory performance in practice. In this work, a new procedure which combines the previous approaches (smoothing and excess mass) is presented and compared with the existing methods, showing a superior behaviour. A real data example on philatelic data is also included for illustration purposes.
\end{abstract}

\noindent%
\vfill

\newpage
\spacingset{1.45} 
\section{Introduction}
\label{sec:intro}

Simple distribution models, such as the Gaussian density, may fail to capture the stochastic underlying structure driving certain mechanism in applied sciences. Complex measurements in geology, neurology, economics, ecology or astronomy exhibit some characteristics that cannot be reflected by unimodal densities. In addition, the identification of the (unknown) number of \emph{peaks} or modes is quite common in these fields. Some examples include the study of the percentage of silica in chondrite meteors \citep{GoodGask80}, the analysis of the macaques neurons when performing an attention--demanding task \citep{Mitchell07}, the distribution of household incomes of the United Kingdom \citep{MarSch92}, the study of the body--size in endangered fishes \citep{Oldenetal07} or the analysis of the velocity at which galaxies are moving away from ours \citep{Roeder90}. In all these examples, identifying the number (and location) of local maxima of the density function (i.e. modes) is important \emph{per se}, or as a previous step for applying other procedures. 

{An illustrative example which has been extensively considered in mode testing literature can be found in philately (the study of stamps and postal history and other related items). Research in this field has been motivated by the use of stamps for investment purposes. 
The value of stamps depends on its scarcity, and thickness is determinant in this sense. However, in some stamp issues, there is not a differentiation between groups available in stamps catalogs. The importance of establishing an objective criterion specially appears in stamp issues printed on a mixture of paper types, such as the 1872 Hidalgo issue. This particular example has been shown in several references in the literature as a paradigm of the problem of determining the number of modes/groups. In this work, this example will be revisited, recalling previous analysis and comparing results with the ones provided by the new testing procedure presented in this paper.
}

A formal hypothesis test for a null hypothesis of a certain number of modes can be stated as follows. Let $f$ be the density function of a real random variable $X$ and denote by $j$ the number of modes. For $k\in\mathbb Z^+$, the testing problem on the number of modes can be formulated as:
\begin{equation}
H_0:\, j= k\quad\mbox{vs.}\quad H_a:\, j>k.
\label{test}
\end{equation}
There have been quite a few proposals in the statistical literature for solving (\ref{test}) and the different techniques can be classified in two groups: a first group of tests based on or using a critical bandwidth, introduced by \citet{Silverman81}, further studied by \citet{HallYork01} and also used by \citet{FisMar01}; and a second group of tests based on the \emph{excess mass}, such as those ones proposed by \citet{Hartigan85}, \citet{MulSaw91} and \citet{ChengHall98}. These methods are briefly revised and compared in this paper, where a new proposal gathering strength from both areas is also introduced, outperforming the existing procedures, in testing unimodality and more general hypotheses.

\begin{figure}
\centering
    \includegraphics[width=0.45\textwidth]{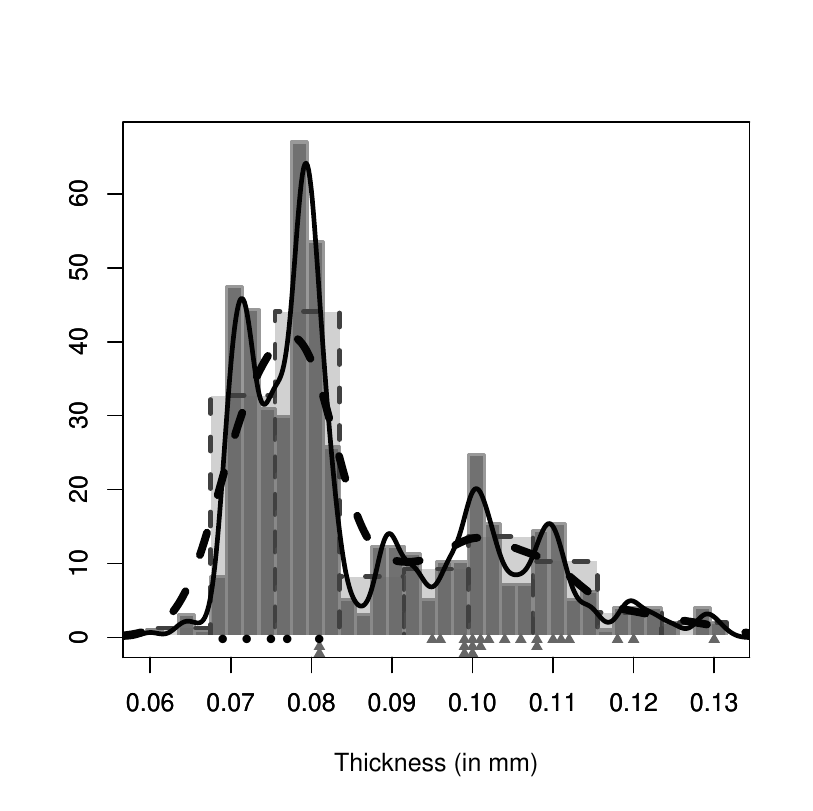}
    \includegraphics[width=0.45\textwidth]{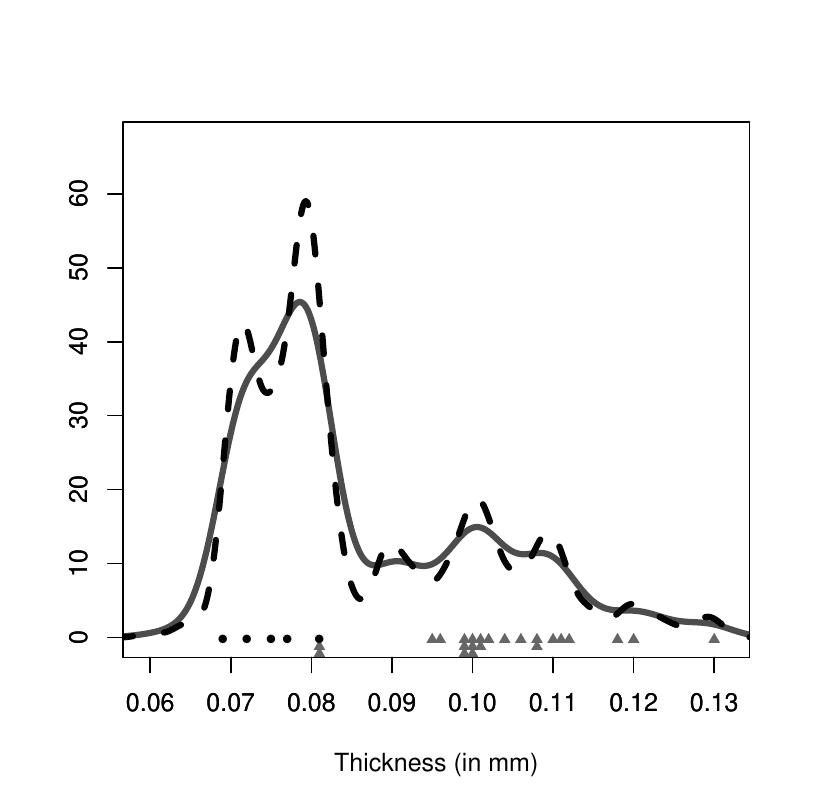} \\
 \caption{Sample of 485 stamps from the 1872 Hidalgo Issue of Mexico. {Points: stamps watermarked with \textit{LA+-F} (circles) and \textit{Papel sellado} (triangles). Kernel density estimators with Gaussian kernel and different bandwidths; left panel: $h=0.003910$ (rule of thumb -solid line-) and $h=0.001205$ \citep[plug--in rule -dashed line-, see][Ch. 3]{wandjones}; right panel: critical bandwidths $h_4=0.002831$ (solid line) and $h_7=0.001487$ (dashed line). Left: histograms with different bin widths (0.002 -continuous border- and 0.008 -dashed border-).}}
 \label{fig1}
\end{figure}

Apart from the formal testing procedures, and as a complementary tool for them, a first step when confronting the problem of identifying modes in a data distribution is the exploration of a nonparametric estimator of the underlying probability density, which can be done by kernel methods. Classical kernel density estimation \citep[][Ch. 2]{wandjones} allows for the reconstruction of the data density structure without imposing parametric restrictions (only subjected to mild regularity assumptions) but at the expense of choosing an appropriate bandwidth parameter, which controls the degree of smoothing. A direct observation of a kernel estimator may lead to inaccurate or even wrong conclusions about the mode density structure. {This can be noticed from the plots shown in Figure \ref{fig1}, where, with the kernel density estimator for the stamp dataset, different conclusions can be drawn about the number of modes with different bandwidths}. Based on this estimator, from an exploratory perspective, there are several alternatives for identifying modes such as the SiZer map \citep{ChMar99}, the mode tree and the random forest \citep{MinSco93,Minetal98}. Although these tools are helpful in supporting the results of formal testing procedures on the number and location of modes, apart from giving some insight on the global mode structure, the interpretation of the outputs from these procedures requires an \emph{expert eye}.

This paper presents a new testing procedure combining the use of a critical bandwidth and an excess mass statistic, which can be applied to solve (\ref{test}) in a general setting. In Section 2, a review on mode testing methods is presented, considering both tests based on critical bandwidth and on excess mass, jointly with the new proposal. A simulation study comparing all the procedures, in terms of empirical size and power, is included in Section 3. Section 4 is devoted to data analysis, revising the stamp dataset and presenting new results. Some final comments and discussion are given in Section 5. Details on the simulated models, the technical proofs, a modification of the proposal when the modes and antimodes lie in a known closed interval, a more flexible testing scenario, the computation of the proposed test and further details of the example analysed in Section 4 are provided as Supplementary Material available from the journal website.

\section{A review on multimodality tests}
\label{background}
Different proposals for multimodality tests will be briefly revised in this section. Section \ref{background:critical} includes a review on the methods using the critical bandwidth, and excess mass approaches are detailed in Section \ref{background:mass}. A new proposal, borrowing strength from both alternatives, is presented in Section \ref{background:new}: an excess mass statistic will be calibrated from a modified nonparametric kernel density estimator using a critical bandwidth. 

\subsection{Tests based on the critical bandwidth}
\label{background:critical}
For a certain number of modes $k\in\mathbb Z^+$, the critical bandwidth \citep{Silverman81} is the smallest bandwidth such that the kernel density estimator has at most $k$ modes:
\begin{equation*}
h_k=\inf\{h :\hat{f_h} \mbox{ has at most } k \mbox{ modes}\},
\end{equation*}
where $\hat{f_h}$ denotes the kernel density estimator, computed from a random sample $\mathcal{X}=(X_1,...,X_n)$, with kernel $K$ and bandwidth $h$:
\begin{equation}\label{kernel}
\widehat{f}_h(x)=\frac{1}{n h} \overset {n} {\underset {i=1} \sum} K\left(\frac{x-X_i}{h}\right).
\end{equation}
\citet{Silverman81} proposed to use the critical bandwidth with the Gaussian kernel as a statistic to test $H_0:j\leq k$ vs. $H_a:j>k$, being its use justified by the fact that, with a Gaussian kernel, the number of modes of $\hat{f_h}$ is a nonincreasing function of $h$. Hence, $H_0$ is rejected for large values of $h_k$, whose distribution is approached using bootstrap procedures. Specifically, the proposed methodology consists in obtaining $B$ samples $\mathcal{Z}^{*b}=(Z^{*b}_1, \cdots ,Z^{*b}_n)$ with $b=1, \cdots , B$, where $Z^{*b}_i=(1+h_k^2/\hat{\sigma}^2)^{-1/2} X^{*b}_i$, being $\hat{\sigma}^2$ the sample variance and $X^{*b}_i$ generated from $\hat{f}_{h_k}$. By computing the critical bandwidth, $h^{*b}_k$, from each sample $\mathcal{Z}^{*b}$, given a significance level $\alpha$, the null hypothesis is rejected if $\mathbb{P}(h_{k}^* \leq h_{k} | \mathcal{X}) \geq 1- \alpha$. When $k=1$, the testing problem tackled by \citet{Silverman81} coincides with (\ref{test}). However, for a general $k$, the null hypothesis in (\ref{test}) is more restrictive than the one considered by \citet{Silverman81}, but asymptotic consistency of the test is only derived for $j=k$ \citep[for a deeper insight see][or Section \ref{testbimuni} in Supplementary Material]{HallYork01}.

\citet{HallYork01} proved that the previous bootstrap algorithm does not provide a consistent approximation of the test statistic  distribution under the null hypothesis and suggested a way for accurate calibration when $k=1$. Given a closed interval $I$ where the null hypothesis is tested ($f$ has a single mode in $I$), if both the support of $f$ an the interval $I$ are unbounded then properties of $h_1$ (critical bandwidth when $k=1$) are generally determined by extreme values in the sample, not by the modes of $f$. To avoid this issue, the testing problem (\ref{test}) is reformulated as follows:
\begin{equation}
\resizebox{.98 \textwidth}{!} 
{ 
$H_0:\, j=1 \mbox{ in the interior of a given closed interval } I \mbox{ and no local minimum in } I,$}
\label{testhy}
\end{equation}
and the critical bandwidth is redefined accordingly as:
\begin{equation}
\label{bw_hy}
h_{\tiny{\mbox{HY}}}=\inf\{h :\hat{f_h} \mbox{ has exactly one mode in } I\}.
\end{equation}
An issue that should be kept in mind in the computation of this critical bandwidth is that even if $K$ is the Gaussian kernel, the number of modes of $\hat{f_h}$ inside $I$ is not necessarily a monotone function of $h$. But under relatively general conditions \citep[see][]{HallYork01}, the probability that the number of modes is monotone in $h$ converges to 1 for such a kernel. \citet{HallYork01} proposed using $h_{\tiny{\mbox{HY}}}$ as a statistic to test (\ref{testhy}). The null distribution of $h_{\tiny{\mbox{HY}}}$ is approximated by bootstrap, generating bootstrap samples from $\hat{f}_{\tiny{\mbox{HY}}}$. 

Unfortunately, the critical bandwidths for the bootstrap samples $h_{\tiny{\mbox{HY}}}^{*b}$, are smaller than $h_{\tiny{\mbox{HY}}}$, so for an $\alpha$--level test, a correction factor $\lambda_{\alpha}$ to compute the p--value $\mathbb{P}(h_{\tiny{\mbox{HY}}}^* \leq \lambda_\alpha h_{\tiny{\mbox{HY}}} | \mathcal{X}) \geq 1- \alpha$ must be considered. Two different methods were suggested for computing this factor $\lambda_{\alpha}$, the first one based on a polynomial approximation and a second one using Monte Carlo techniques considering a simple unimodal distribution.

The previous proposal could be extended, as mentioned by \citet{HallYork01}, to test that $f$ has exactly $k$ modes in $I$, against the alternative that it has $(k+1)$ or more modes there, extending the critical bandwidth in (\ref{bw_hy}) for $k$ modes, namely $h_{\tiny{\mbox{HY}},k}$. Nevertheless, in this scenario, the bootstrap test cannot be directly calibrated under the hypothesis that $f$ has $k$ modes and $(k-1)$ antimodes, since it depends on the $(2k-2)$ unknowns $(c_i/c_1)$, where $c_i=f^{1/5}(t_i)/|f''(t_i)|^{2/5}$ (assuming $f''(t_i)\neq 0$ for all $i$), and $t_i$ being the ordered turning points of $f$ in $I$ with $i=1,\cdots,(2k-1)$; which notably complicates the computations.

Finally, it should be also commented that the use of the critical bandwidth for testing (\ref{test}) is not limited to its use as a test statistic. Consider a Cram\'er--von Mises test statistic:
\begin{equation}
T=n \int_{-\infty}^{\infty} {[F_n(x)-F_0(x)]^2dF_0 (x)}= \overset {n} {\underset {i=1} \sum} \left( F_0(X_{(i)}) - \frac{2i-1}{2n} \right)^2 + \frac{1}{12 n},
\label{cvm}
\end{equation}
where $F_0$ is a given continuous distribution function, $\{X_{(1)} \leq \cdots \leq X_{(n)}\}$ denotes the ordered sample and $F_n$ is the empirical distribution function. \citet{FisMar01} proposed the use of (\ref{cvm}) for solving the general problem of testing $k$ modes ($H_0:j\leq k$) by taking $F_0(x)=\hat{F}_{h_k}(x)=\int_{-\infty}^x {\hat{f}_{h_k}(t)dt}$ and derived the statistic:
\begin{equation}
T_k=\overset {n} {\underset {i=1} \sum} \left( \hat{F}_{h_k}(X_{(i)}) - \frac{2i-1}{2n} \right)^2 + \frac{1}{12n},
\label{fmstatistic}
\end{equation}
where the null hypothesis is rejected for large values of $T_k$. To approximate the distribution of the test statistic (\ref{fmstatistic}) under the null hypothesis, a bootstrap procedure is also proposed. It will be seen in Section \ref{simulation} that the behaviour of the \cite{FisMar01} proposal is far from satisfactory.

\subsection{Tests based on excess mass}
\label{background:mass}
\citet{MulSaw91} confront the testing problem (\ref{test}), employing a different perspective, under the following premise: a mode is present where an excess of probability mass is concentrated. Specifically, given a continuous real density function $f$ and a constant $\lambda$, the excess mass is defined as:
\begin{equation*}
E(\mathbb{P}_X,\lambda)=\mathbb{P}_X(C(\lambda))-\lambda ||C(\lambda)||=\int_{C(\lambda)}{f(x)dx}-\lambda ||C(\lambda)||,
\end{equation*}
{where $C(\lambda)=\{x : f(x)\geq \lambda\}$, and $||C(\lambda)||$ denotes the measure of $C(\lambda)$. If $f$ has $k$ modes, independently on $\lambda$, it can be divided in at most $k$ disjoint connected sets over the support of $f$, called $\lambda$--clusters. If $f$ has $k$ $\lambda$--clusters, then the excess mass can be defined as:}
\begin{equation}
E_{k}(\mathbb{P}_X,\lambda)=\underset{C_1(\lambda),...,C_k(\lambda)}{\mbox{sup}} \left\{ \overset {k} {\underset {m=1} \sum} \left(\mathbb{P}_X(C_m(\lambda))-\lambda ||C_m(\lambda)|| \right)\right\},
\label{elems}
\end{equation}  
where the supremum is taken over all families $\{C_m(\lambda) : m = 1, \cdots, k\}$ of $\lambda$--clusters. Under the assumption that $f$ has $k$ $\lambda$--clusters, the excess mass defined in (\ref{elems}) can be empirically estimated with $E_{n,k}(\mathbb{P}_n,\lambda)$ in the following way
\begin{equation*}
E_{n,k}(\mathbb{P}_n,\lambda)=\underset{\hat{C}_1(\lambda),...,\hat{C}_k(\lambda)}{\mbox{sup}} \left\{ \overset {k} {\underset {m=1} \sum} \mathbb{P}_n(\hat{C}_m(\lambda))-\lambda ||\hat{C}_m(\lambda)||\right\},
\end{equation*}
where the empirical sets $\{\hat{C}_m(\lambda) : m = 1, \cdots, k\}$ are closed intervals with endpoints at data points, and $\mathbb{P}_n(\hat{C}_m(\lambda))=(1/n) \sum_{i=1}^{n} \mathcal{I}(X_i \in \hat{C}_m(\lambda))$, being $\mathcal{I}$ the indicator function. The difference $D_{n,k+1}(\lambda)=E_{n,k+1}(\mathbb{P}_n,\lambda) - E_{n,k}(\mathbb{P}_n,\lambda)$ measures the plausibility of the null hypothesis, that is, large values of $D_{n,k+1}(\lambda)$ would indicate that $H_0$ is false. Using these differences, \citet{MulSaw91} proposed the following test statistic:
\begin{equation} 
\label{msstatistic}
\Delta_{n,k+1}=  \underset{\lambda}{\max } \{D_{n,k+1}(\lambda) \},
\end{equation}
rejecting the null hypothesis that $f$ has $k$ modes for large values of $\Delta_{n,k+1}$. Note that just the sample is needed for computing the value of the excess mass test statistic. \citet{MulSaw91} showed that this statistic is an extension of the \textit{dip} test introduced by \citet{Hartigan85}, just valid for the unimodal case, since both quantities (dip and excess mass) coincide up to a factor, for the unimodality case. In addition, the proposal of \citet{MulSaw91} for testing unimodality is the same as that one of \citet{Hartigan85} and considers a Monte Carlo calibration, generating resamples from the uniform distribution.

In view of the extremely conservative behaviour of the calibration of the previous proposals (see Section \ref{simulation} for results), \citet{ChengHall98} designed a calibration procedure based on the following result: for large samples and under the hypothesis that $f$ is unimodal, the distribution of $\Delta_{n,2}$ is independent of unknowns except for a factor $c=\left({f^3(x_0)}/{|f''(x_0)|}\right)^{1/5},$ where $x_0$ denotes the unique mode of $f$. Using this fact, for the case $k=1$, \citet{ChengHall98} approximated the distribution of $\Delta_{n,2}$ employing the values of $\Delta_{n,2}^*$ obtained from the samples generated from a parametric calibration distribution $\Psi(\cdot,\beta)$, being $\beta$ a certain parameter. Depending on the value of $d=c^{-5}$, different parametric distributions were suggested by the authors: a normal ($d=2\pi$), a beta distribution ($d<2\pi$) or a rescaled Student $t$ ($d>2\pi$). For estimating $d$, if $\widehat{x}_0$ denotes the largest mode of $\hat{f_h}$, then $\hat{d}={|\hat{f}_{h'}''(\widehat{x}_0)|}/{\hat{f}_h^3(\widehat{x}_0)},$ is used, were $\hat{f}''$ and $\hat{f}$ are kernel estimators with a Gaussian kernel and $h'$ and $h$ are their respective asymptotically optimal global bandwidths, replacing the unknown quantities for the ones associated with a $N(0,\hat{\sigma}^2)$. The methodology proposed by \citet{ChengHall98} consists in generating samples from $\Psi(\cdot,\hat{\beta})$, where $\hat{\beta}$ and the distribution family are chosen using $\hat{d}$. The excess mass statistic given in (\ref{msstatistic}) when $k=1$, that is $\Delta_{n,2}^*$, is computed from the resamples and, for a given significance level $\alpha$, the null hypothesis is rejected if $\mathbb{P}(\Delta_{n,2}^*\leq \Delta_{n,2}|\mathcal{X})\geq 1- \alpha$.

\subsection{A new proposal}
\label{background:new}
The previous tests show some limitations for practical applications: first, just the proposals of \citet{Silverman81} and \citet{FisMar01} allow to test (\ref{test}) for $k>1$. Despite the efforts of \citet{ChengHall98} and \citet{HallYork01} for providing good calibration algorithms, it will be shown in Section \ref{simulation} that the behaviour of all the proposals is far from satisfactory. Specifically, the test presented by \citet{Silverman81} is very conservative in general (although sometimes can show the opposite behaviour) and the proposal of \citet{FisMar01} does not have a good level accuracy. The new method proposed in this work overcomes these drawbacks by considering an excess mass statistic, as the one proposed by \citet{MulSaw91} with bootstrap calibration. Unlike \citet{ChengHall98}, a completely data-driven procedure will be designed, using the critical bandwidth under $H_0:j=k$, $k\in\mathbb Z^+$.

\textbf{The proposal, in a nutshell.} Consider the testing problem (\ref{test}) and take the excess mass statistic given in (\ref{msstatistic}), under the null hypothesis. Given $\mathcal{X}$, generate $B$ resamples $\mathcal{X}^{*b}$ ($b=1,\ldots, B$) of size $n$ from a modified version of $\hat{f}_{h_k}$, namely the \textit{calibration function} and subsequently denoted by $g$. For a significance level $\alpha$, the null hypothesis will be rejected if $\mathbb{P}(\Delta_{n,k+1}^*\leq \Delta_{n,k+1}|\mathcal{X})\geq 1- \alpha$, where $\Delta_{n,k+1}^*$ is the excess mass statistic obtained from the generated samples. It should be also noted that the procedure can be easily adapted to handle \citet{HallYork01} scenario: to test the null hypothesis that $f$ has at most $k$ modes in the interior of a given closed interval $I$, if $I$ is known, use (a modified version of) $\hat{f}_{h_{\tiny{\mbox{HY}},k}}$ to generate the samples. From this brief description, two questions arise: How is this \emph{modified} version of $\hat{f}_{h_k}$ constructed? Does the procedure guarantee a correct calibration of the test? In fact, the construction of the calibration function as a modification of the kernel density estimator ensures the correct calibration, under some regularity conditions.
\paragraph{Regularity conditions}\label{cond} (RC1) The density function $f$ is bounded with continuous derivative. (RC2) There exist $t_1$ and $t_2$, such that $f$ is monotone in $(-\infty,t_1)$ and in $(t_2,\infty)$. (RC3) {There are $(2j-1)$ points satisfying} $\{x : f'(x)=0 \mbox{ and } f(x)\neq 0\}$, which are the modes and antimodes of $f$, denoted as $x_i$, with $i=1,\ldots,(2j-1)$; and $f''(x_i)\neq 0$. (RC4) $f''$ exists and is H\"older continuous in a neighbourhood of each $x_i$.

Define $d_i={|f''(x_i)|}/{f^3(x_i)}$. {To guarantee the asymptotic correct behaviour of the test, $f$ must satisfy the regularity conditions (RC1)--(RC4) and the calibration function $g$ is going to be build in order to preserve them, and to ensure the convergence, in probability, of the values $\widehat{d_i}={|g''(\widehat{x_i})|}/{g^3(\widehat{x_i})}$ to $d_i$, in the modes and antimodes of $g$, namely $\widehat{x_i}$, as $n \rightarrow \infty$, for $i=1,\ldots,(2j-1)$.} As mentioned before, the calibration function $g$ from which the bootstrap resamples are generated is obtained by modifying $\hat{f}_{h_j}$. Function $g$ is constructed preserving the regularity conditions (RC1)--(RC4), by modifying $\widehat f_{h_j}$ in a neighbourhood of $\{x : \widehat f_{h_j}'(x)=0\}$, being such values a finite collection \citep[see][]{Silverman81}, having positive estimated density. This modification also ensures that the only points that satisfy $\{x : \widehat g'(x)=0\}$ are the modes and antimodes of $g$. The estimator of ${d_i}$ will be equal to the following ratio,

\begin{equation}\label{distar} 
\widehat{d_i}={|\widehat{f}''_{h_{\tiny{\mbox{PI}}}}(\widehat{x_i})|}/{\widehat{f}_{h_j}^3(\widehat{x_i})} \mbox{, being } h_{\tiny{\mbox{PI}}} \mbox{ a plug--in bandwidth,}
\end{equation} 
where, in this work, the plug--in rule for the second derivative will be obtained deriving the asymptotic mean integrated squared error and replacing $f$ in its expression using a two--step procedure \citep[see, for example,][Ch. 3]{wandjones}. {Employing this calibration function $g$, which complete expression is given in (\ref{gfunc}), the assumptions over $g$ of the Theorem~\ref{th1} ({the proofs of this result and Proposition~\ref{th2} are provided as Supplementary Material}) will be satisfied.}

\begin{theorem}\label{th1}
Let $g$ be a modified version $f_{h_j}$, having $j$ modes and satisfying that ${|g''(\widehat{x_i})|}/{g^3(\widehat{x_i})}$ converges in probability to ${|f''(x_i)|}/{f^3(x_i)}$, where $x_i$ and $\widehat{x_i}$ are respectively the modes and antimodes of $f$ and $g$, for $i=1,\ldots,(2j-1)$. If both, $f$ and $g$, satisfy conditions (RC1)--(RC4), then the limiting bootstrap distribution of $\Delta^*_{n,j+1}$ (calculated from the resamples associated to $g$) is identical to the asymptotic distribution of $\Delta_{n,j+1}$ (calculated from the sample associated to $f$), and so the test $\mathbb{P}(\Delta_{n,j+1}^*\leq \Delta_{n,j+1}|\mathcal{X})\geq 1- \alpha$ has an asymptotic level $\alpha$.
\end{theorem}

\begin{remark}
Following \cite{ChengHall98}, a parametric family having the desired values of $d_i$, for $i=1,\ldots,(2j-1)$, could be used as calibration function $g$ when $j>1$. Two issues appear related with their calibration procedure. First, it is not an easy task to construct this family. In addition, the second--order limiting properties of the test depend on the form of the density function. Then, a better behaviour is expected if the calibration function is ``more similar'' to the real density function. Our method deals with these two issues to get a test having a good performance in the finite--sample case and allowing to solve the general problem of testing $k$ modes.
\end{remark}

As mentioned before, our calibration function is constructed by modifying $\widehat f_{h_k}$ in a neighbourhood of the points $\{x : \widehat f_{h_k}'(x)=0\}$. Depending on the nature of these points, two modifications in their neighbourhood will be done. If the point is a mode or an antimode of $\widehat f_{h_k}$, namely $\widehat{x_i}$, in its neighbourhood, $\widehat f_{h_k}$ will be replaced by the function $J$, described in (\ref{jfunc}). This modification will preserve the location $\widehat{x_i}$, its estimated density value and it will satisfy that $g''(\widehat{x_i})=\widehat{f}''_{h_{\tiny{\mbox{PI}}}}(\widehat{x_i})$\footnote{Note that, although asymptotically the sign of $\widehat{f}''_{h_{\tiny{\mbox{PI}}}}(\widehat{x_i})$ is always correct (under the assumptions of Theorem~\ref{th1}), in the finite--sample case, it may not be negative in the modes or positive in the antimodes. In that case, an abuse of notation will be done, denoting as $h_{\tiny{\mbox{PI}}}$ to the critical or other plug--in bandwidth in order to guarantee that the sign of this second derivative remains correct.}. In fact, this procedure guarantees the correct estimation of $d_i$ and (RC4). The second modification, achieved by the function $L$, defined in (\ref{Lgranfunc}), will remove the $t$ saddle points of $\widehat f_{h_k}$, denoted as $\zeta_p$, with $p= \{1,\ldots, t\}$. This modification is done in order to satisfy (RC3). Since all the modifications are made in bounded neighbourhoods, condition (RC2) will continue to be fulfilled and the modifications of the functions $J$ and $L$ will be carried out preserving condition (RC1). {The calibration function $g$ for $k$ modes will be constructed as follows, to ensure that the assumptions of Theorem~\ref{th1} are satisfied (see Proposition~\ref{th2})}. 

\begin{figure}
\centering
    \includegraphics[width=0.44\textwidth]{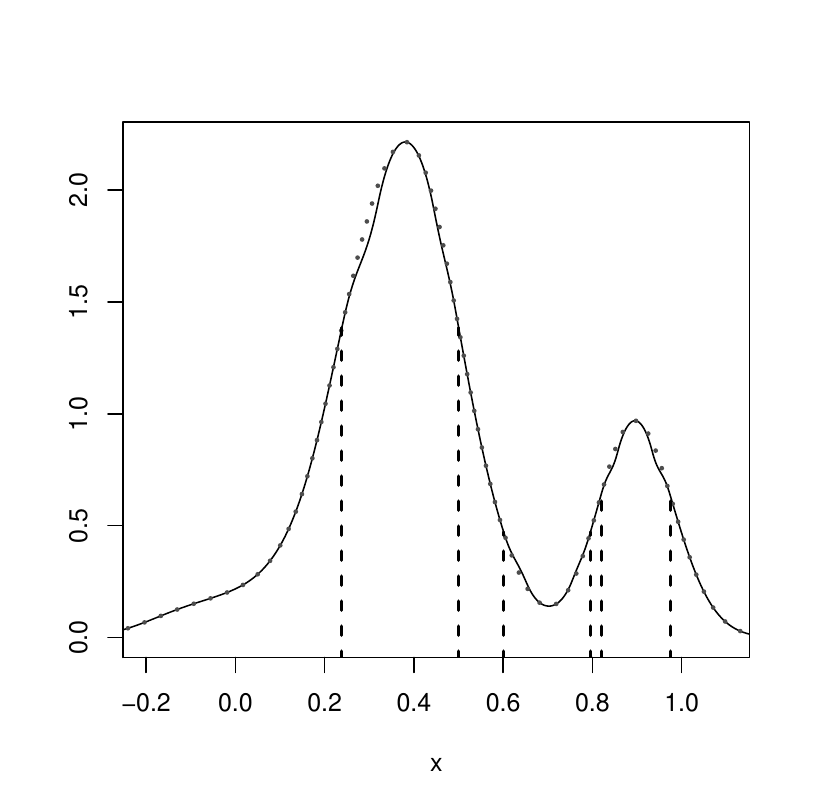}
    \includegraphics[width=0.44\textwidth]{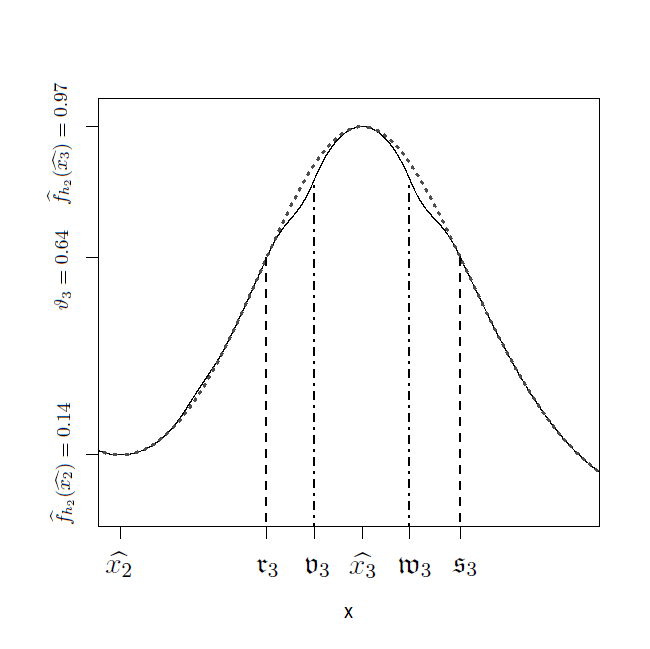} \\
 \caption{Sample of $n=1000$ observations from model M16. Dotted grey line: $\widehat{f}_{h_2}$. Solid line: $g(\cdot;h_k,h_{\tiny{\mbox{PI}}},(0.4,0.4,0.4))$. Dashed line: support of $J(\cdot;\widehat{x_i},h_2,h_{\tiny{\mbox{PI}}},0.4)$, with $i=1,2,3$, Dot--dashed line: support of $\mathcal{K}(\cdot;\boldsymbol{\rho}_3,\eta_3)$. Left: in the support $(-0.2,1.1)$. Right: in a neighbourhood of the mode $\widehat{x}_{3}$.}
 \label{figjfunc}
\end{figure}

{\footnotesize
\begin{equation}\label{gfunc}
g(x;h_k,h_{\tiny{\mbox{PI}}},\boldsymbol{\varsigma})= 
  \begin{cases}
	J(x;\widehat{x_i},h_k,h_{\tiny{\mbox{PI}}},\varsigma_i) &\mbox{ if } x\in (\mathfrak{r}_i,\mathfrak{s}_i) \mbox{ for some } i \in \{1,\ldots, (2k-1)\},  \\
	L(x;z_{(2p-1)},z_{(2p)},h_k) &\mbox{ if } x\in (z_{(2p-1)},z_{(2p)}) \mbox{ for some } p \in \{1,\ldots, t\},\\
		&\mbox{ and }  \zeta_p \notin (\mathfrak{r}_i,\mathfrak{s}_i) \mbox{ for any } i \in \{1,\ldots, (2k-1)\}, \\
	\widehat{f}_{h_k}(x) &\mbox{ otherwise.}\\
	\end{cases}
\end{equation}
}
\hspace*{-1.8mm}In (\ref{gfunc}), $\boldsymbol{\varsigma}$ has $k$ elements $\varsigma_i\in (0,1/2)$, with $i=1,\ldots,k$, determining at which height of the kernel density estimation the modification is done. Values of $\varsigma_i$ close to 0 imply a modification in a ``small'' neighbourhood around the mode or antimode. Note that a little abuse of notation was made as $g$ will depend on the function $\widehat{f}_{h_k}$ (not only on $h_k$) and on the values $\widehat{f}_{h_{\tiny{\mbox{PI}}}}''(\widehat{x_i})$, for $i =1,\ldots, (2k-1)$. An example of the effect of $g$ can be seen in Figure \ref{figjfunc}. {As showed in the Proposition~\ref{th2}, from this calibration function $g$, an asymptotic correct behaviour of our test can be obtained if the critical bandwidth satisfies the following condition.}

\paragraph{Critical bandwidth condition}\label{cond} (CBC) The critical bandwidth $h_k$ satisfies that $a_n \leq h_k \leq b_n$, eventually with probability one, being $a_n$ and $b_n$ two sequences of positive numbers such as $b_n\rightarrow 0$ and $n a_n/ \log n \rightarrow \infty$. 

\begin{proposition}\label{th2}
{Let $g$ be defined as in (\ref{gfunc}), and where the functions $J$ and $L$ are defined as in (\ref {jfunc}) and (\ref{Lgranfunc}). If $h_k$ verifies (CBC), then $g$ satisfies the conditions of Theorem \ref {th1}.}
\end{proposition}

\begin{remark}\label{remcbw5}
From the proof of Proposition~\ref{th2}, the reason for not using just a kernel density estimation with the critical bandwidth can be derived. Under some conditions (see Supplementary Material), the critical bandwidth is of order $n^{-1/5}$ and this order is not enough to guarantee that $\widehat{f}_{h_k}''(\widehat{x_i})$ will converge in probability to $f''(x_i)$.
\end{remark}

The remaining part of this section will be devoted to further describe the construction of this calibration function $g$ and two final remarks will be provided. 

Before defining functions $J$ and $L$, to ensure that $g$ has continuous derivative, a link function $l$ must be introduced:
\begingroup\makeatletter\def\f@size{7}\check@mathfonts
\begin{eqnarray}\label{lfunc}
l(x;u,v,a_0,a_1,b_0,b_1) &=& \frac{a_0-a_1}{2} \left( 1+2 \left( \frac{x-u}{v-u} \right)^3 - 3 \left( \frac{x-u}{v-u} \right)^2 \right) \exp \left({\frac{2 (x-u)b_0}{a_0-a_1}}\right) + \nonumber\\
&+&  \frac{a_0-a_1}{2}\left( 2 \left( \frac{x-u}{v-u} \right)^3 - 3 \left( \frac{x-u}{v-u} \right)^2 \right) \exp \left({\frac{2 (v-x)b_1}{a_0-a_1}}\right) + \frac{a_0+a_1}{2}, \nonumber\\
\end{eqnarray}\endgroup
where $a_0 \neq a_1$ and $v>u$. Two issues must be noticed in this function. First, it allows a smooth connection between two functions, being $u$ and $v$ the starting and ending points where the link function is used, $a_0$ and $a_1$ the values of the connected functions on these points and $b_0$ and $b_1$ their first derivative values. Second, if the signs of $b_0$, $b_1$ and $(a_1-a_2)$ are the same, then the first derivative of $l$ will not be equal to 0 for any point inside $[u,v]$.

The form of $J$ is given in equation (\ref{jfunc}) and its construction guarantees that $\widehat{x_i}$ is the unique point in which the derivative is equal to 0 in the neighbourhood where it is defined. The construction of $J$ is achieved with the $\mathcal{K}$ function defined bellow and properly linked with the link function (\ref{lfunc}) to preserve (RC1). The $\mathcal{K}$ function is defined as follows
\begin{equation*}
\mathcal{K}(x; \widehat{x_i},\mathfrak{p}_i,\mathfrak{q}_i,\eta_i)=\mathfrak{p}_i\left( 1+ \delta_i \left( \frac{x-\widehat{x_i}}{\eta_i} \right)^2 \right)^{\eta_i^2 \frac{\delta_i \cdot \mathfrak{q}_i}{2 \mathfrak{p}_i}}, 
\end{equation*}
being $\delta_i$ a value indicating if $\widehat{x_i}$ is a mode ($\delta_i=-1$) or an antimode ($\delta_i=1$). The value $\eta_i$ will be defined later and it will depend on $\varsigma_i$. The second derivative of this function exists and is H\"older continuous in $(\widehat{x_i}-\eta_i/2, \widehat{x_i}+\eta_i/2)$. The following equalities are also satisfied: $\mathcal{K}(\widehat{x_i};\widehat{x_i},\mathfrak{p}_i,\mathfrak{q}_i,\eta_i)=\mathfrak{p}_i$ and $\mathcal{K}''(\widehat{x_i};\widehat{x_i},\mathfrak{p}_i,\mathfrak{q}_i,\eta_i)=\mathfrak{q}_i$. Then, denoting as $\boldsymbol{\rho}_i=(\widehat{x_i},\widehat{f}_{h_k}(\widehat{x_i}),\widehat{f}_{h_{\tiny{\mbox{PI}}}}''(\widehat{x_i}))$, the $J$ function can be defined as follows
\begingroup\makeatletter\def\f@size{6.4}\check@mathfonts
\begin{equation} \label{jfunc}
J(x;\widehat{x_i},h_k,h_{\tiny{\mbox{PI}}},\varsigma_i) = 
  \begin{cases}
     l\left(x;\mathfrak{r}_i,\mathfrak{v}_i, \widehat{f}_{h_k}(\mathfrak{r}_i),\mathcal{K}(\mathfrak{v}_i; \boldsymbol{\rho}_i,\eta_i), \widehat{f}'_{h_k}(\mathfrak{r}_i), \mathcal{K}'(\mathfrak{v}_i;\boldsymbol{\rho}_i,\eta_i) \right) &\mbox{ if } x\in(\mathfrak{r}_i,\mathfrak{v}_i), \\
		\mathcal{K}(x; \boldsymbol{\rho}_i,\eta_i)  &\mbox{ if } x\in[\mathfrak{v}_i, \mathfrak{w}_i], \\
		 l\left(x;\mathfrak{w}_i,\mathfrak{s}_i,\mathcal{K}(\mathfrak{w}_i;\boldsymbol{\rho}_i,\eta_i), \widehat{f}_{h_k}(\mathfrak{s}_i),\mathcal{K}'(\mathfrak{w}_i;\boldsymbol{\rho}_i,\eta_i), \widehat{f}'_{h_k}(\mathfrak{s}_i)\right) &\mbox{ if } x\in(\mathfrak{w}_i,\mathfrak{s}_i), \\
  \end{cases}
\end{equation}\endgroup
being $\mathfrak{v}_i=\widehat{x_i}- \eta_i/2$ and $\mathfrak{w}_i=\widehat{x_i}+ \eta_i/2$. As it was mentioned, the function $J$ described in (\ref{jfunc}) (and hence also the calibration function $g$) depends on the constant $\varsigma_i \in (0,1/2)$. Ordering the modes and denoting as $\widehat{x}_0=-\infty$ and $\widehat{x}_{(2k)}=\infty$, that is $-\infty=\widehat{x}_0<\widehat{x}_1<\ldots<\widehat{x}_{2k-1}<\widehat{x}_{2k}=\infty$, the remaining unknowns values in (\ref{jfunc}) will be obtained as follows. First, it is necessary to decide at which height $\vartheta_i$ the modification in $\widehat{f}_{h_k}$ is done. For values of $\varsigma_i$ close to 0, $\vartheta_i$ will be close to $\widehat{f}_{h_k}(\widehat{x}_i)$; while for values close to $0.5$, $\vartheta_i$ will be in the middle point between $\widehat{f}_{h_k}(\widehat{x}_i)$ and the highest (or lowest if $\widehat{x}_i$ is an antimode) value of $\widehat{f}_{h_k}$ in the two closest modes or antimodes ($\widehat{x}_{i-1}$ and $\widehat{x}_{i+1}$). Second, once the height is decided, $\mathfrak{r}_i$ and $\mathfrak{s}_i$ will be the left and the right closest points to $\widehat{x}_i$ at which the density estimation is equal to $\vartheta_i$. Third, in order to link correctly the $\mathcal{K}$ function, it is necessary to define $\eta_i$ ensuring that $\mathcal{K}(\widehat{x_i}\pm\eta_i/2; \boldsymbol{\rho}_i,\eta_i)$ will be higher (lower in the antimodes) than $\vartheta_i$. With this objective, $\eta_i$ is chosen in such a way that $\mathcal{K}(\widehat{x_i}\pm\eta_i/2; \boldsymbol{\rho}_i,\eta_i)$ is near $\widehat{f}_{h_k}(\widehat{x_i})$ and as close as possible to the middle point between $\vartheta_i$ and $\widehat{f}_{h_k}(\widehat{x_i})$. Also, the value $\eta_i$ will ensure that the neighbourhood $[\mathfrak{v}_i, \mathfrak{w}_i]$ in which $\mathcal{K}$ is defined is inside $(\mathfrak{r}_i,\mathfrak{s}_i)$. Finally, $\widehat{f}'_{h_k}$ must be different to 0 in the four points ($\mathfrak{r}_i$, $\mathfrak{v}_i$, $\mathfrak{w}_i$ and $\mathfrak{s}_i$) where the two link functions are employed. An example of the modifications achieved by the $J$ function in the modes and antimodes of $\widehat{f}_{h_k}$ is shown in Figure \ref{figjfunc} and the complete characterization is provided bellow
\begingroup\makeatletter\def\f@size{9}\check@mathfonts
\begin{eqnarray}\label{variabJ}
 \vartheta_i&=& \widehat{f}_{h_k}(\widehat{x_i}) + \delta_i \cdot \varsigma_i \cdot \min\left(|\widehat{f}_{h_k}(\widehat{x_i}) - \widehat{f}_{h_k}(\widehat{x}_{i-1})|,|\widehat{f}_{h_k}(\widehat{x_i}) - \widehat{f}_{h_k}(\widehat{x}_{i+1})|\right), \nonumber\\
\mathfrak{r}_i&=&\inf\{x: x>\widehat{x}_{i-1}, \delta_i \cdot \widehat{f}_{h_k}(x)\leq  \delta_i \cdot \vartheta_i  \mbox{ and } \widehat{f}'_{h_k}(x)\neq 0\}, \nonumber\\
\mathfrak{s}_i&=&\sup\{x: x<\widehat{x}_{i+1}, \delta_i \cdot \widehat{f}_{h_k}(x)\leq \delta_i \cdot \vartheta_i \mbox{ and } \widehat{f}'_{h_k}(x)\neq 0 \} , \nonumber\\
\eta_i&=& \sup \{\gamma: \gamma \in (0,\min(\widehat{x_{i}}-\mathfrak{r}_i,\mathfrak{s}_i-\widehat{x_{i}})), \delta_i \mathcal{K}(\widehat{x_i}+\gamma/2; \boldsymbol{\rho}_i,\gamma) \leq \delta_i (\widehat{f}_{h_k}(\widehat{x_i}) +\vartheta_i)/2  \nonumber\\
&&\quad \quad \mbox{and }  \widehat{f}'_{h_k}(\widehat{x_i}\pm\gamma/2)\neq 0\}. \nonumber\\
\end{eqnarray}
\endgroup

In order to proceed with the modification achieved with the $L$ function, assume that this estimator has $t$ saddle points $\zeta_p$, with $p=1,\ldots,t$. Define as $\xi=\min \{|x-y|: x,y \in (\zeta_1,\ldots,\zeta_t) \cup (\mathfrak{r}_1,\mathfrak{s}_1,\ldots, \mathfrak{r}_{2k-1},\mathfrak{s}_{2k-1}) \}$. Then, if $\zeta_p$ is not inside the interval where the $J$ funtions are defined, the neighbourhood used to remove the stationary and turning points will be delimited by $z_{(2p-1)}=\zeta_p-\varpi  \xi$ and $z_{(2p)}=\zeta_p+\varpi  \xi$, with $\varpi \in (0,1/4)$. In the simulation study, the value of $\varpi$ will be taken close enough to 0 to avoid an impact in the value of the integral associated to $g$. Once these points are calculated, the saddle points can be removed from $g$ with the link function by taking $L$ equal to
\begingroup\makeatletter\def\f@size{7}\check@mathfonts
\begin{equation}\label{Lgranfunc}
L(x;z_{(2p-1)},z_{(2p)},h_k)=l(x;z_{(2p-1)},z_{(2p)},\widehat{f}_{h_k}(z_{(2p-1)}),\widehat{f}_{h_k}(z_{(2p)}),\widehat{f}'_{h_k}(z_{(2p-1)}),\widehat{f}'_{h_k}(z_{(2p)})).
\end{equation} \endgroup

To construct the calibration function, first, values of $\varsigma_i \in (0,1/2)$, for $i \in \{1,\ldots,(2k-1)\}$ must be fixed. Then, using the $J$ function (\ref{jfunc}) with the values given in (\ref{variabJ}) and the $L$ function (\ref{Lgranfunc}), the function $g$ defined in (\ref{gfunc}) satisfies the specified regularity conditions and ${|g''(\widehat{x_i};h_k,h_{\tiny{\mbox{PI}}},\boldsymbol{\varsigma})|}/{g^3(\widehat{x_i};h_k,h_{\tiny{\mbox{PI}}},\boldsymbol{\varsigma})}$ converges in probability to $d_i$. With this modification the calibration function also preserves the structure of the data under the hypothesis that $f$ has $k$ modes. However, this calibration function $g$ may not be a density function since
\begin{equation}
q(\boldsymbol{\varsigma})=  \int_{-\infty}^{\infty} g(x;h_k,h_{\tiny{\mbox{PI}}},\boldsymbol{\varsigma}) dx
\end{equation}
may not be equal to 1. To ensure that $g$ is indeed a density, a possible approach consists in proceeding with a search of values for $\boldsymbol{\varsigma}$ such that $q(\boldsymbol{\varsigma})$ is equal to 1. {It can be seen that the convergence of this algorithm is guaranteed just considering ``small enough'' neighbourhoods (where the $J$ function is applied), so that the calibration function is ``almost equal'' to the kernel density estimation which integral over the entire space is equal to one, that is, }
$$
\lim_{\varsigma_i\rightarrow 0^+; \forall i\in (1,\ldots,2k-1)} q(\boldsymbol{\varsigma})=\int_{-\infty}^{\infty} \widehat{f}_{h_k}(x) dx =1.
$$ 
For convenience, in the simulation study, the employed approach will be followed using $\varsigma_i$ close enough to 0 ($\forall i\in \{1,\ldots,2k-1\}$) in order to avoid an impact on the integral value.

\begin{remark}
Under some regularity conditions, when $f$ is is twice continuously differentiable, a sufficient condition for the convergence in distribution of $n^{1/5}h_j$ is obtained when $f$ has a bounded support or when employing \cite{HallYork01} critical bandwidth (see Remark \ref{MHY} in Supplementary Material). Then, ``better'' asymptotic results are expected when using their critical bandwidth. Although our proposal presents satisfactory results even if the support is unbounded (as it can be seen in  Section \ref{simulation}), if the modes and antimodes lie in a known closed interval $I$, ${h_{\tiny{\mbox{HY}},k}}$ can be employed. An alternative approach for this case is given in Section \ref{approachA3} in Supplementary Material. After obtaining a conclusion about the number of modes, when the objective is to estimate their location, it should noted that, under some regularity conditions, the modes and antimodes of $\hat{f}_{h_{\tiny{\mbox{HY}},k}}$ will provide a good estimation of their locations.
\end{remark}

\begin{remark}
It should be also reminded that, unlike \cite{Silverman81} and \cite{FisMar01} proposals, the one presented in this paper considers $H_0:j=k$ instead of $H_0:j \leq k$. A deeper insight is presented in Section \ref{testbimuni} in Supplementary Material. 
\end{remark}

\section{Simulation study}
\label{simulation}
The aim of the following simulation study is to compare the different proposals presented in Section \ref{background}. Samples of size $n=50$, $n=200$ and $n=1000$ ($n=100$ instead of $n=1000$ in power studies) were drawn from twenty five different distributions, ten of them unimodal (M1--M10), ten bimodal (M11--M20) and five trimodal (M21--M25) (see Section \ref{modelos} in Supplementary Material). For each choice of sampling distribution and sample size, 500 realizations of the sample were generated. Conditionally on each of those samples, for testing purposes, 500 resamples of size $n$ were drawn from the population. Tables \ref{estsim2a}--\ref{estsim6} report the percentages of rejections for significance levels $\alpha=0.01$, $\alpha=0.05$ and $\alpha=0.10$ under different scenarios: testing unimodality vs. multimodality (Tables \ref{estsim2a} and \ref{estsim2b}); testing bimodality against more than two modes (Table \ref{estsim5}) and power analysis (respectively Tables \ref{estsim3} and \ref{estsim6}). The procedures considered include the proposals by \citet{Silverman81} (SI), \citet{FisMar01} (FM), \citet{HallYork01} (HY), \cite{Hartigan85} (HH), \citet{ChengHall98} (CH) and the new proposal (NP) in this paper. Note that for testing $H_0:\,j = 2$, only SI, FM and NP can be compared. For the critical bandwidth test HY, the two proposed methods for computing $\lambda_{\alpha}$ have been tried, with very similar results. The ones reported in this section correspond to a polynomial approximation for $\lambda_{\alpha}$. $I=[0,1]$ is used both for HY and for NP, when the interval containing the modes is assumed to be known (Table \ref{estsim7}). Further computational details are included in Section \ref{numericas} in Supplementary Material.

\textbf{Testing unimodality vs. multimodality.} From the results reported in Tables \ref{estsim2a} and \ref{estsim2b}, it can be concluded that SI is quite conservative: even for high sample sizes, the percentage of rejections is below the significance level, and quite close to 0 even for $\alpha=0.10$. Regarding FM, a systematic behaviour cannot be concluded: the percentage of rejections is above the significance level for models M1, M5, M7, M9 or M10, but it can be also below the true level for M2, M4 or M8. 


The behaviour of HY is quite good when using $I=[0,1]$ for the different distributions and large sample sizes. For $n=1000$, the percentage of rejections is quite close to $\alpha$, except for model M5 (for $\alpha=0.05$, below level) and for models M6 and M7 (for $\alpha=0.10$, above level). However, the percentage of rejections is usually below the significance level for small sample sizes. Exceptions to this general pattern are found for model M1 ($n=200$), M3 ($n=50$) and M10 ($n=200$), where percentage of rejections is close to $\alpha$ and models M3 ($n=200$), M6 ($n=200$) and M7 with percentages above $\alpha$. Nevertheless, it should be kept in mind that the support where unimodality is tested must be known. Similarly to SI, the results obtained with HH are quite conservative. For instance, for $n=1000$, even taking $\alpha=0.10$, the percentage of rejections is always below $0.002$.

Calibration seems correct in simple models for CH, although slightly conservative in some cases, such as for models M4 ($n=1000$), M5 ($n=200$) and M8 ($n=200$). As expected, the parametric calibration distributions do not capture, for example, the skewness and this affects the second--order properties in more complex models. This effect is reflected in the asymmetric M3, M7 and M10 ($n=1000$), or model M9, where the percentage of rejections is below $\alpha$, and for M6 where is considerably higher than the significance level.  

Finally, regarding the new proposal NP, it can be concluded that the calibration is quite satisfactory, even for complicated models, with a slightly conservative performance for M3 ($n=200$), M4 ($n=1000$) or M7 ($n=200$), being this effect more clear for model M9. The only scenario where the percentage of rejections is above $\alpha$ is for M6 with $n=200$, but this behaviour is corrected when increasing the sample size. Although the performance is better for higher sample sizes, in some cases, such as M9 or M16 (in the bimodal case), it can be seen that even for $n=1000$, a percentage of rejections close to $\alpha$ is hard to get. In this difficult cases, the knowledge of the support can be used for obtaining better results as it was reported in Table \ref{estsim7}, where the percentage of rejections is close to $\alpha$ for the sample sizes $n=200$ and $n=1000$.

Regarding power behaviour (and just commenting on the three methods which exhibit a correct calibration), results are reported in Table \ref{estsim3}: none of the proposals is clearly more powerful. For instance, for M13, HY clearly detects the appearance of the second small mode, whereas the other approaches do not succeed in doing so. For M11, M12, M14 and M15 ($n=50$), CH presents the highest empirical power and HY shows the lowest one.

\textbf{Assessing bimodality.} For testing $H_0: j= 2$, \citet{HallYork01} prove that, even knowing the density support, SI cannot be consistently calibrated by a bootstrap procedure, similar to the one used for the unimodality test. The conservative behaviour, observed in the unimodality test, is also perceived in most cases. But also, when testing bimodality, there is a model where the percentage of rejections is considerably higher than the significance level, M20, being this bimodal scenario similar to the conservative M19, just generating some outliers. FM presents again an erratic behaviour: for M17 (except for $n=200$), M18 or M19, the percentage of rejections is below $\alpha$, whereas the opposite happens for M11, M12, M15, M16 or M20 (except for $n=50$).

For testing bimodality, NP presents good results. The percentage of rejections is close to the significance level, except for M12 ($n=200$) and M13 ($n=200$), slightly below $\alpha$, and M11 ($n=50$), M15 ($n=50$, $n=200$), M16 and M19 ($n=50$), slightly above $\alpha$. For $n=1000$, all the results are good except for M16, but the calibration problem is corrected (as seen in Table \ref{estsim7}) applying NP with known support, taking for that purpose $I=[0,1]$. So, just the new proposal presents a correct calibration. Hence, power results reported in Table \ref{estsim6} are only judged for the new proposal: power increases with sample size, detecting that all the alternative distributions do not satisfy the null hypothesis, except for M21 ($n=50$) and M24 ($n=50$).

\setlength{\tabcolsep}{2.2pt}
\setlength\extrarowheight{0.3pt}

\begin{table}
\begin{center}
\scalebox{0.55}{
\begin{tabular}{|c|c |c c c | c c c |c c c |}
\hline
   &$\alpha$ & 0.01 & 0.05& 0.10& 0.01 & 0.05& 0.10&0.01 & 0.05& 0.10\\ \hline
M1 & &\multicolumn{3}{c|}{SI}& \multicolumn{3}{c|}{FM}& \multicolumn{3}{c|}{HY} \\  \cline{3-11}
 \multirow{7}{*}{\includegraphics[width=23mm]{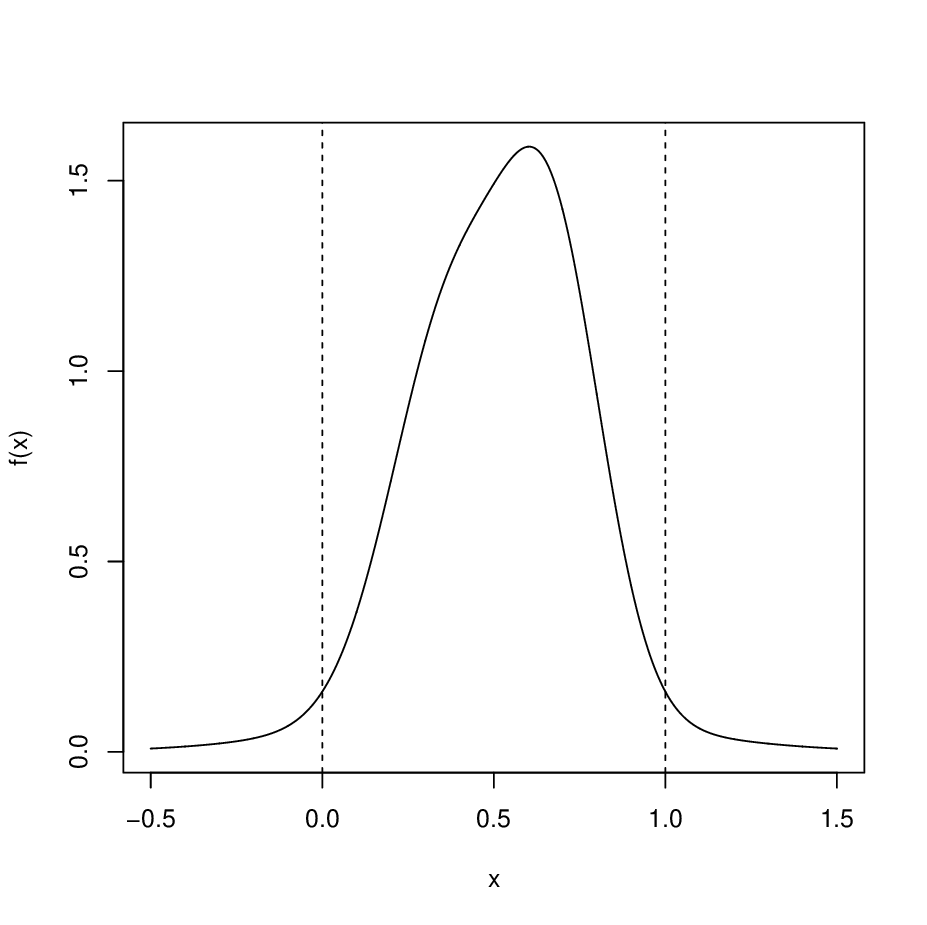}} &  $n=50$ &  0(0) & 0(0) & 0(0) & 0.010(0.009) & 0.076(0.023) & 0.178(0.034) & 0(0) & 0.022(0.013) & 0.050(0.019) \\ 
&  $n=200$ & 0(0) & 0(0) & 0(0) & 0.056(0.020) & 0.162(0.032) & 0.262(0.039) & 0.002(0.004) & 0.046(0.018) & 0.090(0.025) \\ 
&  $n=1000$ & 0(0) & 0(0) & 0(0) & 0.036(0.016) & 0.126(0.029) & 0.210(0.036) & 0.002(0.004) & 0.052(0.019) & 0.096(0.026) \\   \cline{2-11}
 & & \multicolumn{3}{c|}{HH}& \multicolumn{3}{c|}{CH}&\multicolumn{3}{c|}{NP} \\ \cline{3-11}

&  $n=50$ & 0(0) & 0.006(0.007) & 0.022(0.013) & 0.022(0.013) & 0.072(0.023) & 0.140(0.030) & 0.010(0.009) & 0.064(0.021) & 0.120(0.028)   \\ 
&  $n=200$ & 0(0) & 0.002(0.004) & 0.002(0.004) & 0.014(0.010) & 0.058(0.020) & 0.122(0.029) & 0.010(0.009) & 0.044(0.018) & 0.120(0.028)  \\ 
&  $n=1000$ & 0(0) & 0(0) & 0(0) & 0.006(0.007) & 0.048(0.019) & 0.104(0.027) & 0.008(0.008) & 0.052(0.019) & 0.108(0.027) \\ \hline

M2 & &\multicolumn{3}{c|}{SI}& \multicolumn{3}{c|}{FM}& \multicolumn{3}{c|}{HY} \\ \cline{3-11}
 \multirow{7}{*}{\includegraphics[width=23mm]{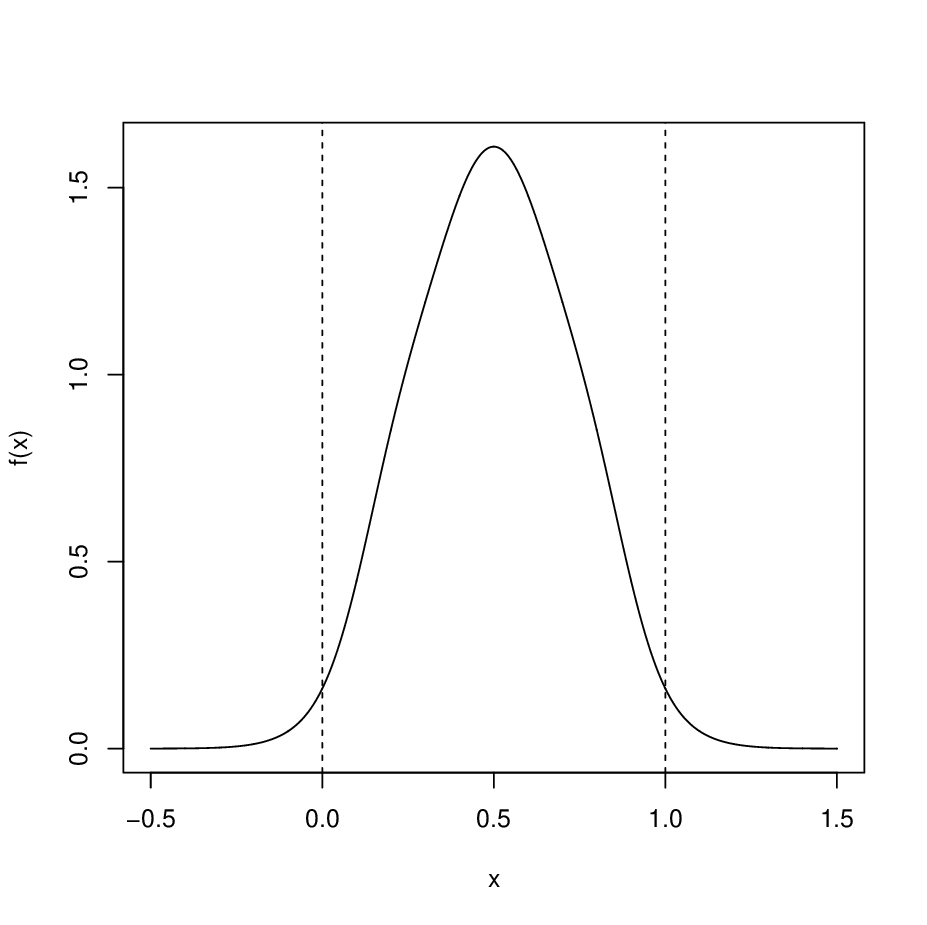}} & $n=50$ & 0(0) & 0(0) & 0(0) & 0.004(0.006) & 0.040(0.017) & 0.076(0.023) & 0(0) & 0.024(0.013) & 0.068(0.022) \\ 
&  $n=200$ & 0(0) & 0(0) & 0(0) & 0(0) & 0.006(0.007) & 0.056(0.020) & 0(0) & 0.030(0.015) & 0.082(0.024) \\ 
&  $n=1000$ & 0(0) & 0(0) & 0.004(0.006) & 0(0) & 0.014(0.010) & 0.040(0.017) & 0.004(0.006) & 0.038(0.017) & 0.080(0.024)  \\  \cline{2-11}
 & & \multicolumn{3}{c|}{HH}& \multicolumn{3}{c|}{CH}&\multicolumn{3}{c|}{NP} \\ \cline{3-11}

&  $n=50$   &  0(0) & 0.012(0.010) & 0.028(0.014) & 0.046(0.018) & 0.100(0.026) & 0.140(0.030) & 0.016(0.011) & 0.070(0.022) & 0.122(0.029) \\ 
&  $n=200$  & 0(0) & 0.002(0.004) & 0.004(0.006) & 0.020(0.012) & 0.074(0.023) & 0.164(0.032) & 0.004(0.006) & 0.050(0.019) & 0.114(0.028)  \\ 
&  $n=1000$ & 0(0) & 0(0) & 0(0) & 0.008(0.008) & 0.032(0.015) & 0.092(0.025) & 0.006(0.007) & 0.030(0.015) & 0.082(0.024) \\  \hline

 M3& &\multicolumn{3}{c|}{SI}& \multicolumn{3}{c|}{FM}& \multicolumn{3}{c|}{HY} \\ \cline{3-11}
\multirow{7}{*}{\includegraphics[width=23mm]{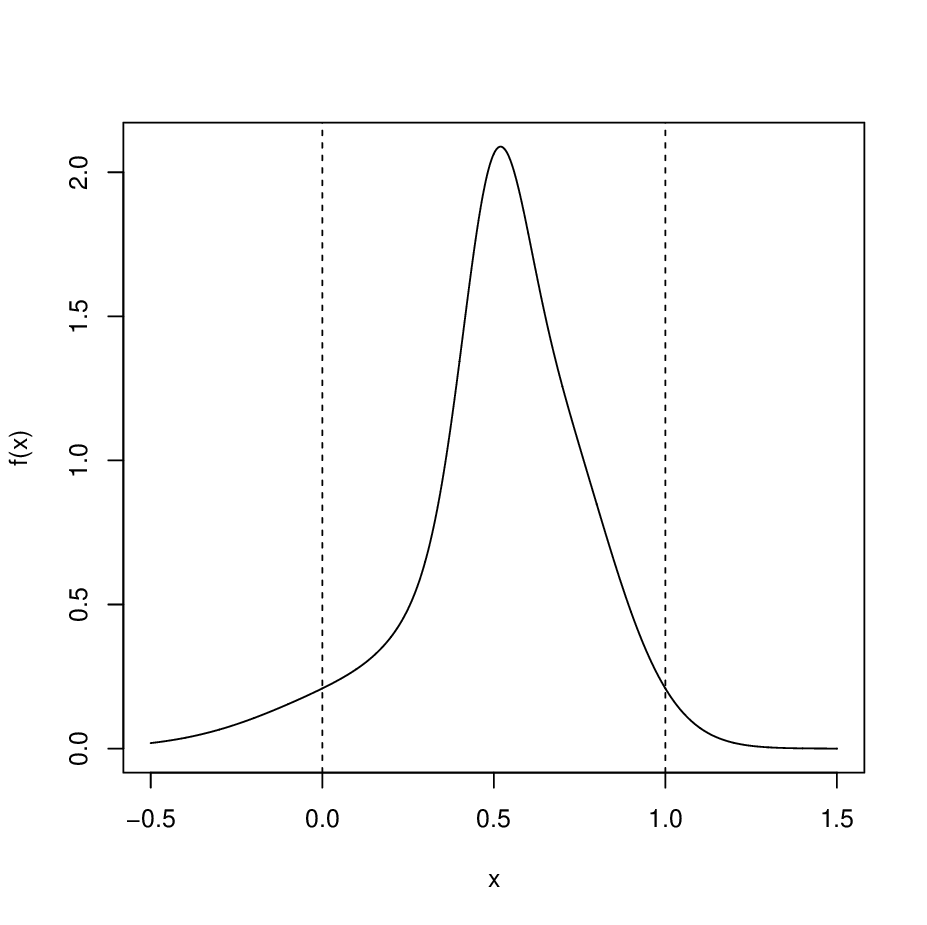}}& $n=50$   & 0(0) & 0(0) & 0(0) & 0.026(0.014) & 0.112(0.028) & 0.222(0.036) & 0.008(0.008) & 0.066(0.022) & 0.108(0.027) \\ 
&  $n=200$ & 0(0) & 0(0) & 0(0) & 0.014(0.010) & 0.072(0.023) & 0.146(0.031) & 0.030(0.015) & 0.088(0.025) & 0.146(0.031) \\ 
&  $n=1000$ & 0(0) & 0(0) & 0(0) & 0.002(0.004) & 0.050(0.019) & 0.128(0.029) & 0.018(0.012) & 0.070(0.022) & 0.120(0.028) \\  \cline{2-11}
 & &  \multicolumn{3}{c|}{HH}& \multicolumn{3}{c|}{CH}&\multicolumn{3}{c|}{NP} \\ \cline{3-11}

&  $n=50$  & 0(0) & 0(0) & 0.004(0.006) & 0.002(0.004) & 0.032(0.015) & 0.056(0.020) & 0.004(0.006) & 0.042(0.018) & 0.078(0.024) \\ 
&  $n=200$  & 0(0) & 0(0) & 0.002(0.004) & 0.002(0.004) & 0.004(0.006) & 0.030(0.015) & 0.002(0.004) & 0.022(0.013) & 0.054(0.020) \\ 
&  $n=1000$ & 0(0) & 0(0) & 0(0) & 0.002(0.004) & 0.012(0.010) & 0.032(0.015) & 0.006(0.007) & 0.032(0.015) & 0.082(0.024) \\  \hline

 M4 &&\multicolumn{3}{c|}{SI}& \multicolumn{3}{c|}{FM}& \multicolumn{3}{c|}{HY} \\ \cline{3-11}
\multirow{7}{*}{\includegraphics[width=23mm]{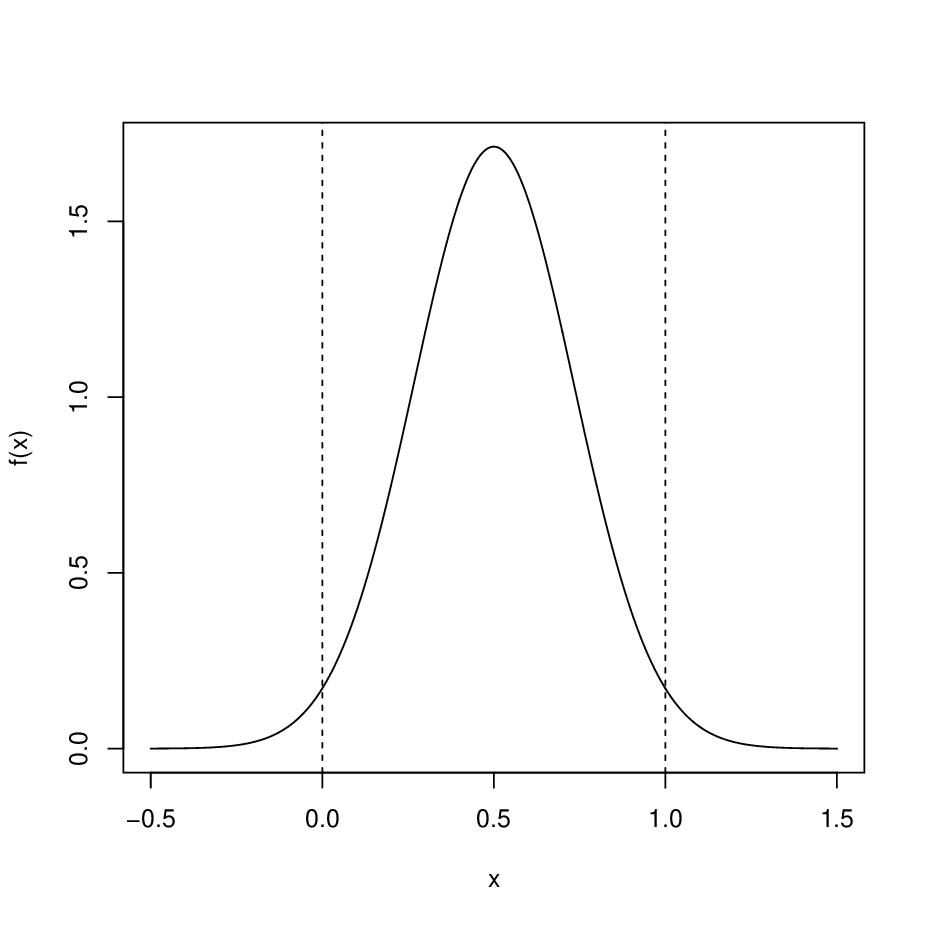}}& $n=50$  & 0(0) & 0(0) & 0(0) & 0.002(0.004) & 0.018(0.012) & 0.060(0.021) & 0(0) & 0.020(0.012) & 0.050(0.019) \\
&  $n=200$ & 0(0) & 0(0) & 0(0) & 0(0) & 0.012(0.010) & 0.044(0.018) & 0.004(0.006) & 0.026(0.014) & 0.074(0.023) \\
&  $n=1000$ & 0(0) & 0.002(0.004) & 0.002(0.004) & 0(0) & 0.010(0.009) & 0.046(0.018) & 0.008(0.008) & 0.052(0.019) & 0.090(0.025) \\  \cline{2-11}
 & &  \multicolumn{3}{c|}{HH}& \multicolumn{3}{c|}{CH}&\multicolumn{3}{c|}{NP} \\ \cline{3-11}

&  $n=50$ & 0(0) & 0.004(0.006) & 0.018(0.012) & 0.016(0.011) & 0.064(0.021) & 0.118(0.028) & 0.014(0.010) & 0.050(0.019) & 0.102(0.027) \\ 
&  $n=200$ & 0(0) & 0(0) & 0(0) & 0.008(0.008) & 0.032(0.015) & 0.082(0.024) & 0.004(0.006) & 0.030(0.015) & 0.080(0.024) \\ 
&  $n=1000$  & 0(0) & 0(0) & 0(0) & 0.004(0.006) & 0.034(0.016) & 0.066(0.022) & 0(0) & 0.028(0.014) & 0.066(0.022)\\  \hline
	
 M5& &\multicolumn{3}{c|}{SI}& \multicolumn{3}{c|}{FM}& \multicolumn{3}{c|}{HY} \\ \cline{3-11}
\multirow{7}{*}{\includegraphics[width=23mm]{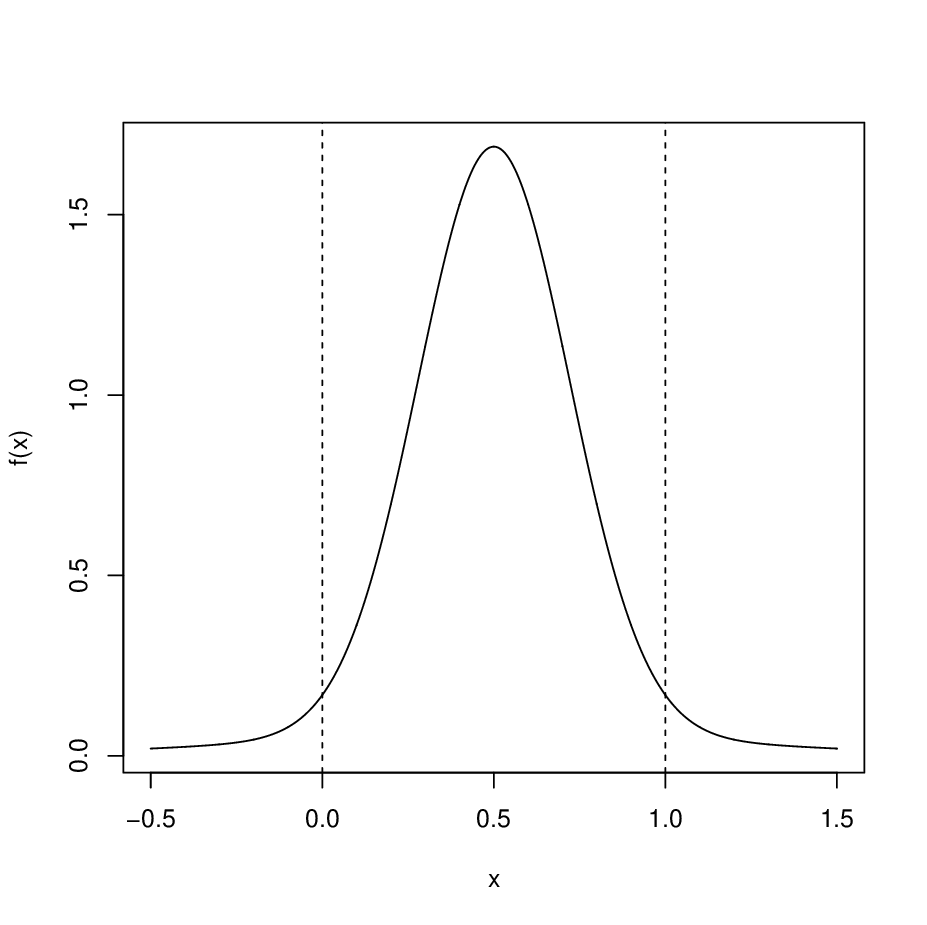}}& $n=50$   & 0(0) & 0(0) & 0(0) & 0.186(0.034) & 0.366(0.042) & 0.494(0.044) & 0(0) & 0.006(0.007) & 0.038(0.017) \\   
&  $n=200$ & 0(0) & 0(0) & 0(0) & 0.268(0.039) & 0.500(0.044) & 0.612(0.043) & 0.002(0.004) & 0.030(0.015) & 0.074(0.023) \\  
&  $n=1000$ & 0(0) & 0(0) & 0(0) & 0.210(0.036) & 0.380(0.043) & 0.504(0.044) & 0.006(0.007) & 0.028(0.014) & 0.080(0.024) \\  \cline{2-11}
 & &  \multicolumn{3}{c|}{HH}& \multicolumn{3}{c|}{CH}&\multicolumn{3}{c|}{NP} \\ \cline{3-11}

&  $n=50$ & 0(0) & 0(0) & 0.006(0.007) & 0.004(0.006) & 0.052(0.019) & 0.084(0.024) & 0.006(0.007) & 0.062(0.021) & 0.106(0.027) \\ 
&  $n=200$ & 0(0) & 0(0) & 0(0) & 0.010(0.009) & 0.034(0.016) & 0.064(0.021) & 0.012(0.010) & 0.050(0.019) & 0.092(0.025)  \\ 
&  $n=1000$ & 0(0) & 0(0) & 0(0) & 0.006(0.007) & 0.022(0.013) & 0.082(0.024) & 0.006(0.007) & 0.052(0.019) & 0.106(0.027) \\  \hline

\hline

\end{tabular}
}
\caption{Percentages of rejections for testing $H_0:j=1$, with $500$ simulations ($1.96$ times their estimated standard deviation in parenthesis) and $B = 500$ bootstrap samples.}
\label{estsim2a}
\end{center}
\end{table}

\begin{table}
\begin{center}
\scalebox{0.55}{
\begin{tabular}{|c|c |c c c | c c c |c c c |}
\hline
  &$\alpha$ & 0.01 & 0.05& 0.10& 0.01 & 0.05& 0.10&0.01 & 0.05& 0.10\\ \hline
M6 & &\multicolumn{3}{c|}{SI}& \multicolumn{3}{c|}{FM}& \multicolumn{3}{c|}{HY} \\ \cline{3-11}
\multirow{7}{*}{\includegraphics[width=23mm]{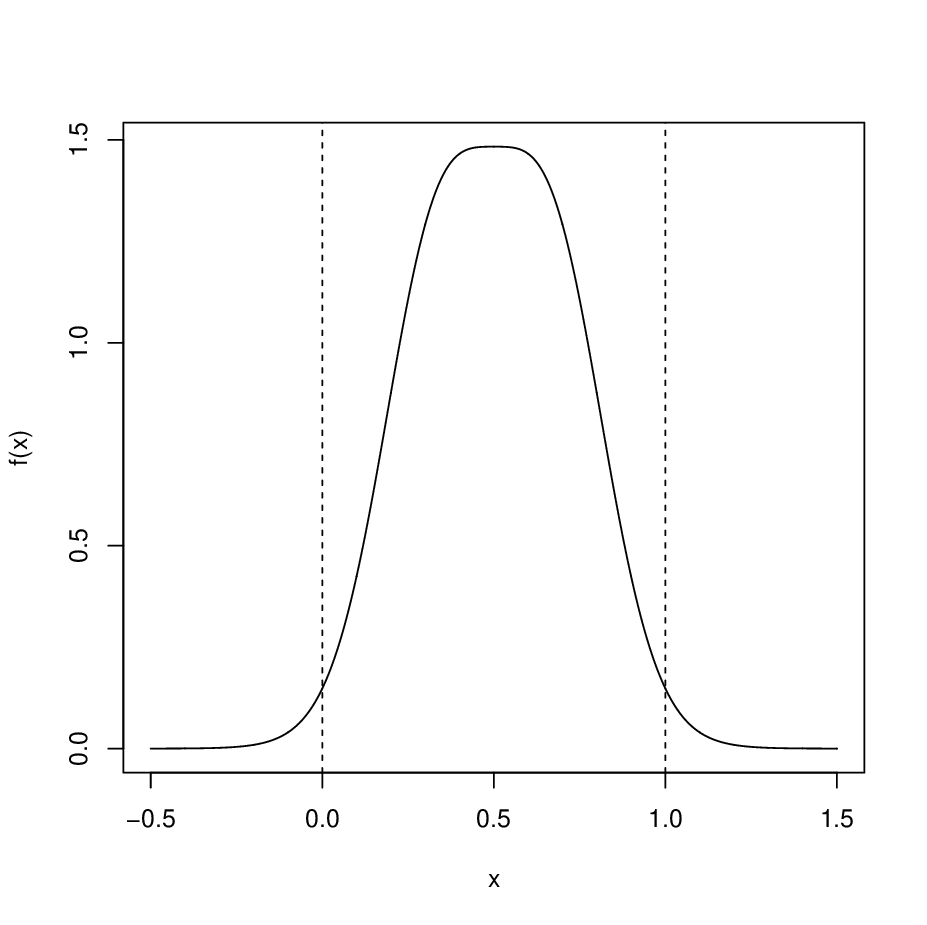}}&  $n=50$  & 0(0) & 0(0) & 0(0) & 0.004(0.006) & 0.040(0.017) & 0.082(0.024) & 0.002(0.004) & 0.022(0.013) & 0.074(0.023) \\ 
&  $n=200$ & 0(0) & 0.004(0.006) & 0.008(0.008) & 0.010(0.009) & 0.064(0.021) & 0.122(0.029) & 0.012(0.010) & 0.110(0.027) & 0.196(0.035) \\ 
&  $n=1000$ & 0(0) & 0.008(0.008) & 0.028(0.014) & 0.008(0.008) & 0.042(0.018) & 0.100(0.026) & 0.048(0.019) & 0.118(0.028) & 0.216(0.036)  \\ 
\cline{2-11}
 & & \multicolumn{3}{c|}{HH}& \multicolumn{3}{c|}{CH}&\multicolumn{3}{c|}{NP} \\ \cline{3-11}
&  $n=50$  & 0(0) & 0.006(0.007) & 0.012(0.010) & 0.028(0.014) & 0.092(0.025) & 0.168(0.033) & 0.008(0.008) & 0.050(0.019) & 0.112(0.028) \\ 
&  $n=200$ & 0(0) & 0.008(0.008) & 0.012(0.010) & 0.050(0.019) & 0.136(0.030) & 0.236(0.037) & 0.018(0.012) & 0.088(0.025) & 0.160(0.032) \\ 
&  $n=1000$ & 0(0) & 0.002(0.004) & 0.002(0.004) & 0.038(0.017) & 0.112(0.028) & 0.202(0.035) & 0.016(0.011) & 0.046(0.018) & 0.116(0.028) \\ 

\hline

M7& &\multicolumn{3}{c|}{SI}& \multicolumn{3}{c|}{FM}& \multicolumn{3}{c|}{HY} \\ \cline{3-11}

\multirow{7}{*}{\includegraphics[width=23mm]{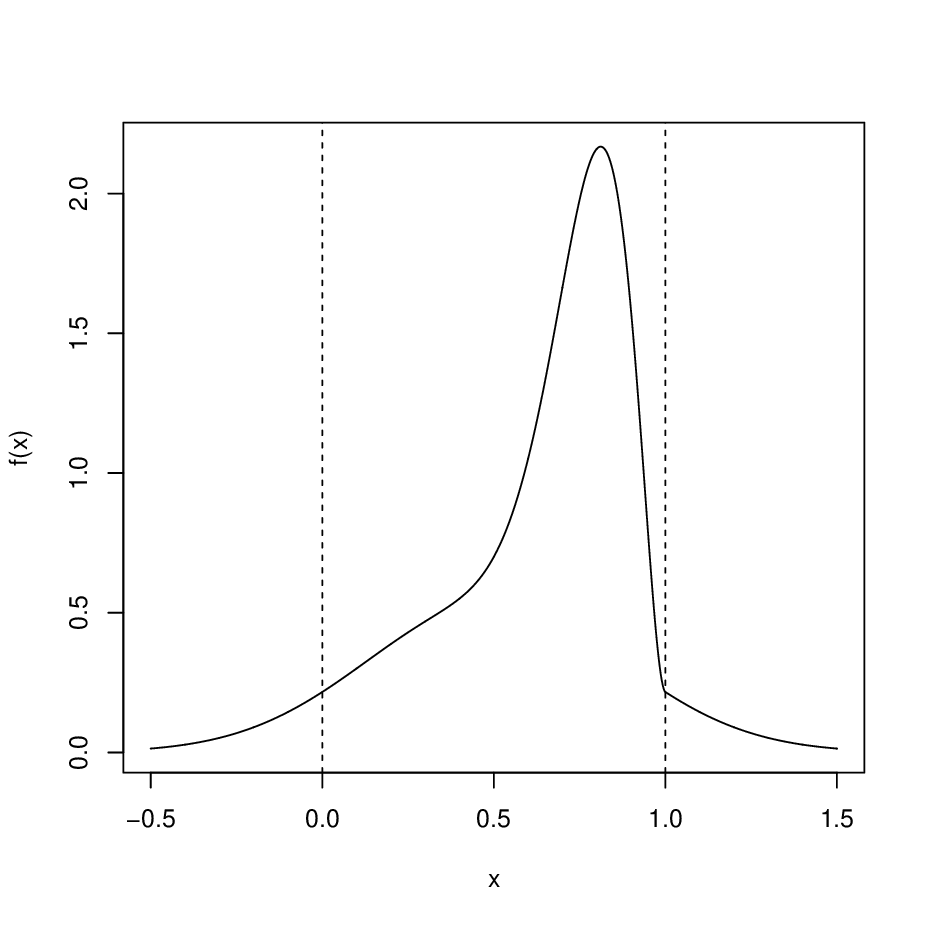}}&  $n=50$ & 0(0) & 0(0) & 0(0) & 0.072(0.023) & 0.246(0.038) & 0.378(0.043) & 0.012(0.010) & 0.072(0.023) & 0.146(0.031) \\ 
&  $n=200$ & 0(0) & 0(0) & 0(0) & 0.064(0.021) & 0.210(0.036) & 0.368(0.042) & 0.016(0.011) & 0.078(0.024) & 0.144(0.031) \\ 
&  $n=1000$ & 0(0) & 0(0) & 0(0) & 0.060(0.021) & 0.214(0.036) & 0.346(0.042) & 0.004(0.006) & 0.072(0.023) & 0.134(0.030)  \\   
\cline{2-11}
 & & \multicolumn{3}{c|}{HH}& \multicolumn{3}{c|}{CH}&\multicolumn{3}{c|}{NP} \\ \cline{3-11}
&  $n=50$  & 0(0) & 0(0) & 0.010(0.009) & 0.008(0.008) & 0.026(0.014) & 0.082(0.024) & 0.006(0.007) & 0.032(0.015) & 0.084(0.024)  \\ 
&  $n=200$ & 0(0) & 0(0) & 0(0) & 0(0) & 0.014(0.010) & 0.042(0.018) & 0.002(0.004) & 0.028(0.014) & 0.070(0.022) \\ 
&  $n=1000$ & 0(0) & 0(0) & 0(0) & 0(0) & 0.012(0.010) & 0.036(0.016) & 0.004(0.006) & 0.042(0.018) & 0.094(0.026)  \\ 

\hline

M8&  &\multicolumn{3}{c|}{SI}& \multicolumn{3}{c|}{FM}& \multicolumn{3}{c|}{HY} \\ \cline{3-11}

\multirow{7}{*}{\includegraphics[width=23mm]{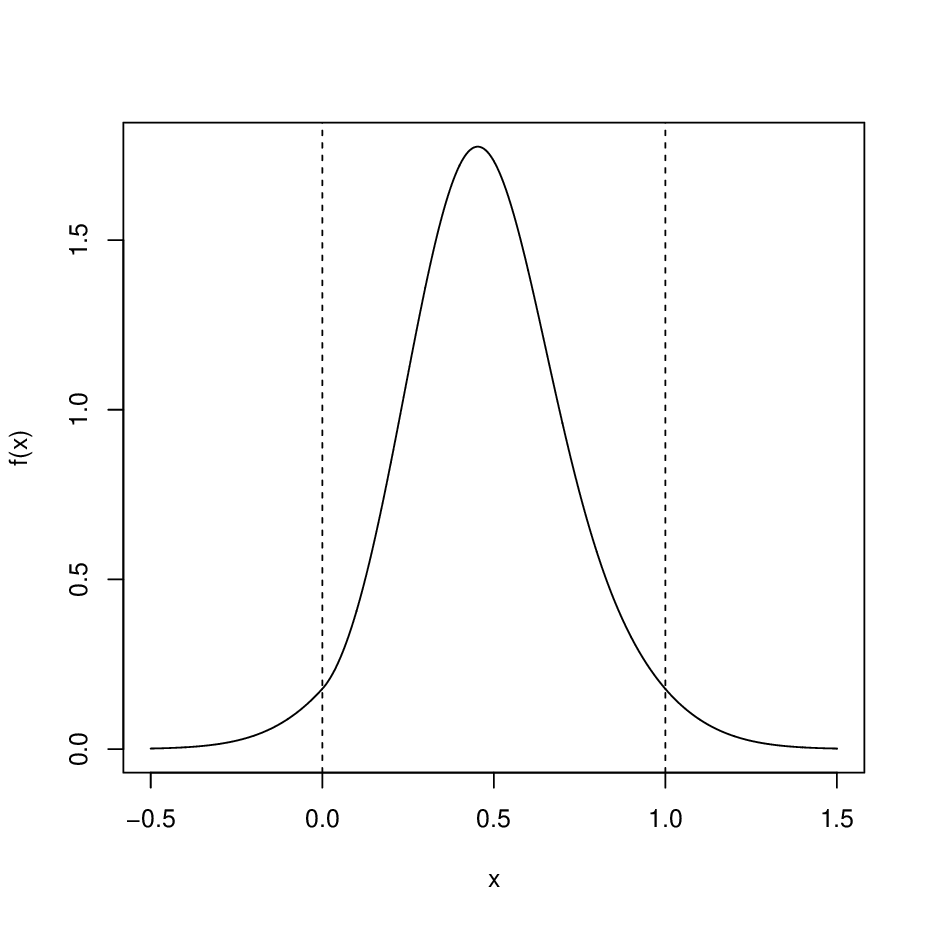}}&  $n=50$ & 0(0) & 0(0) & 0(0) & 0.006(0.007) & 0.024(0.013) & 0.062(0.021) & 0(0) & 0.012(0.010) & 0.046(0.018) \\ 
&  $n=200$ & 0(0) & 0(0) & 0(0) & 0.002(0.004) & 0.022(0.013) & 0.064(0.021) & 0(0) & 0.024(0.013) & 0.054(0.020) \\ 
&  $n=1000$ & 0(0) & 0(0) & 0(0) & 0.002(0.004) & 0.018(0.012) & 0.048(0.019) & 0.010(0.009) & 0.054(0.020) & 0.092(0.025) \\ 
\cline{2-11}
 & & \multicolumn{3}{c|}{HH}& \multicolumn{3}{c|}{CH}&\multicolumn{3}{c|}{NP} \\ \cline{3-11}

&  $n=50$ & 0(0) & 0(0) & 0.006(0.007) & 0.006(0.007) & 0.034(0.016) & 0.078(0.024) & 0.006(0.007) & 0.032(0.015) & 0.076(0.023) \\ 
&  $n=200$ & 0(0) & 0(0) & 0(0) & 0.004(0.006) & 0.026(0.014) & 0.066(0.022) & 0.006(0.007) & 0.028(0.014) & 0.088(0.025)  \\ 
&  $n=1000$ & 0(0) & 0(0) & 0.002(0.004) & 0.016(0.011) & 0.038(0.017) & 0.082(0.024) & 0.014(0.010) & 0.044(0.018) & 0.088(0.025) \\ 
\hline

M9 & &\multicolumn{3}{c|}{SI}& \multicolumn{3}{c|}{FM}& \multicolumn{3}{c|}{HY} \\ \cline{3-11}
\multirow{7}{*}{\includegraphics[width=23mm]{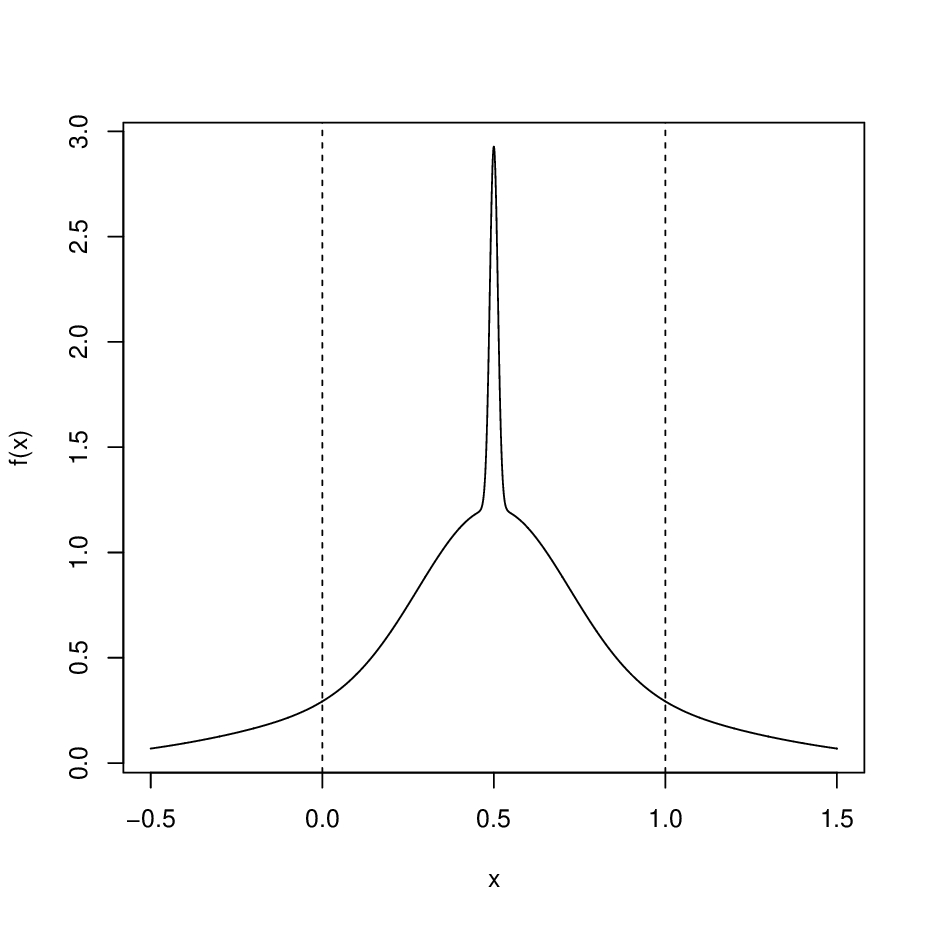}}&  $n=50$  &  0(0) & 0(0) & 0(0) & 0.018(0.012) & 0.084(0.024) & 0.198(0.035) & 0(0) & 0.006(0.007) & 0.014(0.010) \\ 
&  $n=200$  & 0(0) & 0(0) & 0(0) & 0.048(0.019) & 0.182(0.034) & 0.328(0.041) & 0.002(0.004) & 0.022(0.013) & 0.060(0.021) \\ 
&  $n=1000$  & 0(0) & 0(0) & 0(0) & 0.014(0.010) & 0.160(0.032) & 0.318(0.041) & 0.012(0.010) & 0.048(0.019) & 0.086(0.025) \\ 
\cline{2-11}
 & & \multicolumn{3}{c|}{HH}& \multicolumn{3}{c|}{CH}&\multicolumn{3}{c|}{NP} \\ \cline{3-11}
&  $n=50$  & 0(0) & 0(0) & 0(0) & 0.002(0.004) & 0.016(0.011) & 0.032(0.015) & 0.004(0.006) & 0.026(0.014) & 0.068(0.022) \\ 
&  $n=200$  & 0(0) & 0(0) & 0(0) &  0.002(0.004) & 0.018(0.012) & 0.042(0.018) & 0.010(0.009) & 0.046(0.018) & 0.084(0.024) \\ 
&  $n=1000$ & 0(0) & 0(0) & 0(0)  &0.002(0.004) & 0.010(0.009) & 0.014(0.010) & 0.004(0.006) & 0.020(0.012) & 0.062(0.021) \\ 
\hline

M10 & &\multicolumn{3}{c|}{SI}& \multicolumn{3}{c|}{FM}& \multicolumn{3}{c|}{HY} \\ \cline{3-11}
 \multirow{7}{*}{\includegraphics[width=23mm]{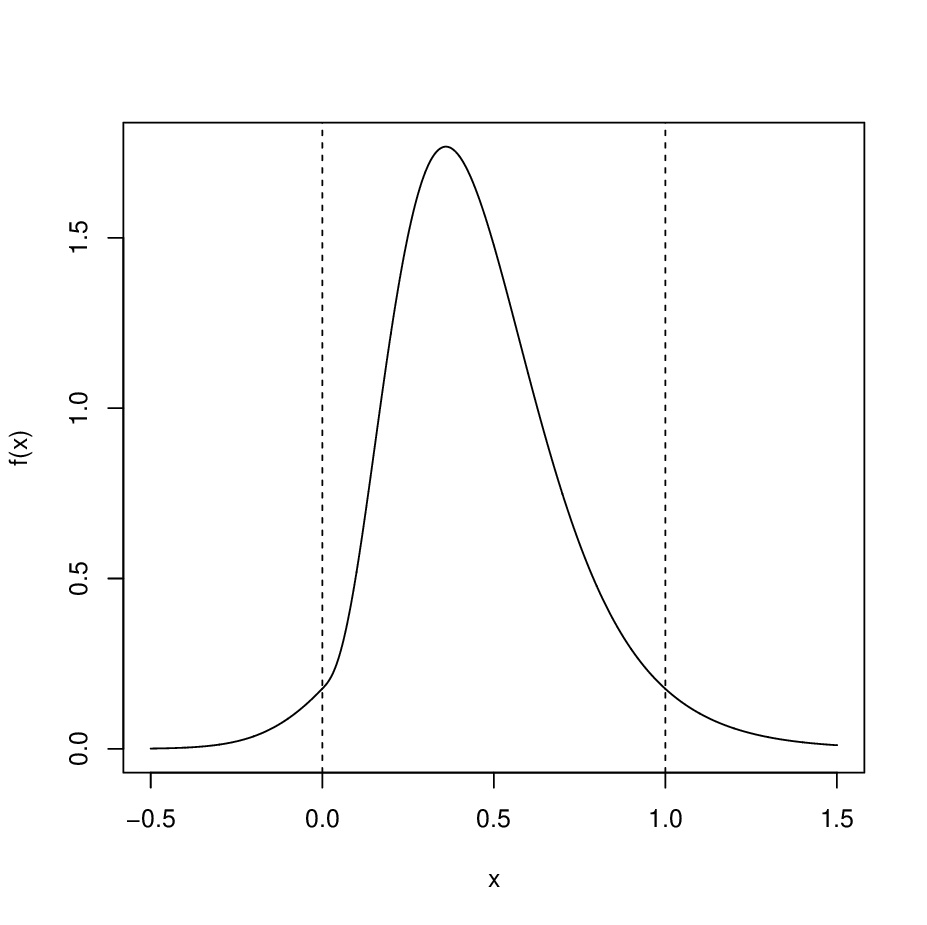}} & $n=50$   & 0(0) & 0(0) & 0(0) & 0.016(0.011) & 0.054(0.020) & 0.116(0.028) & 0(0) & 0.014(0.010) & 0.050(0.019)\\ 
 & $n=200$& 0(0) & 0(0) & 0(0) & 0.018(0.012) & 0.092(0.025) & 0.182(0.034) &  0.006(0.007) & 0.038(0.017) & 0.086(0.025) \\ 
 & $n=1000$ & 0(0) & 0(0) & 0(0) & 0.022(0.013) & 0.094(0.026) & 0.168(0.033) & 0.010(0.009) & 0.050(0.019) & 0.096(0.026) \\ 

\cline{2-11}
 & & \multicolumn{3}{c|}{HH}& \multicolumn{3}{c|}{CH}&\multicolumn{3}{c|}{NP} \\ \cline{3-11}

 & $n=50$  & 0(0) & 0.002(0.004) & 0.008(0.008) & 0.004(0.006) & 0.046(0.018) & 0.086(0.025) & 0.012(0.010) & 0.044(0.018) & 0.094(0.026)\\ 
 & $n=200$ & 0(0) & 0(0) & 0(0) & 0.014(0.010) & 0.042(0.018) & 0.078(0.024) & 0.010(0.009) & 0.062(0.021) & 0.094(0.026) \\ 
 & $n=1000$ & 0(0) & 0(0) & 0(0) & 0.008(0.008) & 0.028(0.014) & 0.074(0.023) & 0.008(0.008) & 0.040(0.017) & 0.104(0.027) \\ 

\hline

\end{tabular}
}
\caption{Percentages of rejections for testing $H_0:j=1$, with $500$ simulations ($1.96$ times their estimated standard deviation in parenthesis) and $B = 500$ bootstrap samples.}
\label{estsim2b}
\end{center}
\end{table}

\begin{table}
\begin{center}
\scalebox{0.55}{
\begin{tabular}{|c |c|c c c | c c c |c c c |}
\hline
   &$\alpha$ & 0.01 & 0.05& 0.10& 0.01 & 0.05& 0.10&0.01 & 0.05& 0.10\\ \hline

M11& &\multicolumn{3}{c|}{SI}& \multicolumn{3}{c|}{FM}& \multicolumn{3}{c|}{HY} \\ \cline{3-11}
\multirow{7}{*}{\includegraphics[width=23mm]{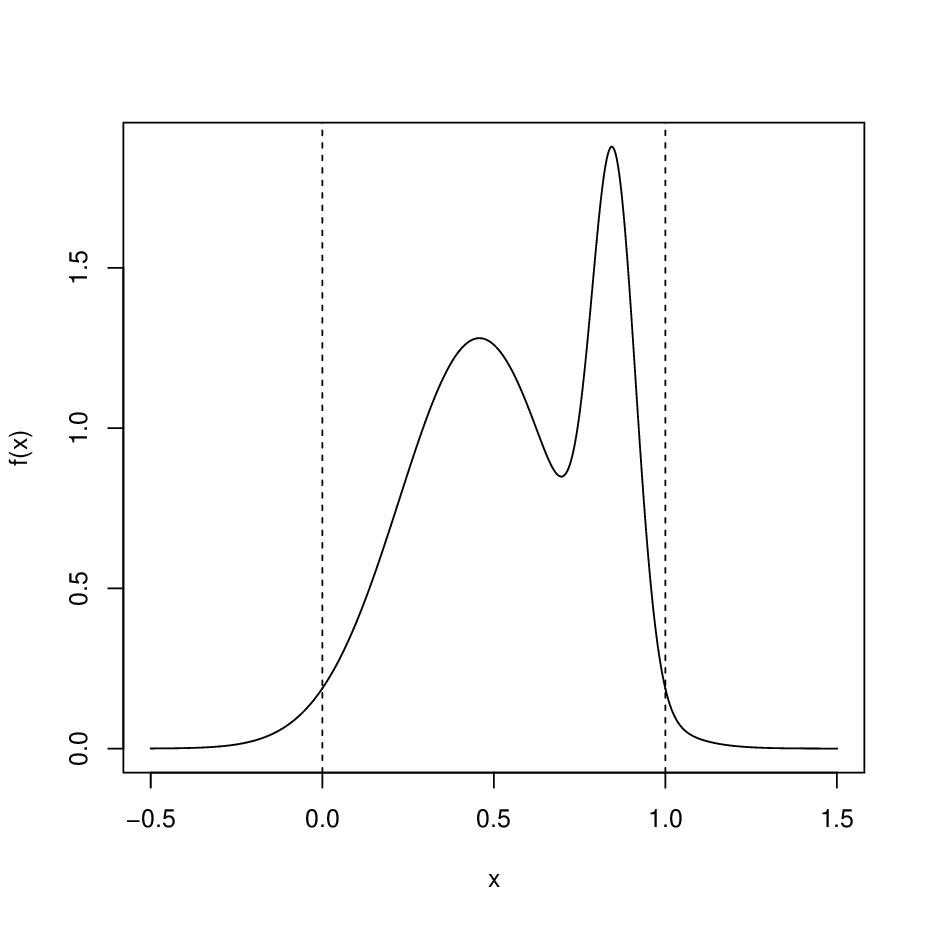}}& $n=50$  & 0(0) & 0(0) & 0.002(0.004) & 0.084(0.024) & 0.250(0.038) & 0.394(0.043) & 0.008(0.008) & 0.168(0.033) & 0.330(0.041) \\ 
& $n=100$  & 0(0) & 0.002(0.004) & 0.066(0.022) & 0.374(0.042) & 0.640(0.042) & 0.738(0.039) & 0.168(0.033) & 0.502(0.044) & 0.630(0.042) \\ 
& $n=200$  & 0(0) & 0.066(0.022) & 0.260(0.038) & 0.600(0.043) & 0.788(0.036) & 0.860(0.030) & 0.378(0.043) & 0.604(0.043) & 0.714(0.040) \\ 
\cline{2-11}
 & &  \multicolumn{3}{c|}{HH}& \multicolumn{3}{c|}{CH}&\multicolumn{3}{c|}{NP} \\ \cline{3-11}

&  $n=50$  & 0.012(0.010) & 0.088(0.025) & 0.156(0.032) & 0.196(0.035) & 0.370(0.042) & 0.494(0.044) & 0.090(0.025) & 0.238(0.037) & 0.376(0.042)  \\ 
& $n=100$  & 0.060(0.021) & 0.182(0.034) & 0.274(0.039) & 0.442(0.044) & 0.630(0.042) & 0.722(0.039) & 0.228(0.037) & 0.418(0.043) & 0.542(0.044) \\ 
& $n=200$  & 0.106(0.027) & 0.238(0.037) & 0.356(0.042) & 0.584(0.043) & 0.768(0.037) & 0.824(0.033) & 0.328(0.041) & 0.506(0.044) & 0.600(0.043) \\ 
  \hline

M12& &\multicolumn{3}{c|}{SI}& \multicolumn{3}{c|}{FM}& \multicolumn{3}{c|}{HY} \\ \cline{3-11}
\multirow{7}{*}{\includegraphics[width=23mm]{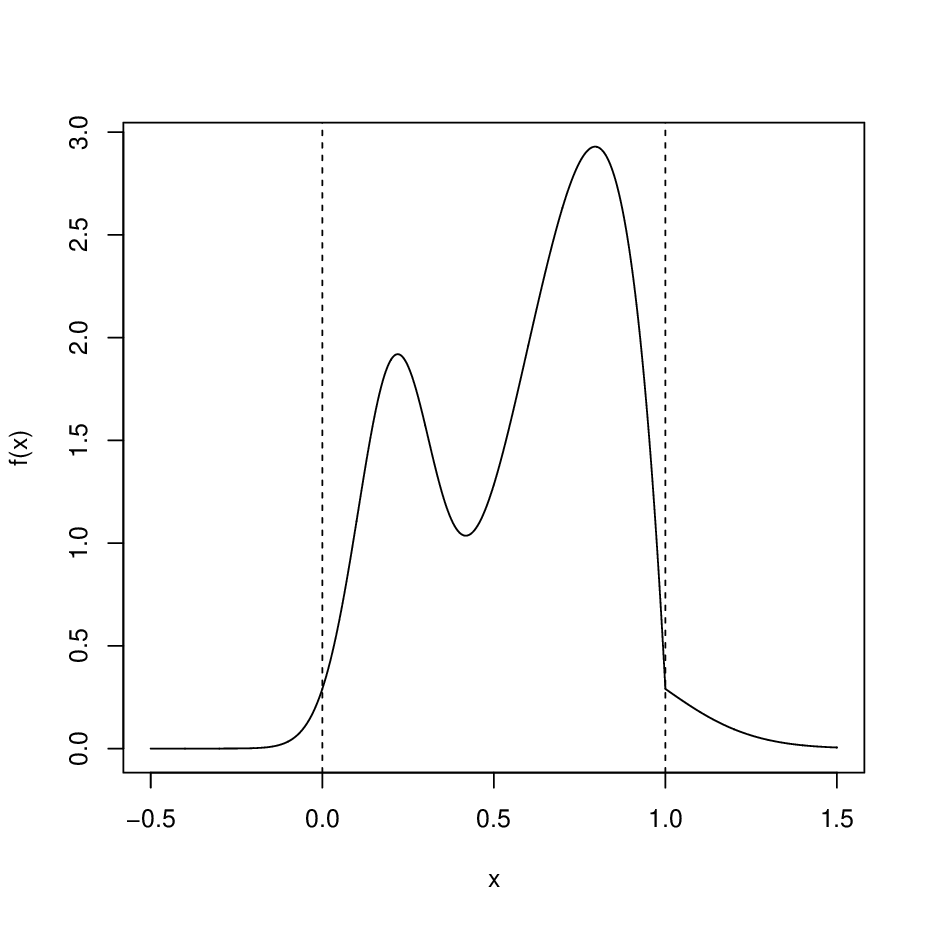}}& $n=50$    & 0(0) & 0(0) & 0.006(0.007) & 0.332(0.041) & 0.584(0.043) & 0.716(0.040) & 0(0) & 0.174(0.033) & 0.422(0.043) \\ 
& $n=100$  & 0(0) & 0(0) & 0.042(0.018) & 0.662(0.041) & 0.838(0.032) & 0.896(0.027) & 0.224(0.037) & 0.642(0.042) & 0.802(0.035) \\ 
& $n=200$  & 0(0) & 0.026(0.014) & 0.312(0.041) & 0.914(0.025) & 0.966(0.016) & 0.986(0.010) & 0.584(0.043) & 0.912(0.025) & 0.950(0.019) \\ 
 \cline{2-11}
 &  & \multicolumn{3}{c|}{HH}& \multicolumn{3}{c|}{CH}&\multicolumn{3}{c|}{NP} \\ \cline{3-11}

&  $n=50$  & 0.042(0.018) & 0.118(0.028) & 0.202(0.035) & 0.268(0.039) & 0.480(0.044) & 0.594(0.043) & 0.100(0.026) & 0.266(0.039) & 0.400(0.043)  \\ 
& $n=100$  & 0.158(0.032) & 0.346(0.042) & 0.460(0.044) & 0.636(0.042) & 0.788(0.036) & 0.844(0.032) & 0.346(0.042) & 0.578(0.043) & 0.680(0.041) \\ 
& $n=200$  & 0.262(0.039) & 0.490(0.044) & 0.622(0.043) & 0.818(0.034) & 0.926(0.023) & 0.956(0.018) & 0.524(0.044) & 0.752(0.038) & 0.844(0.032) \\ 
 \hline

M13& &\multicolumn{3}{c|}{SI}& \multicolumn{3}{c|}{FM}& \multicolumn{3}{c|}{HY} \\ \cline{3-11}
\multirow{7}{*}{\includegraphics[width=23mm]{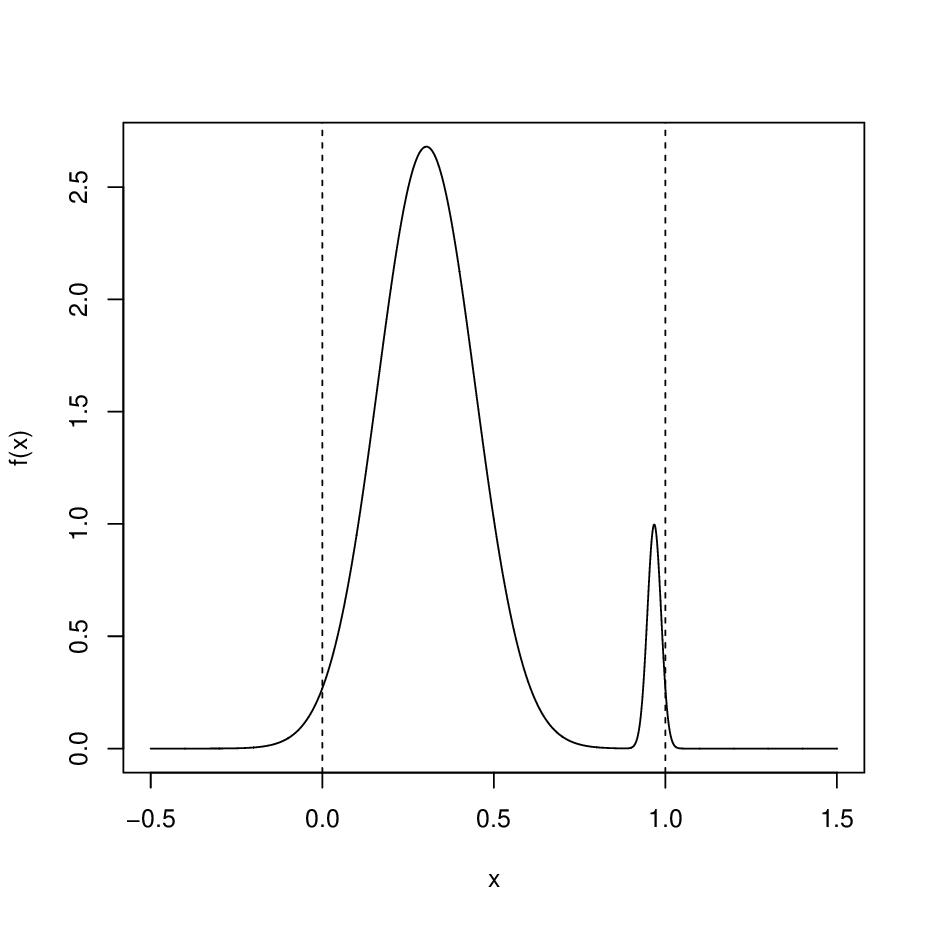}}&  $n=50$  & 0(0) & 0(0) & 0(0) & 0.064(0.021) & 0.492(0.044) & 0.776(0.037) & 0.634(0.042) & 0.858(0.031) & 0.892(0.027) \\ 
& $n=100$  & 0(0) & 0.004(0.006) & 0.066(0.022) & 0.914(0.025) & 0.998(0.004) & 1(0) & 0.994(0.007) & 1(0) & 1(0) \\ 
& $n=200$  & 0.010(0.009) & 0.372(0.042) & 0.784(0.036) & 1(0) & 1(0) & 1(0) & 1(0) & 1(0) & 1(0) \\ 
  \cline{2-11}
 & &  \multicolumn{3}{c|}{HH}& \multicolumn{3}{c|}{CH}&\multicolumn{3}{c|}{NP} \\ \cline{3-11}

&  $n=50$  & 0(0) & 0.004(0.006) & 0.014(0.010) & 0.014(0.010) & 0.052(0.019) & 0.096(0.026) & 0.016(0.011) & 0.058(0.020) & 0.112(0.028) \\ 
& $n=100$  & 0(0) & 0(0) & 0(0)  & 0(0) & 0.032(0.015) & 0.062(0.021) & 0.008(0.008) & 0.050(0.019) & 0.102(0.027) \\ 
& $n=200$  & 0(0) & 0(0) & 0.002(0.004) & 0.006(0.007) & 0.048(0.019) & 0.366(0.042) & 0.016(0.011) & 0.276(0.039) & 0.758(0.038) \\  \hline

M14& &\multicolumn{3}{c|}{SI}& \multicolumn{3}{c|}{FM}& \multicolumn{3}{c|}{HY} \\ \cline{3-11}
\multirow{7}{*}{\includegraphics[width=23mm]{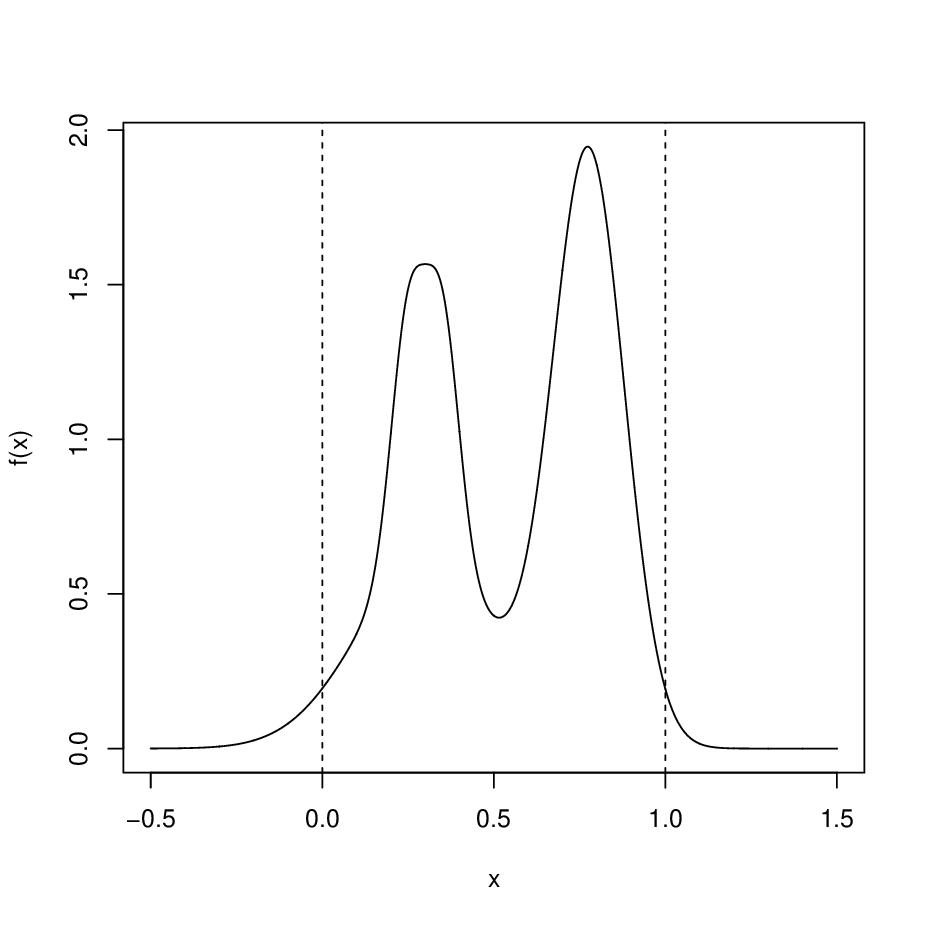}}&  $n=50$ & 0(0) & 0(0) & 0.022(0.013) & 0.662(0.041) & 0.864(0.030) & 0.922(0.024) & 0.050(0.019) & 0.576(0.043) & 0.830(0.033) \\ 
& $n=100$  & 0(0) & 0.056(0.020) & 0.314(0.041) & 0.990(0.009) & 1(0) & 1(0) & 0.920(0.024) & 1(0) & 1(0) \\ 
& $n=200$  & 0.018(0.012) & 0.398(0.043) & 0.902(0.026) & 1(0) & 1(0) & 1(0) & 1(0) & 1(0) & 1(0) \\  \cline{2-11}
 & &  \multicolumn{3}{c|}{HH}& \multicolumn{3}{c|}{CH}&\multicolumn{3}{c|}{NP} \\ \cline{3-11}

&  $n=50$ & 0.226(0.037) & 0.514(0.044) & 0.646(0.042) &  0.718(0.039) & 0.860(0.030) & 0.924(0.023) & 0.460(0.044) & 0.716(0.040) & 0.806(0.035) \\ 
& $n=100$  & 0.784(0.036) & 0.946(0.020) & 0.978(0.013) & 0.994(0.007) & 0.996(0.006) & 1(0) & 0.950(0.019) & 0.992(0.008) & 0.994(0.007)  \\ 
& $n=200$  & 1(0) & 1(0) & 1(0) & 1(0) & 1(0) & 1(0) & 1(0) & 1(0) & 1(0) \\ 
 \hline

M15& &\multicolumn{3}{c|}{SI}& \multicolumn{3}{c|}{FM}& \multicolumn{3}{c|}{HY} \\ \cline{3-11}
\multirow{7}{*}{\includegraphics[width=23mm]{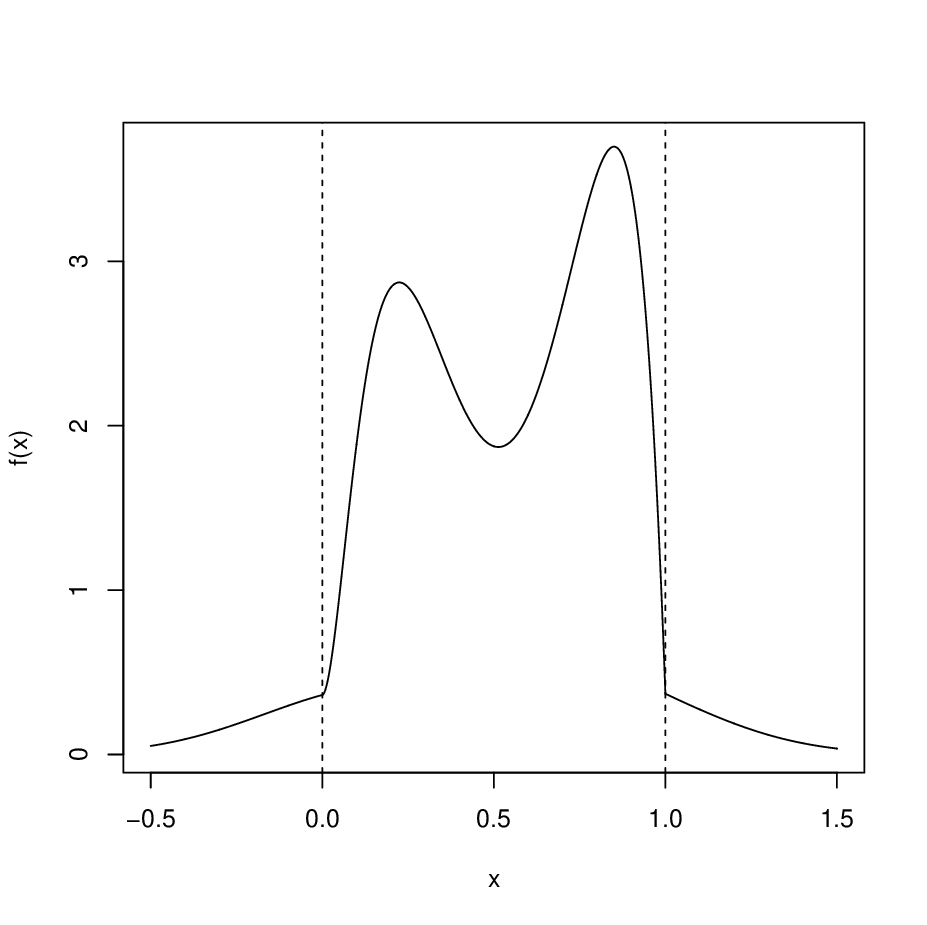}}&  $n=50$  & 0(0) & 0(0) & 0.002(0.004) & 0.042(0.018) & 0.118(0.028) & 0.228(0.037) & 0.014(0.010) & 0.128(0.029) & 0.268(0.039) \\ 
& $n=100$  & 0(0) & 0(0) & 0.012(0.010) & 0.118(0.028) & 0.334(0.041) & 0.472(0.044) & 0.066(0.022) & 0.342(0.042) & 0.500(0.044) \\ 
& $n=200$  & 0(0) & 0.020(0.012) & 0.094(0.026) & 0.266(0.039) & 0.524(0.044) & 0.636(0.042) & 0.298(0.040) & 0.576(0.043) & 0.736(0.039) \\  \cline{2-11}
 & &  \multicolumn{3}{c|}{HH}& \multicolumn{3}{c|}{CH}&\multicolumn{3}{c|}{NP} \\ \cline{3-11}

&  $n=50$  & 0.010(0.009) & 0.046(0.018) & 0.072(0.023) & 0.098(0.026) & 0.242(0.038) & 0.374(0.042) & 0.054(0.020) & 0.156(0.032) & 0.274(0.039)  \\ 
& $n=100$  & 0.014(0.010) & 0.070(0.022) & 0.150(0.031) & 0.232(0.037) & 0.424(0.043) & 0.542(0.044) & 0.124(0.029) & 0.288(0.040) & 0.400(0.043) \\ 
& $n=200$  & 0.026(0.014) & 0.104(0.027) & 0.194(0.035) & 0.364(0.042) & 0.582(0.043) & 0.690(0.041) & 0.192(0.035) & 0.400(0.043) & 0.548(0.044)  \\ 
 \hline

\end{tabular}
}
\caption{Percentages of rejections for testing $H_0:j=1$, with $500$ simulations ($1.96$ times their estimated standard deviation in parenthesis) and $B = 500$ bootstrap samples.}
\label{estsim3}
\end{center}
\end{table}

\begin{table}
\centering
\scalebox{0.48}{
\begin{tabular}{|c |c| c|c c c |c |c| c|c c c |}
\hline
 & & $\alpha$ & 0.01 & 0.05& 0.10& & & $\alpha$ & 0.01 & 0.05& 0.10\\ 
  \hline 
 M11  &  \multirow{3}{*}{SI}  &  $n=50$  &  0(0) & 0(0) & 0.018(0.012)  &  M16  &  \multirow{3}{*}{SI}  &  $n=50$  &  0(0) & 0.006(0.007) & 0.038(0.017)  \\ 
   \multirow{8}{*}{\includegraphics[width=23mm]{121a.eps}}  &  &  $n=200$   &  0(0) & 0.018(0.012) & 0.042(0.018)  &  \multirow{8}{*}{\includegraphics[width=23mm]{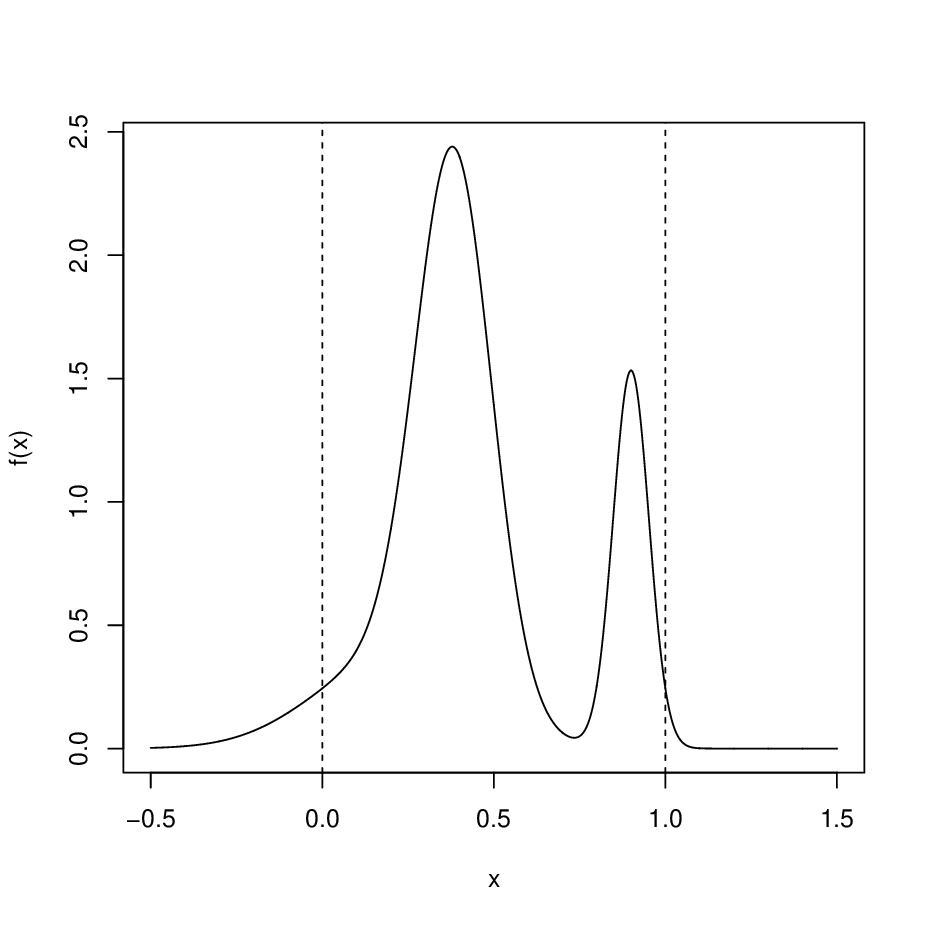}}  &  &  $n=200$   &  0.004(0.006) & 0.020(0.012) & 0.040(0.017)  \\ 
   &  &  $n=1000$   &  0.008(0.008) & 0.020(0.012) & 0.044(0.018)  &  &  &  $n=1000$   &  0(0) & 0.004(0.006) & 0.010(0.009)  \\ \cline{2-6} \cline{8-12}
   &  \multirow{3}{*}{FM}  &  $n=50$  &  0.016(0.011)  &  0.076(0.023)  &  0.152(0.031)  &  &  \multirow{3}{*}{FM}  &  $n=50$  &  0.092(0.025)  &  0.210(0.036)  &  0.284(0.040)  \\ 
   &  &  $n=200$   &  0.092(0.025)  &  0.276(0.039)  &  0.392(0.043)  &  &  &  $n=200$   &  0.058(0.020)  &  0.128(0.029)  &  0.186(0.034)  \\ 
   &  &  $n=1000$   &  0.080(0.024)  &  0.218(0.036)  &  0.316(0.041)  &  &  &  $n=1000$   &  0.038(0.017)  &  0.082(0.024)  &  0.156(0.032)  \\ 
\cline{2-6} \cline{8-12} &  \multirow{3}{*}{NP}  &  $n=50$  &  0.028(0.014)  &  0.088(0.025)  &  0.178(0.034)   &  &  \multirow{3}{*}{NP}  &  $n=50$  &  0.006(0.007)  &  0.064(0.021)  &  0.112(0.028)  \\ 
   &  &  $n=200$   &  0.014(0.010)  &  0.056(0.020)  &  0.108(0.027)   &  &  &  $n=200$   &  0.016(0.011)  &  0.098(0.026)  &  0.200(0.035)   \\ 
   &  &  $n=1000$   &  0.006(0.007)  &  0.046(0.018)  &  0.084(0.024)    &  &  &  $n=1000$   &  0.010(0.009)  &  0.074(0.023)  &  0.150(0.031)   \\ 
 \hline   M12  &  \multirow{3}{*}{SI}  &  $n=50 $  &  0(0) & 0.004(0.006) & 0.012(0.010)  &  M17  &  \multirow{3}{*}{SI}  &  $n=50$  &  0(0) & 0.002(0.004) & 0.006(0.007)  \\ 
   \multirow{8}{*}{\includegraphics[width=23mm]{121b.eps}}  &  &  $n=200$   &  0(0) & 0.012(0.010) & 0.018(0.012)  &  \multirow{8}{*}{\includegraphics[width=23mm]{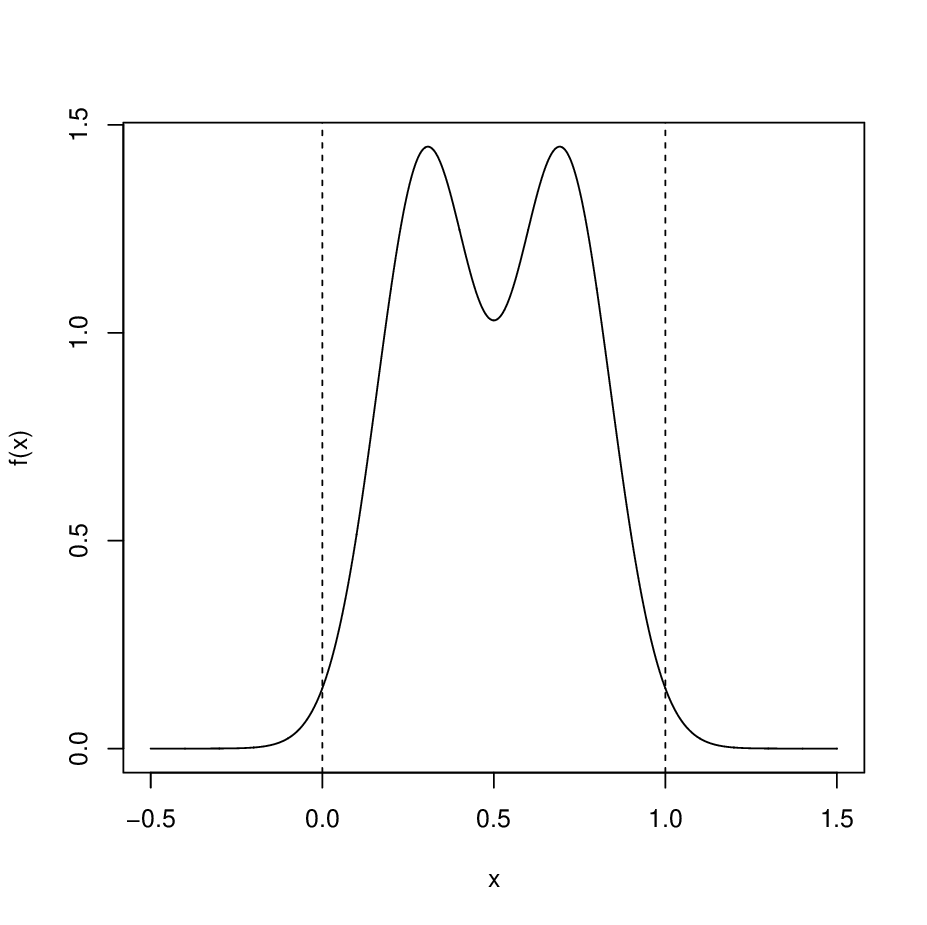}}  &  &  $n=200$   &  0.002(0.004) & 0.008(0.008) & 0.024(0.013)  \\ 
   &  &  $n=1000$   &  0.002(0.004) & 0.006(0.007) & 0.008(0.008)  &  &  &  $n=1000$   &  0.002(0.004) & 0.002(0.004) & 0.012(0.010)  \\ 
\cline{2-6} \cline{8-12} &  \multirow{3}{*}{FM}  &  $n=50 $  &  0.030(0.015)  &  0.100(0.026)  &  0.178(0.034)  &  &  \multirow{3}{*}{FM}  &  $n=50$  &  0.002(0.004)  &  0.028(0.014)  &  0.060(0.021)  \\ 
   &  &  $n=200$   &  0.038(0.017)  &  0.134(0.030)  &  0.184(0.034)  &  &  &  $n=200$   &  0.008(0.008)  &  0.046(0.018)  &  0.096(0.026)  \\ 
   &  &  $n=1000$   &  0.052(0.019)  &  0.094(0.026)  &  0.168(0.033)  &  &  &  $n=1000$   &  0.002(0.004)  &  0.026(0.014)  &  0.048(0.019)  \\ 
\cline{2-6} \cline{8-12} &  \multirow{3}{*}{NP}  &  $n=50 $  &  0.004(0.006)  &  0.034(0.016)  &  0.074(0.023)   &  &  \multirow{3}{*}{NP}  &  $n=50$  &  0.012(0.010)  &  0.060(0.021)  &  0.136(0.030)   \\ 
   &  &  $n=200$   &  0.002(0.004)  &  0.030(0.015)  &  0.076(0.023)   &  &  &  $n=200$   &  0.008(0.008)  &  0.070(0.022)  &  0.106(0.027)   \\ 
   &  &  $n=1000$   &  0.008(0.008)  &  0.046(0.018)  &  0.082(0.024)   &  &  &  $n=1000$   &  0.008(0.008)  &  0.038(0.017)  &  0.074(0.023)   \\ 
  \hline  M13  &  \multirow{3}{*}{SI}  &  $n=50$  &  0(0) & 0(0) & 0.006(0.007)  &  M18  &  \multirow{3}{*}{SI}  &  $n=50$  &  0(0)  &  0(0)  &  0(0)  \\ 
   \multirow{8}{*}{\includegraphics[width=23mm]{121c.eps}}  &  &  $n=200$   &  0(0) & 0.002(0.004) & 0.002(0.004)  &  \multirow{8}{*}{\includegraphics[width=23mm]{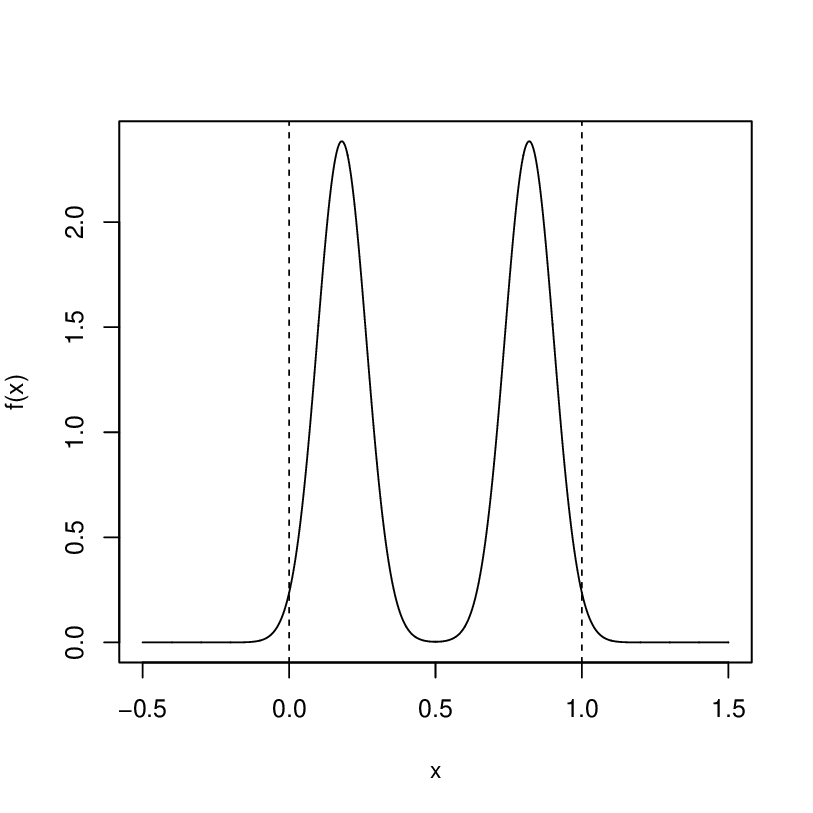}}  &  &  $n=200$   &  0(0)  &  0(0)  &  0.020(0.012)  \\ 
   &  &  $n=1000$   &  0.002(0.004) & 0.014(0.010) & 0.034(0.016)  &  &  &  $n=1000$   &  0(0)  &  0.010(0.009) & 0.022(0.013)  \\ 
\cline{2-6} \cline{8-12} &  \multirow{3}{*}{FM}  &  $n=50$  &  0.006(0.007)  &  0.044(0.018)  &  0.102(0.027)  &  &  \multirow{3}{*}{FM}  &  $n=50$  &  0(0)  &  0.008(0.008)  &  0.038(0.017)  \\ 
   &  &  $n=200$   &  0(0)  &  0.024(0.013)  &  0.056(0.020)  &  &  &  $n=200$   &  0.004(0.006)  &  0.032(0.015)  &  0.050(0.019)  \\ 
   &  &  $n=1000$   &  0.004(0.006)  &  0.036(0.016)  &  0.072(0.023)  &  &  &  $n=1000$   &  0.004(0.006)  &  0.028(0.014)  &  0.066(0.022)  \\ 
\cline{2-6} \cline{8-12} &  \multirow{3}{*}{NP}  &  $n=50$  &  0.006(0.007)  &  0.052(0.019)  &  0.118(0.028)   &  &  \multirow{3}{*}{NP}  &  $n=50$  &  0.004(0.006)  &  0.054(0.020)  &  0.108(0.027)   \\ 
   &  &  $n=200$   &  0.006(0.007)  &  0.028(0.014)  &  0.070(0.022)   &  &  &  $n=200$   &  0.006(0.007)  &  0.048(0.019)  &  0.108(0.027)   \\ 
   &  &  $n=1000$   &  0.010(0.009)  &  0.044(0.018)  &  0.088(0.025)   &  &  &  $n=1000$   &  0.002(0.004)  &  0.034(0.016)  &  0.080(0.024)   \\ 
 \hline   M14  &  \multirow{3}{*}{SI}  &  $n=50$   &  0(0) & 0(0) & 0.004(0.006)  &  M19  &  \multirow{3}{*}{SI}  &  $n=50$  &  0(0)  &  0(0)  &  0(0)  \\ 
   \multirow{8}{*}{\includegraphics[width=23mm]{121d.eps}}  &  &  $n=200$   &  0(0)  &  0(0)  &  0.002(0.004)  &  \multirow{8}{*}{\includegraphics[width=23mm]{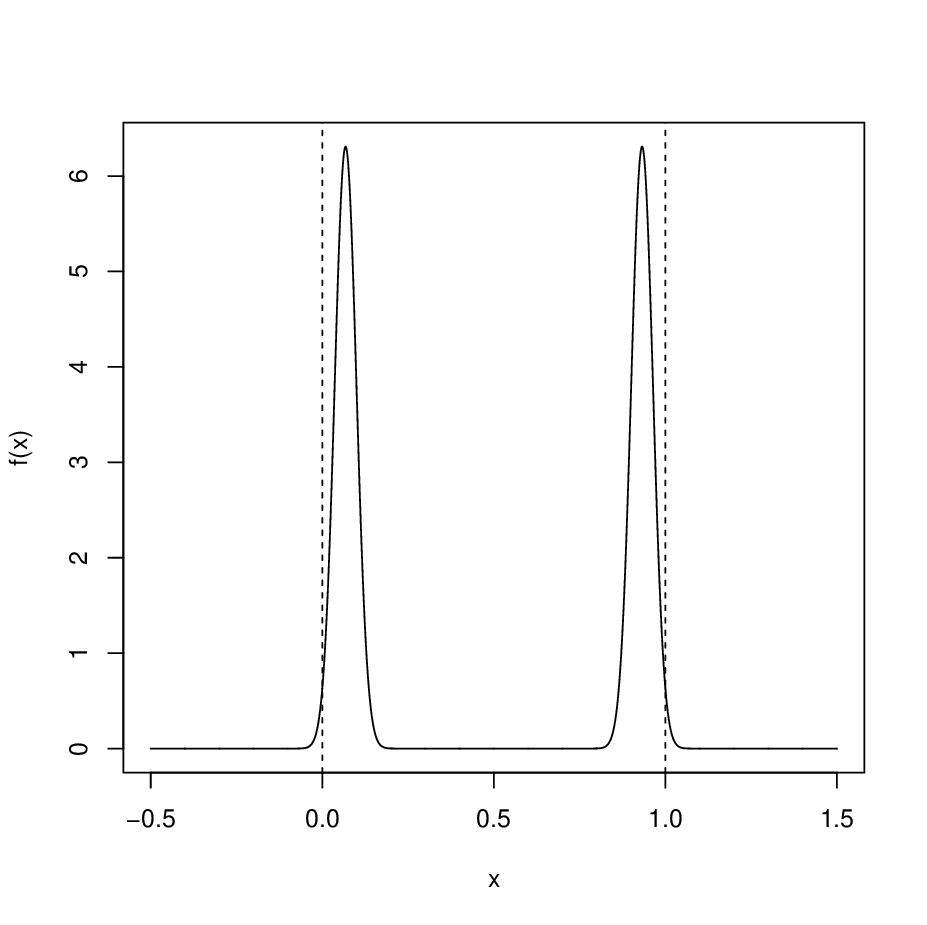}}  &  &  $n=200$   &  0(0)  &  0(0)  &  0(0)  \\ 
   &  &  $n=1000$   &  0(0) & 0(0) & 0.004(0.006)  &  &  &  $n=1000$   &  0(0)  &  0(0)  &  0(0)  \\ 
\cline{2-6} \cline{8-12} &  \multirow{3}{*}{FM}  &  $n=50$   &  0.020(0.012)  &  0.072(0.023)  &  0.132(0.030)  &  &  \multirow{3}{*}{FM}  &  $n=50$  &  0(0)  &  0.004(0.006)  &  0.034(0.016)  \\ 
   &  &  $n=200$   &  0.018(0.012)  &  0.088(0.025)  &  0.152(0.031)  &  &  &  $n=200$   &  0(0)  &  0.008(0.008)  &  0.030(0.015)  \\ 
   &  &  $n=1000$   &  0.024(0.013)  &  0.058(0.020)  &  0.114(0.028)  &  &  &  $n=1000$   &  0(0)  &  0.022(0.013)  &  0.044(0.018)  \\ 
\cline{2-6} \cline{8-12} &  \multirow{3}{*}{NP}  &  $n=50$   &  0.008(0.008)  &  0.034(0.016)  &  0.074(0.023)   &  &  \multirow{3}{*}{NP}  &  $n=50$  &  0.024(0.013)  &  0.070(0.022)  &  0.132(0.030)   \\ 
   &  &  $n=200$   &  0.004(0.006)  &  0.034(0.016)  &  0.088(0.025)   &  &  &  $n=200$   &  0.012(0.010)  &  0.066(0.022)  &  0.118(0.028)   \\ 
   &  &  $n=1000$   &  0.008(0.008)  &  0.056(0.020)  &  0.092(0.025)   &  &  &  $n=1000$   &  0.008(0.008)  &  0.040(0.017)  &  0.100(0.026)   \\ 
 \hline   M15  &  \multirow{3}{*}{SI}  &  $n=50$  &  0(0) & 0.002(0.004) & 0.020(0.012)  &  M20  &  \multirow{3}{*}{SI}  &  $n=50$ &  0.108(0.027) & 0.384(0.043) & 0.506(0.044)  \\ 
   \multirow{8}{*}{\includegraphics[width=23mm]{121e.eps}}  &  &  $n=200$   &  0(0) & 0.004(0.006) & 0.020(0.012)  &  \multirow{8}{*}{\includegraphics[width=23mm]{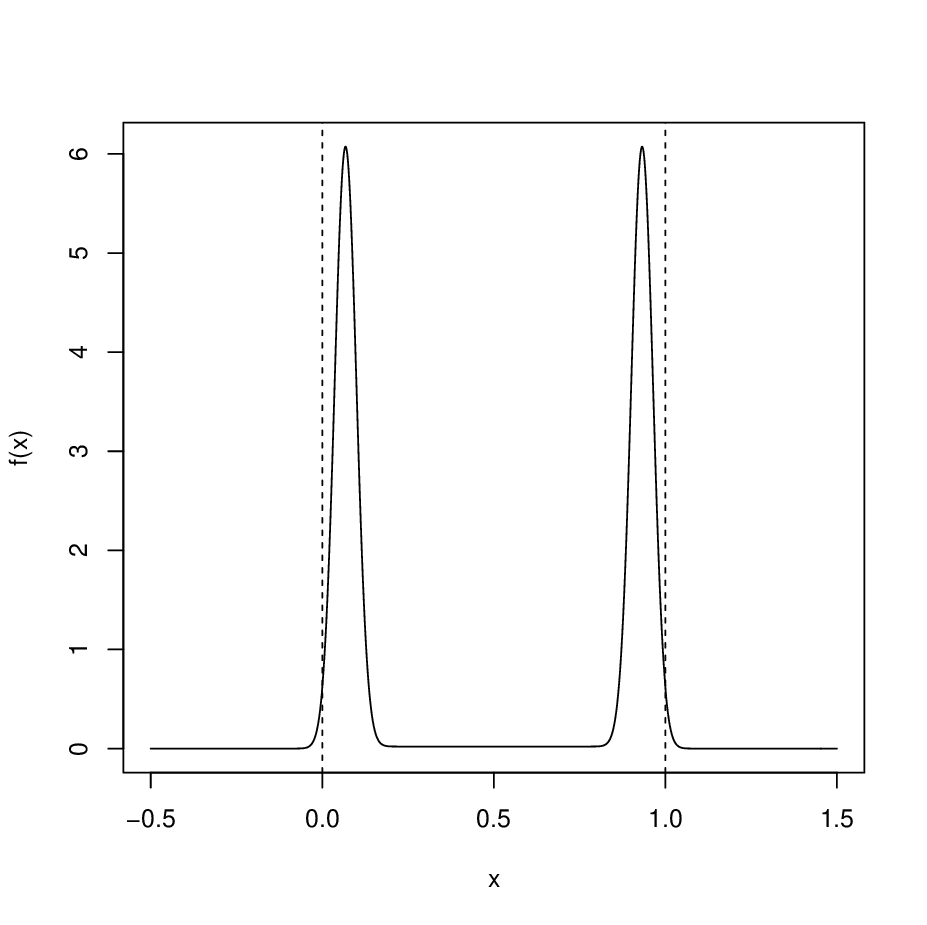}}  &  &  $n=200$   &  0.290(0.040) & 0.412(0.043) & 0.564(0.043)  \\ 
   &  &  $n=1000$   &  0(0) & 0.004(0.006) & 0.032(0.015)  &  &  &  $n=1000$   &  0.358(0.042) & 0.498(0.044) & 0.610(0.043)  \\ 
\cline{2-6} \cline{8-12} &  \multirow{3}{*}{FM}  &  $n=50$  &  0.008(0.008)  &  0.072(0.023)  &  0.136(0.030)  &  &  \multirow{3}{*}{FM}  &  $n=50$ &  0.002(0.004)  &  0.004(0.006)  &  0.022(0.013)  \\ 
   &  &  $n=200$   &  0.032(0.015)  &  0.128(0.029)  &  0.214(0.036)   &  &  &  $n=200$   &  0.958(0.018)  &  0.974(0.014)  &  0.982(0.012)  \\ 
   &  &  $n=1000$   &  0.062(0.021)  &  0.154(0.032)  &  0.224(0.037)  &  &  &  $n=1000$   &  0.976(0.013)  &  0.990(0.009)  &  0.998(0.004)  \\ 
\cline{2-6} \cline{8-12} &  \multirow{3}{*}{NP}  &  $n=50$  &  0.012(0.010)  &  0.078(0.024)  &  0.158(0.032)   &  &  \multirow{3}{*}{NP}  &  $n=50$ &  0.010(0.009)  &  0.068(0.022)  &  0.112(0.028)   \\ 
   &  &  $n=200$   &  0.024(0.013)  &  0.106(0.027)  &  0.200(0.035)   &  &  &  $n=200$   &  0.016(0.011)  &  0.060(0.021)  &  0.128(0.029)   \\ 
   &  &  $n=1000$   &  0.014(0.010)  &  0.048(0.019)  &  0.104(0.027)   &  &  &  $n=1000$   &  0.004(0.006)  &  0.038(0.017)  &  0.096(0.026)   \\ 
   \hline
\end{tabular}
}
\caption{Percentages of rejections for testing $H_0:j = 2$, with $500$ simulations ($1.96$ times their estimated standard deviation in parenthesis) and $B = 500$ bootstrap samples.}
\label{estsim5}
\end{table}

\begin{table}
\scalebox{0.65}{
\centering
\begin{tabular}{|c |c| c|c c c |c |c| c|c c c |}
\hline
 & & $\alpha$ & 0.01 & 0.05& 0.10& & & $\alpha$ & 0.01 & 0.05& 0.10\\ 
\hline M21  &  \multirow{3}{*}{SI}  &  $n=50$   &  0(0) & 0.008(0.008) & 0.028(0.014)  &  M24  &  \multirow{3}{*}{SI}  &  $n=50$   &  0(0) & 0(0) & 0.004(0.006)  \\ 
   \multirow{8}{*}{\includegraphics[width=23mm]{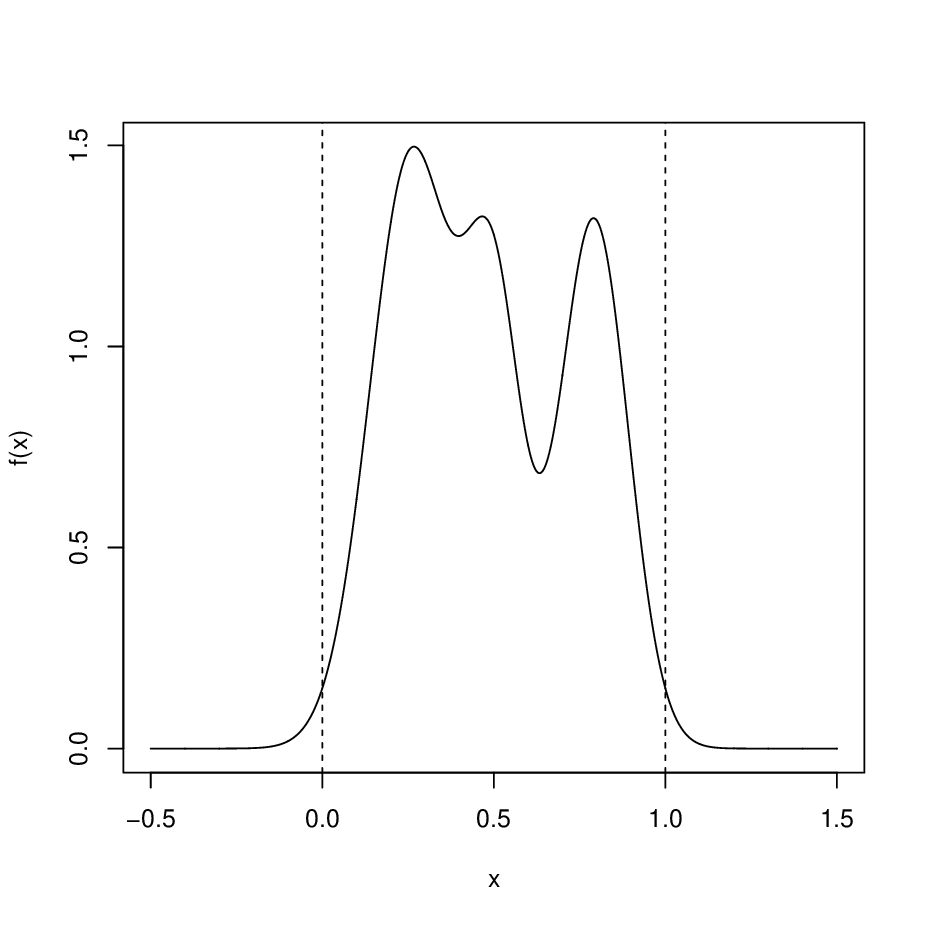}}  &  &  $n=100$   &  0(0) & 0.026(0.014) & 0.062(0.021)  &  \multirow{8}{*}{\includegraphics[width=23mm]{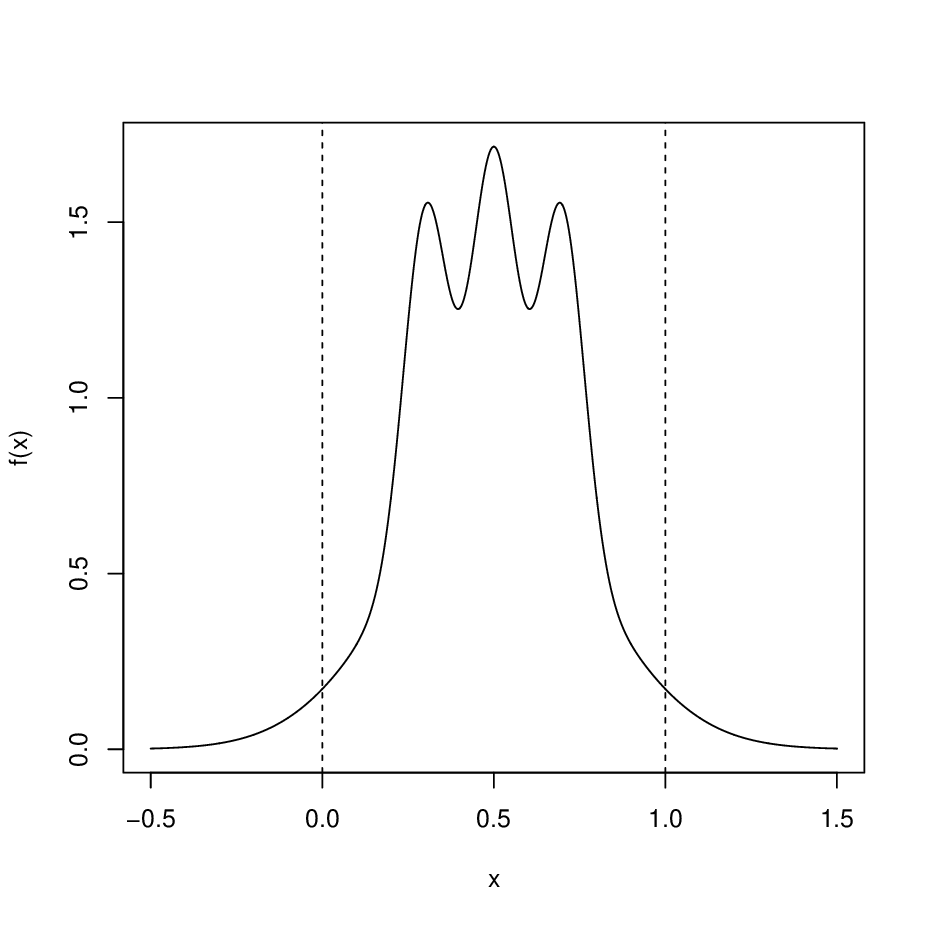}}  &  &  $n=100$   &  0(0) & 0.004(0.006) & 0.008(0.008)  \\ 
   &  &  $n=200$   &  0.004(0.006) & 0.048(0.019) & 0.108(0.027)  &  &  &  $n=200$   &  0(0) & 0.004(0.006) & 0.028(0.014)  \\ 
\cline{2-6} \cline{8-12} &  \multirow{3}{*}{FM}  &  $n=50$   &  0.010(0.009)  &  0.052(0.019)  &  0.112(0.028)  &  &  \multirow{3}{*}{FM}  &  $n=50$   &  0.004(0.006)  &  0.030(0.015)  &  0.082(0.024)  \\ 
   &  &  $n=100$   &  0.044(0.018)  &  0.148(0.031)  &  0.228(0.037)  &  &  &  $n=100$   &  0.012(0.010)  &  0.052(0.019)  &  0.108(0.027)   \\ 
   &  &  $n=200$   &  0.076(0.023)  &  0.202(0.035)  &  0.278(0.039)  &  &  &  $n=200$   &  0.050(0.019)  &  0.160(0.032)  &  0.288(0.040)  \\ 
\cline{2-6} \cline{8-12} &  \multirow{3}{*}{NP}  &  $n=50$   &  0.014(0.010)  &  0.054(0.020)  &  0.112(0.028)   &  &  \multirow{3}{*}{NP}  &  $n=50$   &  0.008(0.008)  &  0.060(0.021)  &  0.134(0.030)   \\ 
   &  &  $n=100$   &  0.020(0.012)  &  0.092(0.025)  &  0.168(0.033)   &  &  &  $n=100$   &  0.034(0.016)  &  0.110(0.027)  &  0.176(0.033)   \\ 
   &  &  $n=200$   &  0.050(0.019)  &  0.134(0.030)  &  0.194(0.035)   &  &  &  $n=200$   &  0.096(0.026)  &  0.232(0.037)  &  0.334(0.041)   \\ 
\hline   M22  &  \multirow{3}{*}{SI}  &  $n=50$   &  0(0) & 0(0) & 0(0)  &  M25  &  \multirow{3}{*}{SI}  &  $n=50$  &  0(0) & 0.012(0.010) & 0.042(0.018)  \\ 
   \multirow{8}{*}{\includegraphics[width=23mm]{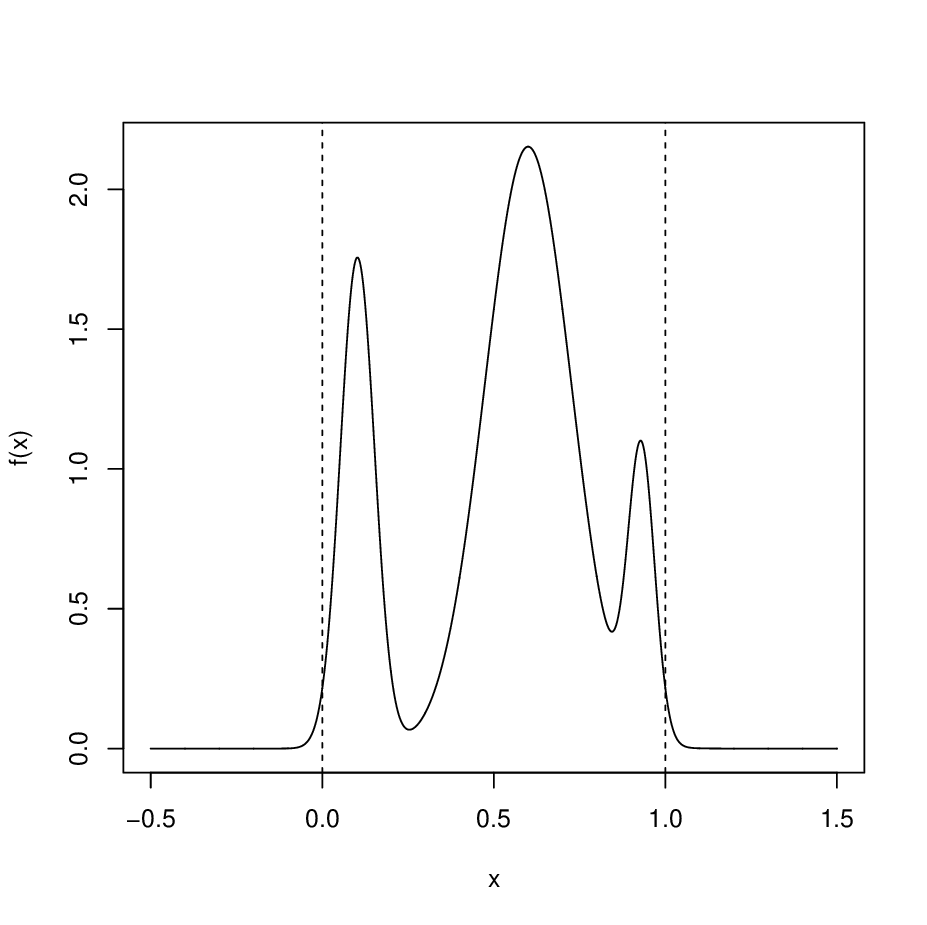}}  &  &  $n=100$   &  0(0) & 0.002(0.004) & 0.036(0.016)  &  \multirow{8}{*}{\includegraphics[width=23mm]{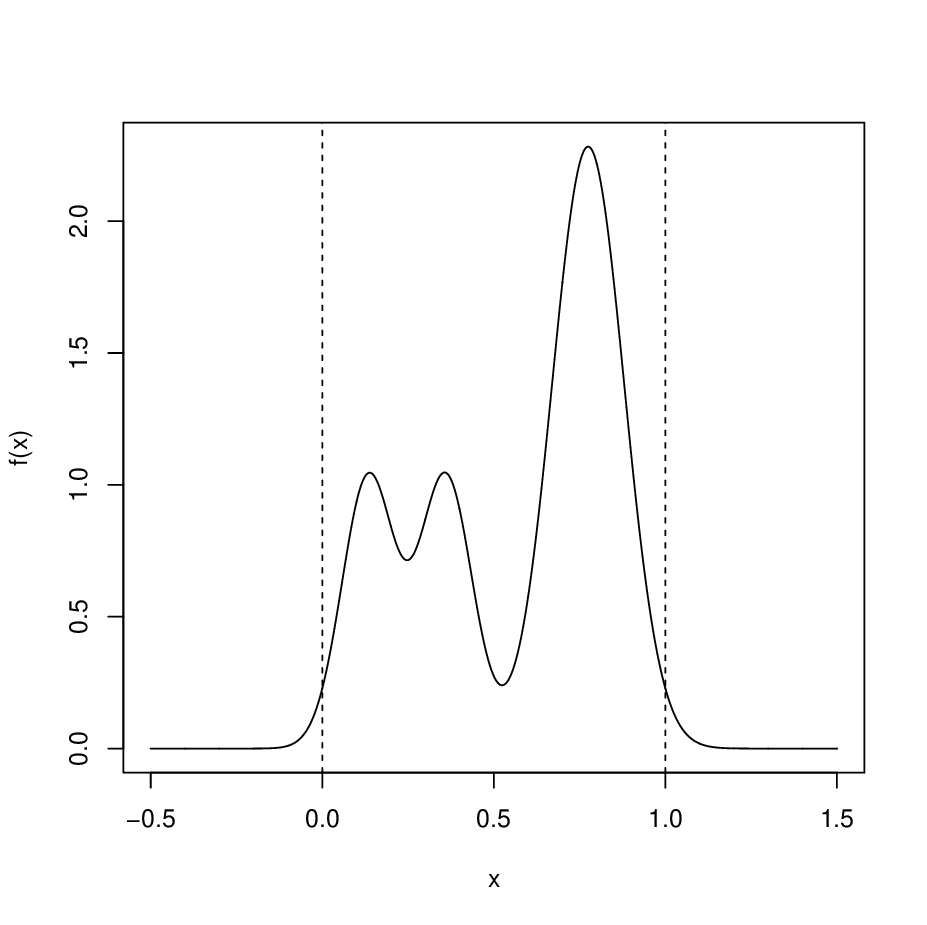}}  &  &  $n=100$   & 0(0) & 0.022(0.013) & 0.096(0.026) \\ 
   &  &  $n=200$   &  0.002(0.004) & 0.144(0.031) & 0.428(0.043)  &  &  &  $n=200$   &  0.008(0.008) & 0.048(0.019) & 0.138(0.030)  \\ 
\cline{2-6} \cline{8-12} &  \multirow{3}{*}{FM}  &  $n=50$   &  0.036(0.016)  &  0.142(0.031)  &  0.300(0.040)  &  &  \multirow{3}{*}{FM}  &  $n=50$  &  0.068(0.022)  &  0.200(0.035)  &  0.312(0.041)  \\ 
   &  &  $n=100$   &  0.248(0.038)  &  0.610(0.043)  &  0.804(0.035)  &  &  &  $n=100$   &  0.170(0.033)  &  0.344(0.042)  &  0.462(0.044)  \\ 
   &  &  $n=200$   &  0.740(0.038)  &  0.960(0.017)  &  0.982(0.012)  &  &  &  $n=200$   &  0.190(0.034)  &  0.404(0.043)  &  0.552(0.044)  \\ 
\cline{2-6} \cline{8-12} &  \multirow{3}{*}{NP}  &  $n=50$   &  0.080(0.024)  &  0.256(0.038)  &  0.402(0.043)   &  &  \multirow{3}{*}{NP}  &  $n=50$  &  0.018(0.012)  &  0.098(0.026)  &  0.148(0.031)   \\ 
   &  &  $n=100$   &  0.266(0.039)  &  0.542(0.044)  &  0.706(0.040)   &  &  &  $n=100$   &  0.098(0.026)  &  0.232(0.037)  &  0.320(0.041)   \\ 
   &  &  $n=200$   &  0.890(0.027)  &  0.956(0.018)  &  0.980(0.012)   &  &  &  $n=200$   &  0.106(0.027)  &  0.248(0.038)  &  0.356(0.042)   \\ \hline
   M23  &  \multirow{3}{*}{SI}  &  $n=50$ & 0(0) & 0.012(0.010) & 0.078(0.024)  & \multicolumn{6}{c}{}  \\ 
   \multirow{8}{*}{\includegraphics[width=23mm]{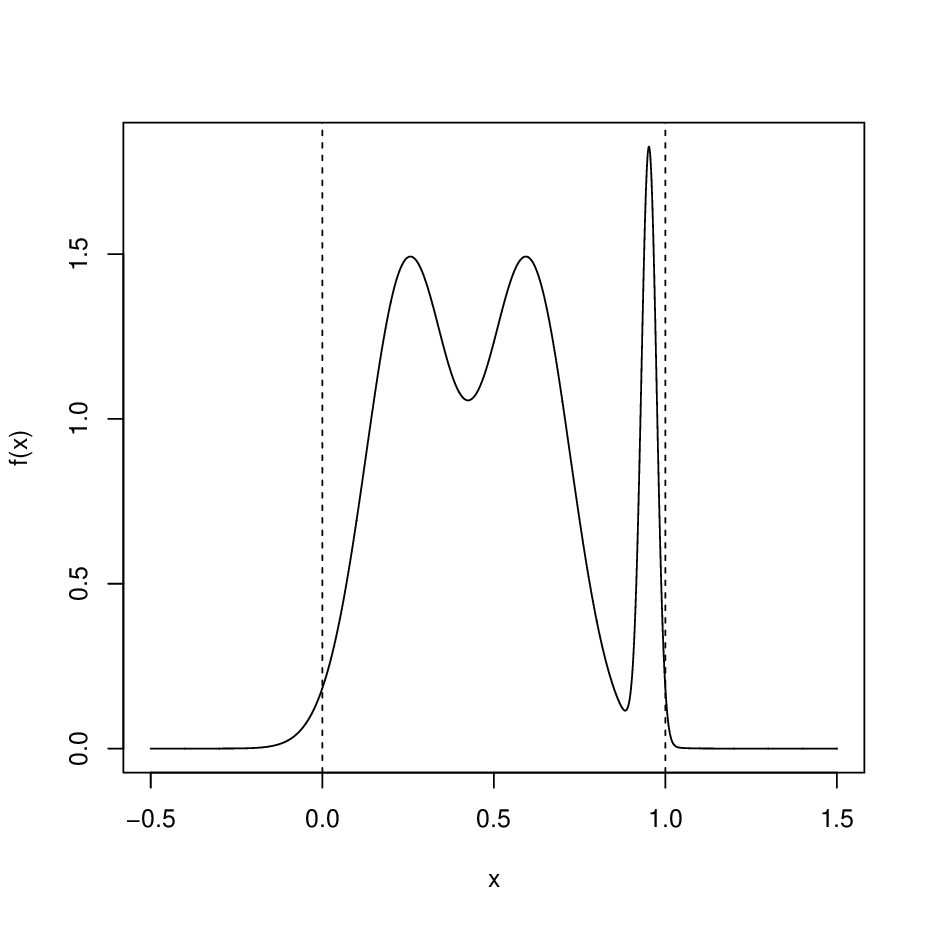}}  &  &  $n=100$   &  0(0) & 0.130(0.029) & 0.330(0.041)  & \multicolumn{6}{c}{}   \\ 
   &  &  $n=200$   &  0.108(0.027) & 0.570(0.043) & 0.752(0.038)  & \multicolumn{6}{c}{}   \\ 
\cline{2-6}  &  \multirow{3}{*}{FM}  &  $n=50$ &  0.054(0.020)  &  0.196(0.035)  &  0.338(0.041)  & \multicolumn{6}{c}{}   \\ 
   &  &  $n=100$   &  0.334(0.041)  &  0.658(0.042)  &  0.780(0.036)  &  \multicolumn{6}{c}{}  \\ 
   &  &  $n=200$   &  0.832(0.033)  &  0.906(0.026)  &  0.938(0.021)  & \multicolumn{6}{c}{}   \\ 
\cline{2-6}  &  \multirow{3}{*}{NP}  &  $n=50$ &  0.050(0.019)  &  0.204(0.035)  &  0.336(0.041)   & \multicolumn{6}{c}{}   \\ 
   &  &  $n=100$   &  0.326(0.041)  &  0.624(0.042)  &  0.746(0.038)   &  \multicolumn{6}{c}{}  \\ 
   &  &  $n=200$   &  0.722(0.039)  &  0.878(0.029)  &  0.934(0.022)   &  \multicolumn{6}{c}{} \\ 
\cline{1-6}

\end{tabular}
}
\caption{Percentages of rejections for testing $H_0:j = 2$, with $500$ simulations ($1.96$ times their estimated standard deviation in parenthesis) and $B = 500$ bootstrap samples.}
\label{estsim6}
\end{table}

\begin{table}
\centering
\scalebox{0.8}{
\begin{tabular}{|c|c|c|c c c|}
  \hline
 &  & $\alpha$ & 0.01 & 0.05& 0.10\\
  \hline
 &  &  &\multicolumn{3}{c|}{NP (known support)}\\ \hline 
$n=50$ & \multirow{3}{*}{M9}& \multirow{3}{*}{\includegraphics[width=20mm]{120i2.eps}} & 0(0) & 0.014(0.010) & 0.034(0.016) \\
$n=200$ & & & 0.012(0.010) & 0.052(0.019) & 0.090(0.025) \\
 $n=1000$ & & & 0.010(0.009) & 0.040(0.017) & 0.092(0.025)\\ \hline
$n=50$ & \multirow{3}{*}{M16}& \multirow{3}{*}{\includegraphics[width=20mm]{121f.eps}} &  0.002(0.004) & 0.024(0.013) & 0.076(0.023) \\ 
$n=200$ & & & 0.008(0.008) & 0.058(0.020) & 0.126(0.029) \\ 
 $n=1000$ & & & 0.016(0.011) & 0.038(0.017) & 0.080(0.024)  \\ 
\hline
	\end{tabular}
}
\caption{Percentages of rejections for testing $H_0:j =1$ (in model M9) and $H_0:j = 2$ (in model M16), with $500$ simulations ($1.96$ times their estimated standard deviation in parenthesis) and $B = 500$ bootstrap samples.}
\label{estsim7}
\end{table}

\section{Real data analysis}
\label{data}

{Before the 1940 decade, stamps images were printed on a variety of paper types and, in general, with a lack of quality control in manufactured paper, which led to important fluctuations in paper thickness, being thin stamps more likely to be produced than thick ones. Given that the price of any stamp depends on its scarcity, the thickness of the paper is crucial for determining its value. However, there is not a standard rule for classifying stamps according to their thickness (not being available such a classification in catalogues), becoming this problem even harder in stamp issues printed on a mixture of paper types with possible differences in their thickness.
For the 1872 Hidalgo stamp issue, it is known that the scarcity of ordinary white wove paper led the utilization of other types of paper (some of them watermarked), such us the white wove paper \textit{Papel Sellado} or the \textit{La Croix--Freres} (\textit{LA+-F}). Some references exploring the number of groups in stamp thickness, and further comments, on this example are given in Section \ref{moredata} in Supplementary Material.}

Taking a sample of 485 stamps, \citet{IzenSom88} revisited this problem previously studied by \citet{Wilson83} who concluded that there were only two kinds of paper (\textit{Papel Sellado} and \textit{La Croix--Freres}) by observing a histogram similar to the one represented in Figure \ref{fig1} (left panel, with dashed border). The same conclusions can be obtained using a kernel density estimator with a rule of thumb bandwidth (left panel, dashed curve). However, both the histogram and the kernel density estimator, depend heavily on the bin width and bandwidth, as it can be seen in Figure \ref{fig1} and different values of these tuning parameters may lead to different conclusions about the number of modes. Given that the exploratory tools did not provide a formal way of determining the number groups, \citet{IzenSom88} employed the multimodality test of \citet{Silverman81}. Results from \citet{IzenSom88}, applying SI with $B=100$, are shown in Table \ref{sellos1}. With a \emph{flexible} rule (due the ``conservative'' nature of this test), these authors concluded that the number of groups in the 1872 Hidalgo Issue is seven. \cite{FisMar01} also analized this example and the p--values obtained in their studio ($B=200$) are shown in Table \ref{sellos1}: it is not clear which conclusion has to be made. They mentioned that their results are consistent with the previous studies, detecting 7 modes. As shown in Section \ref{simulation}, just NP has a good calibration behaviour, even with ``small'' sample sizes, while results of SI and FM are not accurate. Then, NP can be used to figure out how many groups are there in this stamp issue. The p--values obtained with NP are also shown in the Table \ref{sellos1}, with $B=500$. Similar results can be obtained employing the interval $I=[0.04,0.15]$ in NP with known support, as \citet{IzenSom88} noticed that the thickness of the stamps is always in this interval. For a significance level $\alpha=0.05$, it can be observed that the null hypothesis is rejected until $k=4$, and then there is no evidences to reject $H_0$ for larger values of $k$. Then, applying our new procedure, the conclusion is that the number of groups in the 1872 Hidalgo Issue is four.

\begin{table}
\begin{center}
\scalebox{0.8}{
\begin{tabular}{ |c | c c c c c c c c c| }
\hline
$k$ & 1 & 2 & 3 & 4 & 5 & 6 & 7 & 8 & 9 \\ \hline
SI ($B=100$) & 0 & 0.04 & 0.06 & 0.01 & 0 & 0 & 0.44 & 0.31 & 0.82 \\ 
FM ($B=200$) &  0 & 0.04 & 0 & 0 & 0 & 0 & 0.06 & 0.01 & 0.06 \\ 
NP ($B=500$) & 0 & 0.022 & 0.004 & 0.506 & 0.574 & 0.566 & 0.376 & 0.886 & 0.808 \\ \hline
\end{tabular}
}
\caption{P--value obtained using different proposals for testing $k$--modality, with $k$ between 1 and 9. Methods: SI, FM and NP.}
\label{sellos1}
\end{center}
\end{table}

In order to compare the results obtained by \citet{IzenSom88} and the ones derived applying the new proposal, two kernel density estimators, with critical bandwidths $h_4$ and $h_7$ are depicted in Figure \ref{fig1} (right panel). \citet{IzenSom88} argued that the stamps could be divided first in three groups (pelure paper with  mode at 0.072 mm, related with the forged stamps; the medium paper in the point 0.080 mm; and the thick paper at 0.090 mm). Given the efforts made in the new issue in 1872 to avoid forged stamps, it seems quite reasonable to assume that the group associated with the pelure paper had disappeared in this new issue. In that case, the asymmetry in the first group can be attributed to the modifications in the paper made by the manufacturers. Also, this first and asymmetric group, justifies the application of non--parametric techniques to determine the number of groups. It can be seen, in the Section 7 of \citet{IzenSom88}, that the parametric techniques (such as the mixture of Gaussian densities) have problems capturing this asymmetry, and they always determine that there are two modes in this first part of the density, one near the point 0.07 mm and another one near 0.08 mm. The other groups would correspond with stamps produced in 1872, on there it seems that the stamps of 1872 were printed on two different paper types, one with the same characteristics as the unwatermarked white wove paper used in the 1868 issue, and a second much thicker paper that disappeared completely by the end of 1872. Using this explanation, it seems quite reasonable to think that the two final modes using $h_4$, near the points 0.10 and 0.11 mm, correspond to the medium paper and the thick paper in this second block of stamps produced in 1872. Finally, for the two minor modes appearing near 0.12 and 0.13 mm, when $h_7$ is used, \citet{IzenSom88} do not find an explanation and they mention that probably they could be artefacts of the estimation procedure. This seems to confirm the conclusions obtained with our new procedure. The reason of determining more groups than the four obtained with our proposal, seems to be quite similar to that of the model M20 in our simulation study. This possible explanation is that the spurious data in the right tail of the last mode are causing the rejection of $H_0$.

\section{Discussion}
\label{discussion}
Determining the number of modes in a distribution is a relevant practical problem in many applied sciences. The proposal presented in this paper provides a good performance for the testing problem (\ref{test}), being in the case of a general number of modes $k$ the only alternative with a reasonable behaviour. The totally nonparametric testing procedure can be extended to other contexts where a natural nonparametric estimator under the null hypothesis is available. For instance, the method can be adapted for dealing with periodic data, as it happens with the proposal by \cite{FisMar01}.

In practical problems, where a large number of tests must be computed, obtaining a set of p--values is a crucial task. In this setting, such a computation should be accompanied by the application of FDR correction techniques. The proposal in this work, based on the use of critical bandwidth and excess mass ideas, and its combination with FDR, is computationally feasible. With the aim of making this procedure accessible for the scientific community, and therefore, enabling its use in large size practical problems, an R package has been developed.

\newpage

\appendix

\renewcommand{\thesection}{SM\arabic{section}}
\renewcommand{\theremark}{SM\arabic{remark}}
\renewcommand{\theequation}{SM\arabic{equation}}

\section{Models for simulation study}\label{modelos}

\begin{figure}
\begin{multicols}{4}
\centering
\includegraphics[width=1\linewidth]{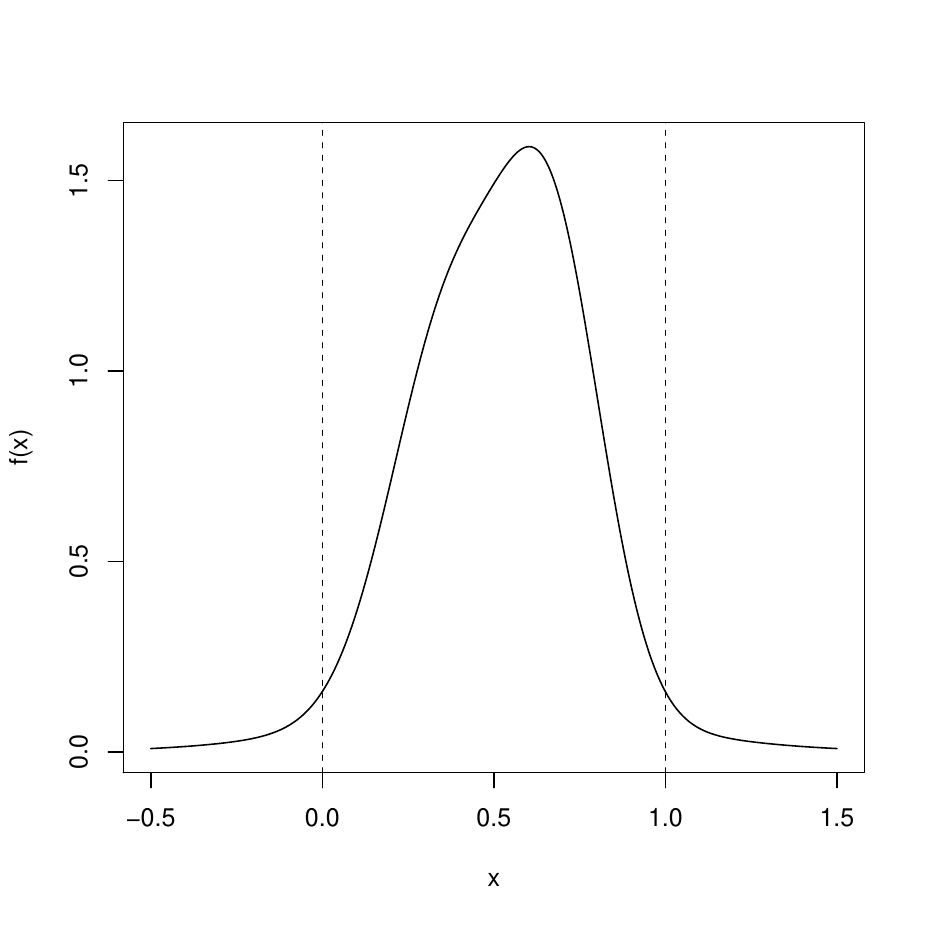}\\
M1
\includegraphics[width=1\linewidth]{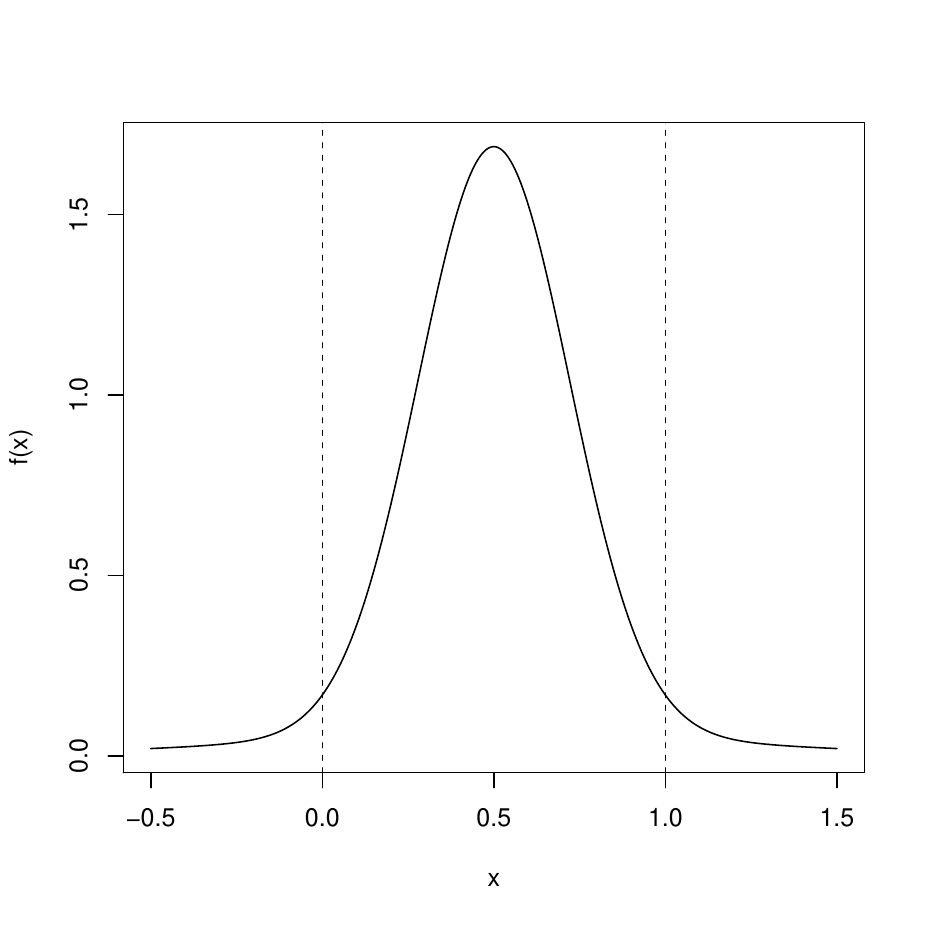}\\
M5
\includegraphics[width=1\linewidth]{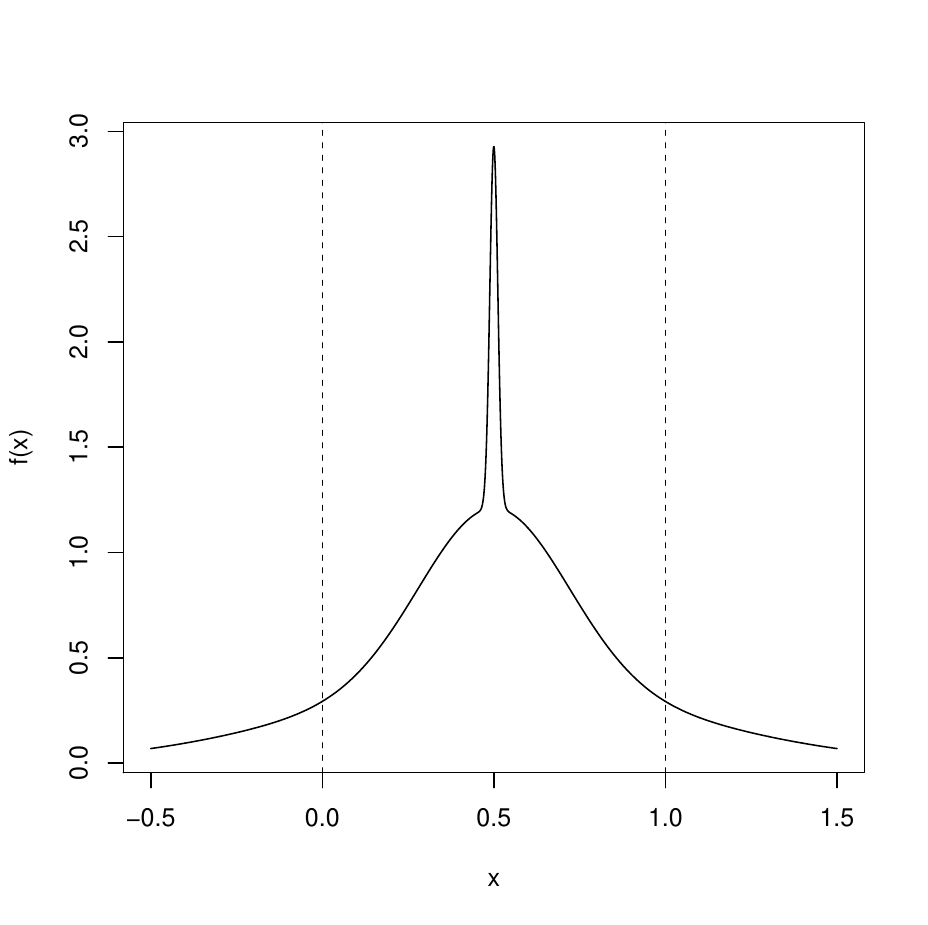}\\
M9
\includegraphics[width=1\linewidth]{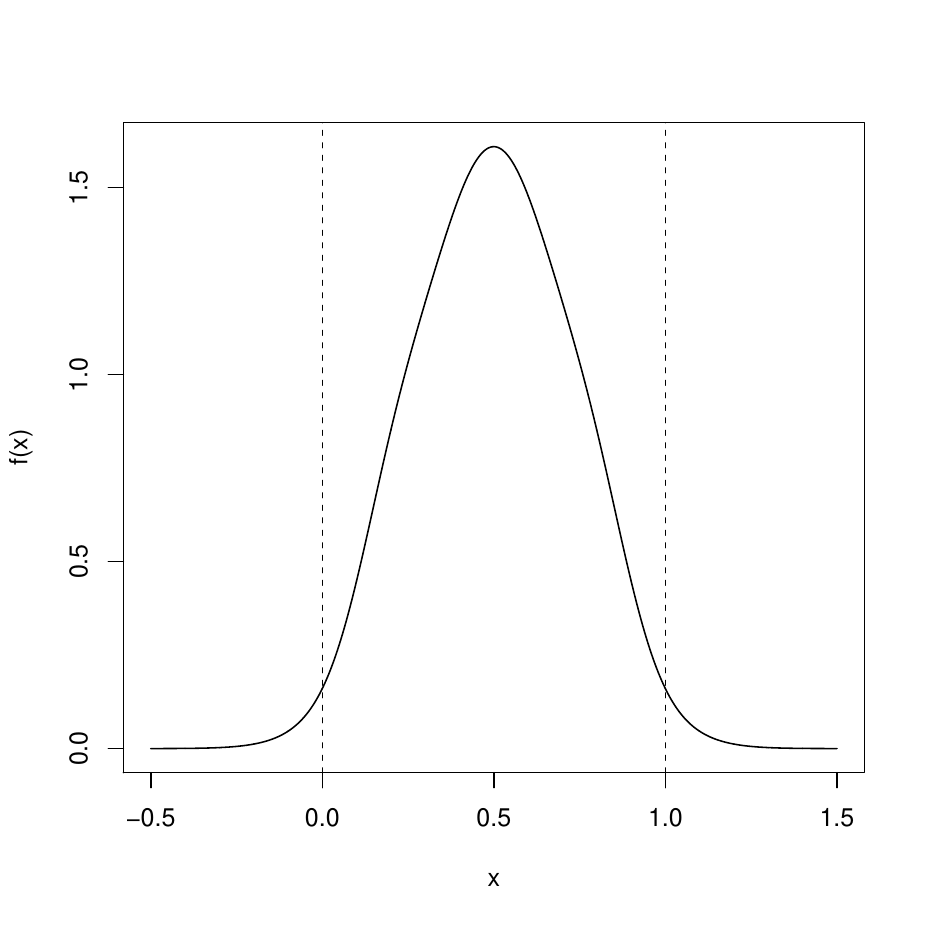}\\
M2
\includegraphics[width=1\linewidth]{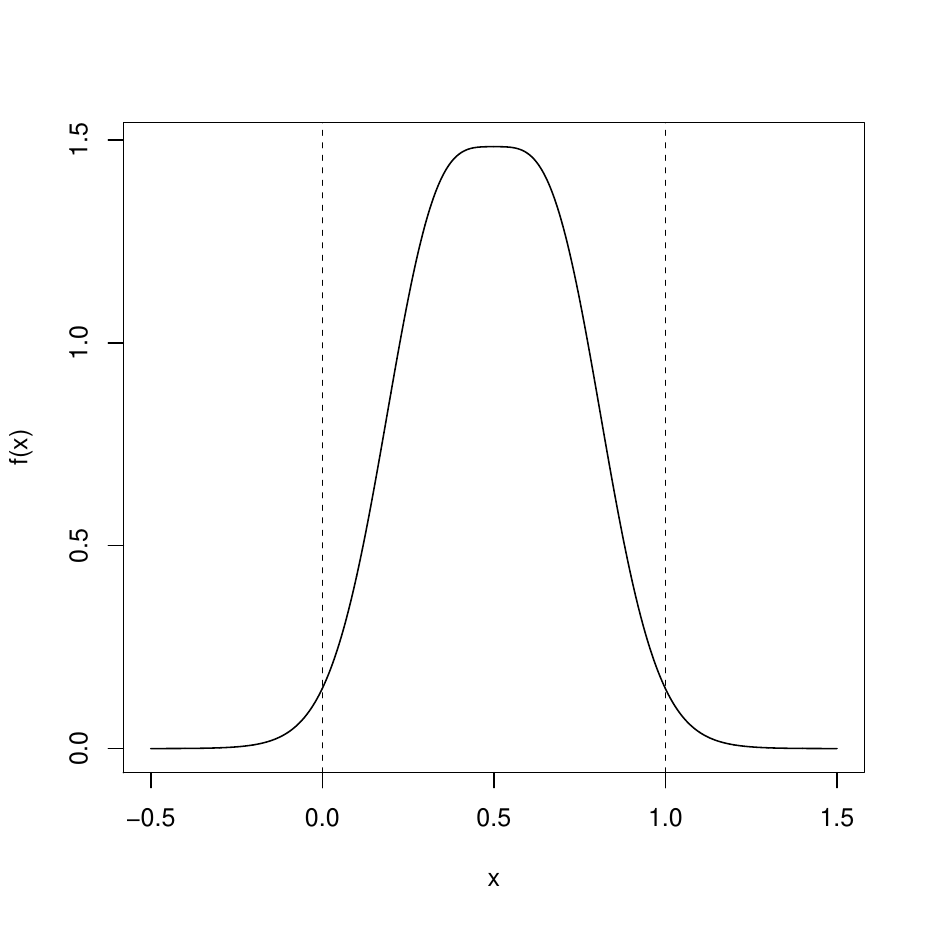}\\
M6
\includegraphics[width=1\linewidth]{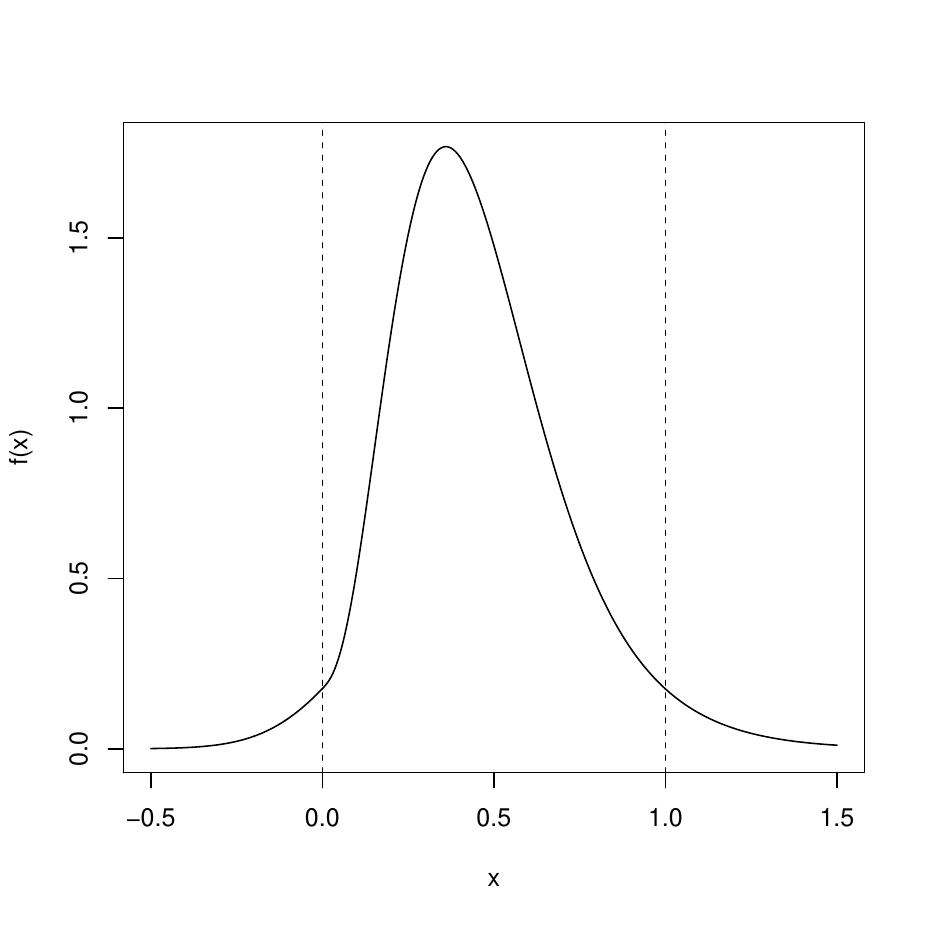}\\
M10
\includegraphics[width=1\linewidth]{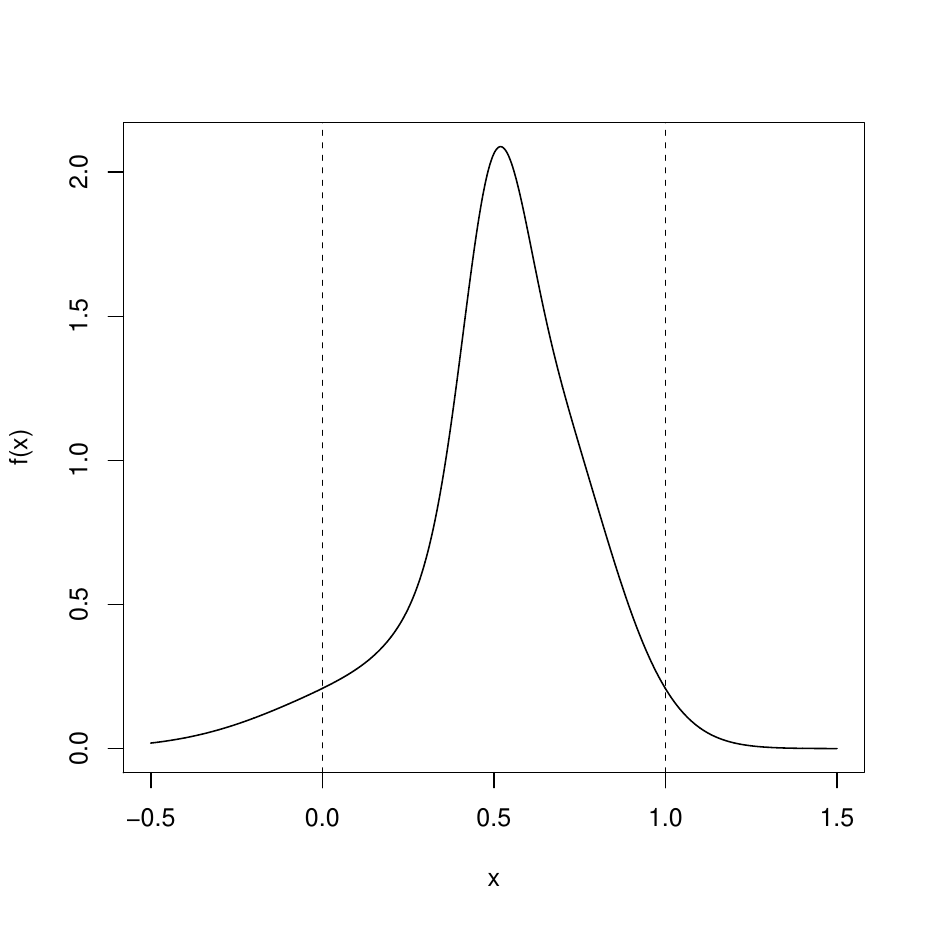}\\
M3
\includegraphics[width=1\linewidth]{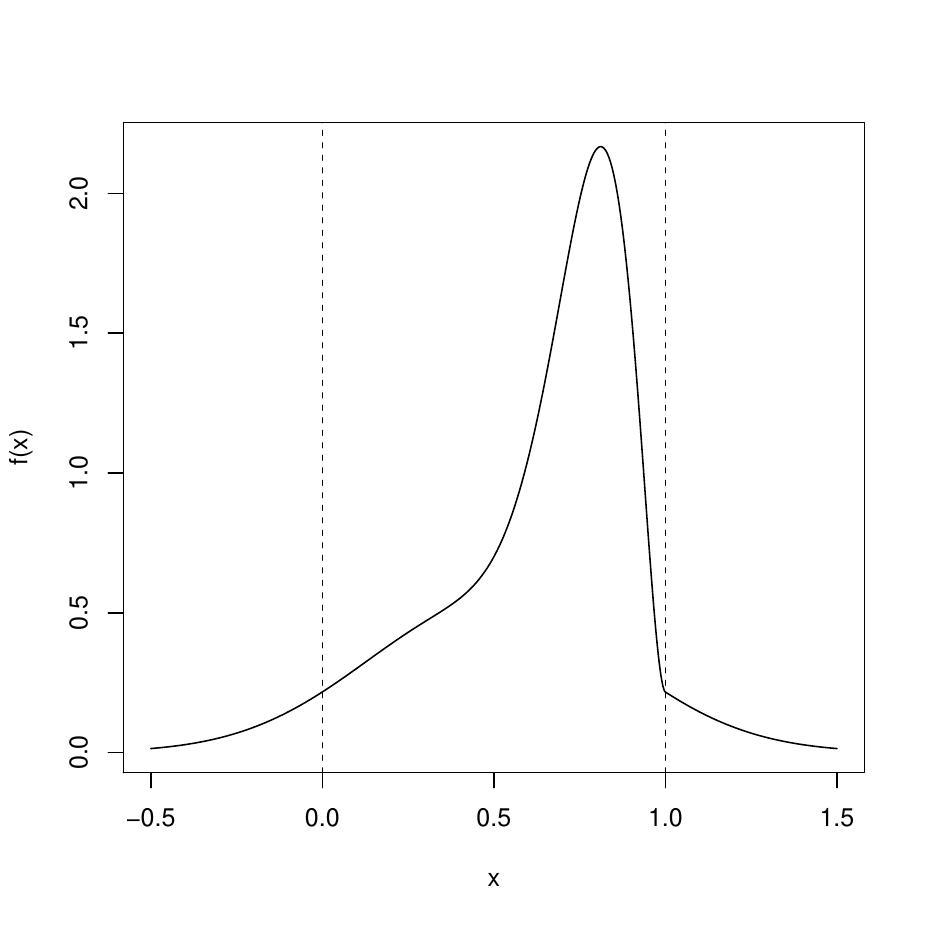}\\
M7
\includegraphics[width=1\linewidth]{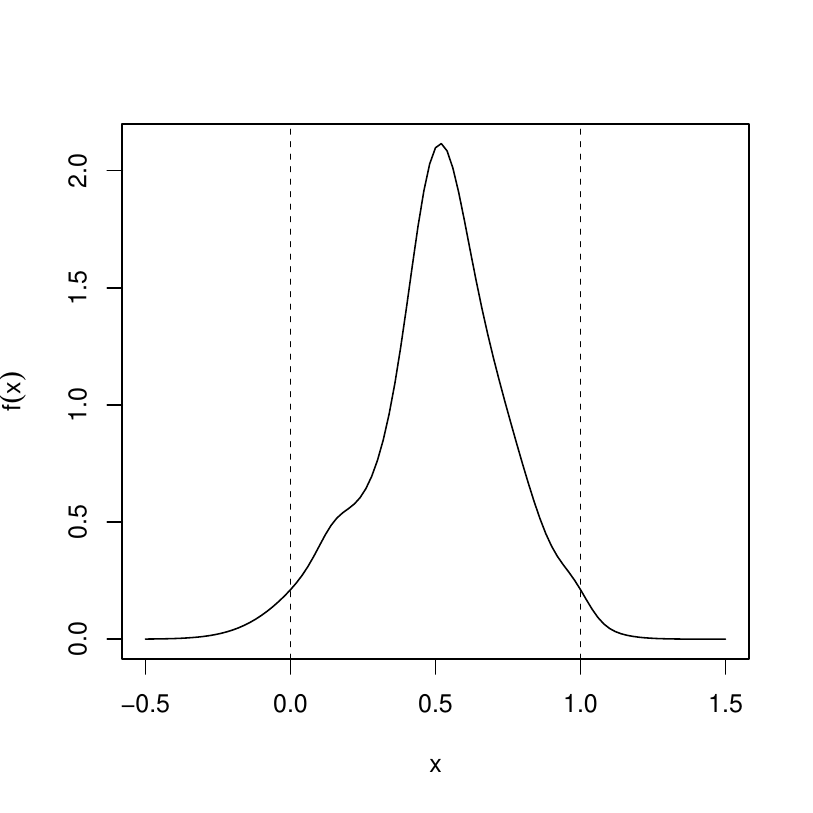}\\
M26
\includegraphics[width=1\linewidth]{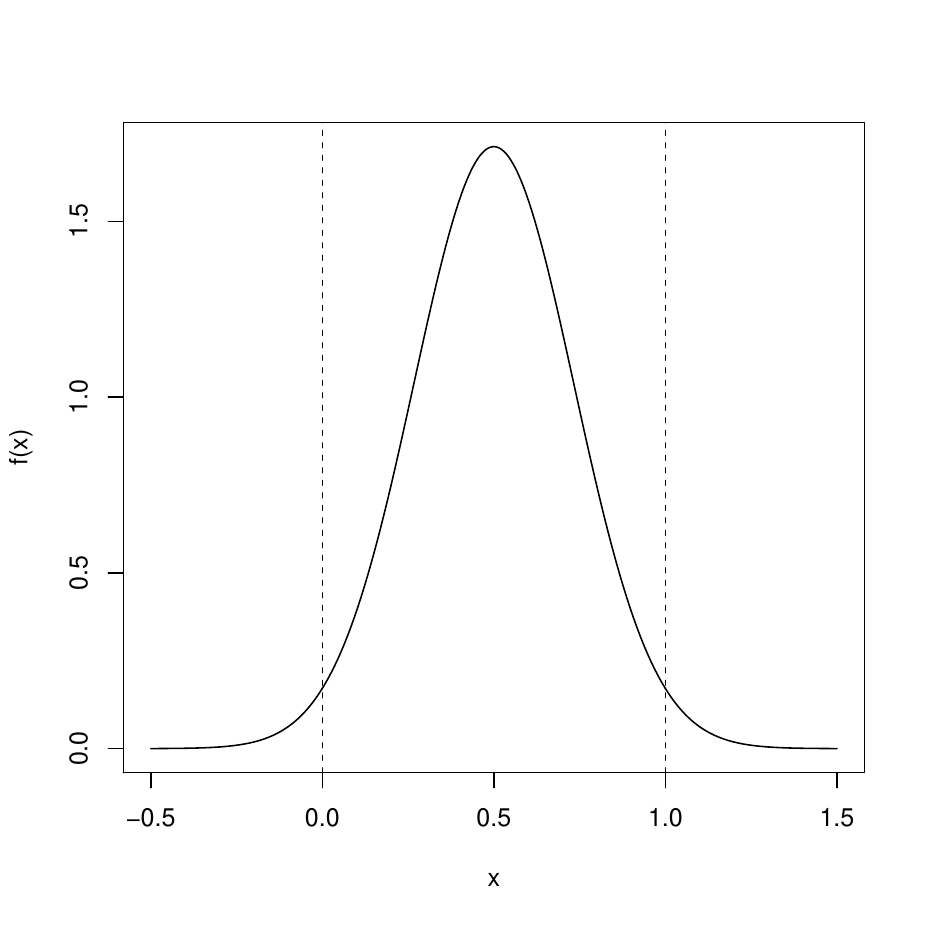}\\
M4
\includegraphics[width=1\linewidth]{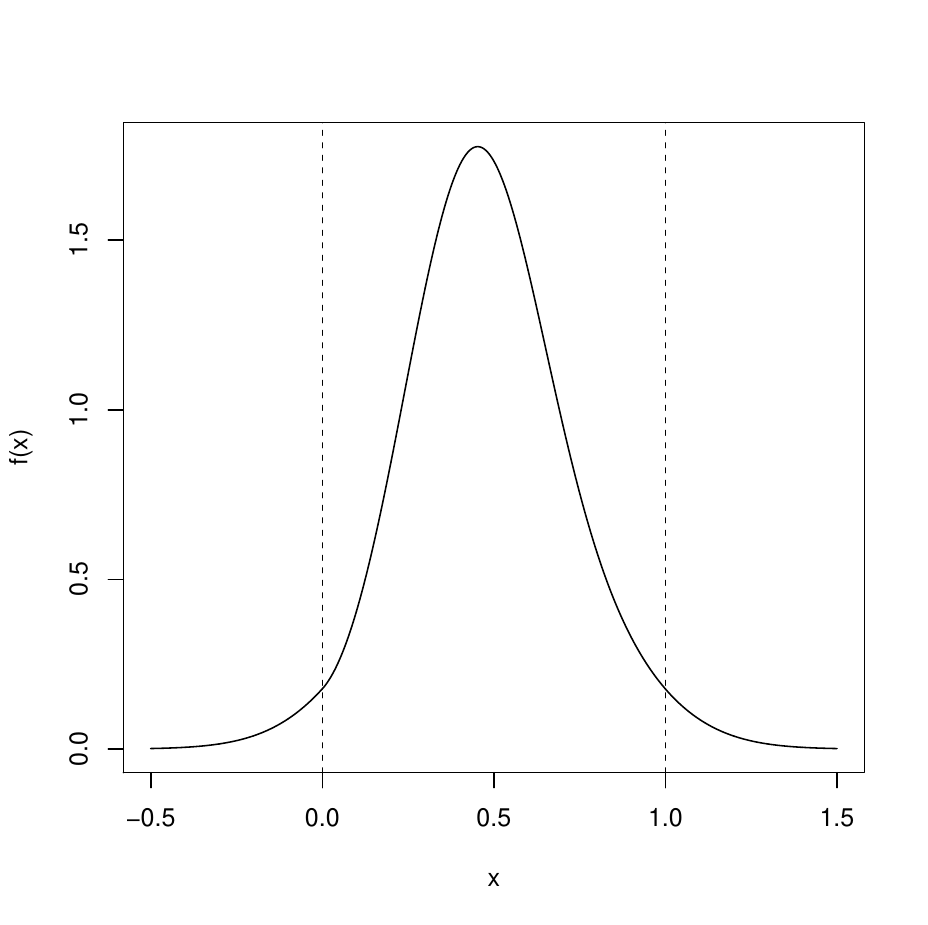}\\
M8
\includegraphics[width=1\linewidth]{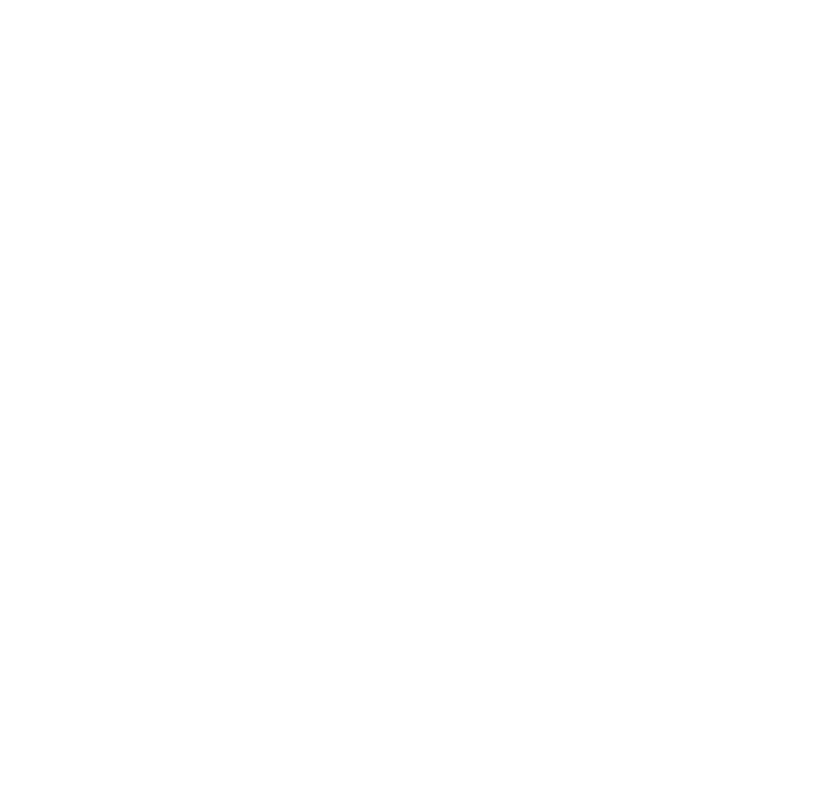}\\
\phantom{M27}
\end{multicols}
\caption{Unimodal density functions: M1--M10 and M26.}
\label{figs1}
\end{figure}

\begin{figure}
\begin{multicols}{4}
\centering
\includegraphics[width=.9\linewidth]{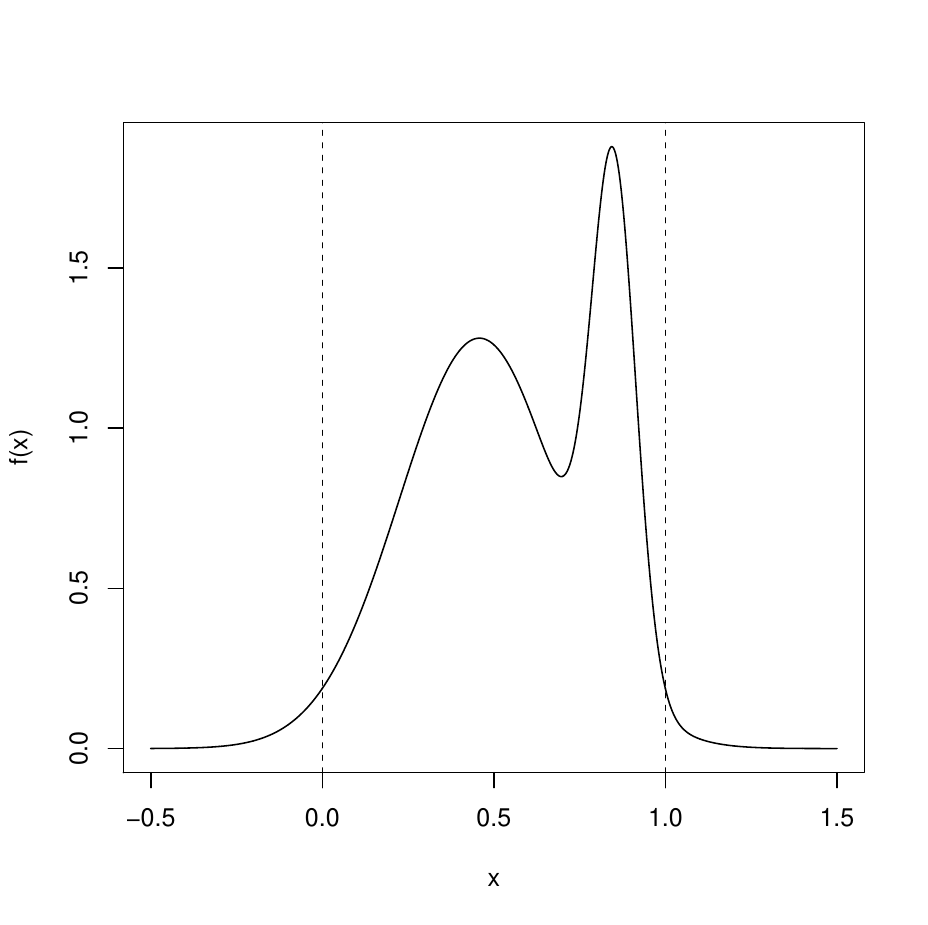}\\
M11
\includegraphics[width=.9\linewidth]{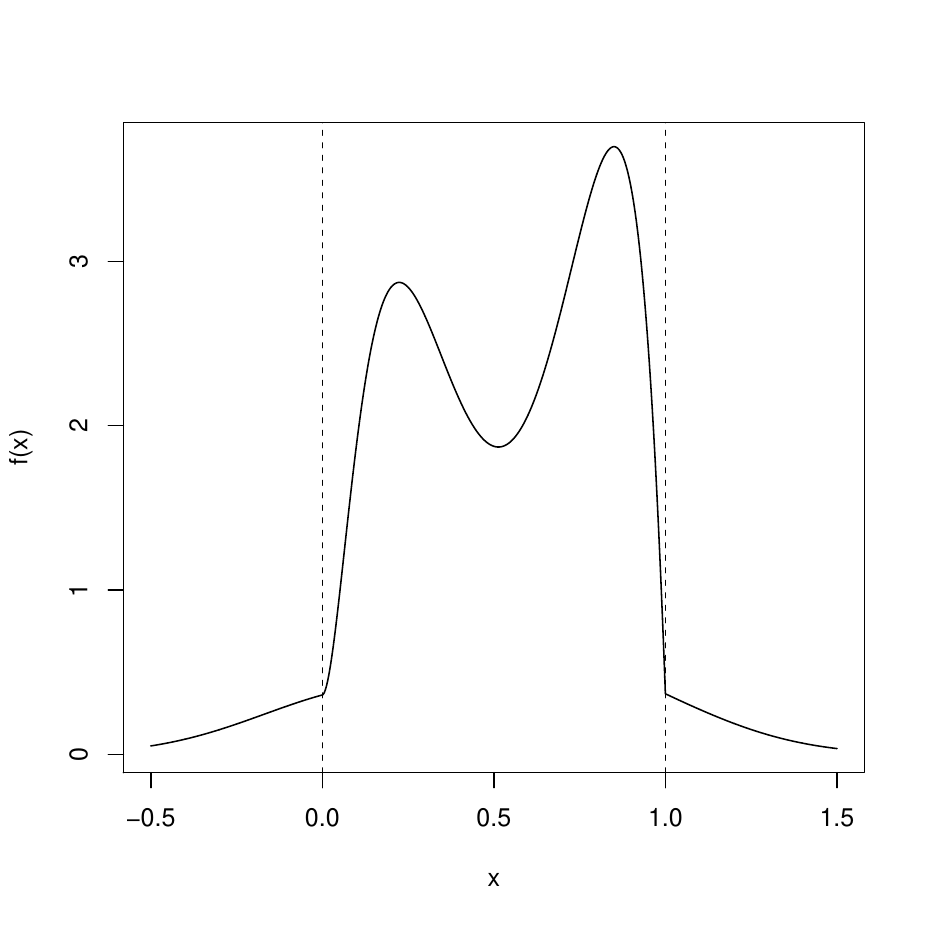}\\
M15
\includegraphics[width=.9\linewidth]{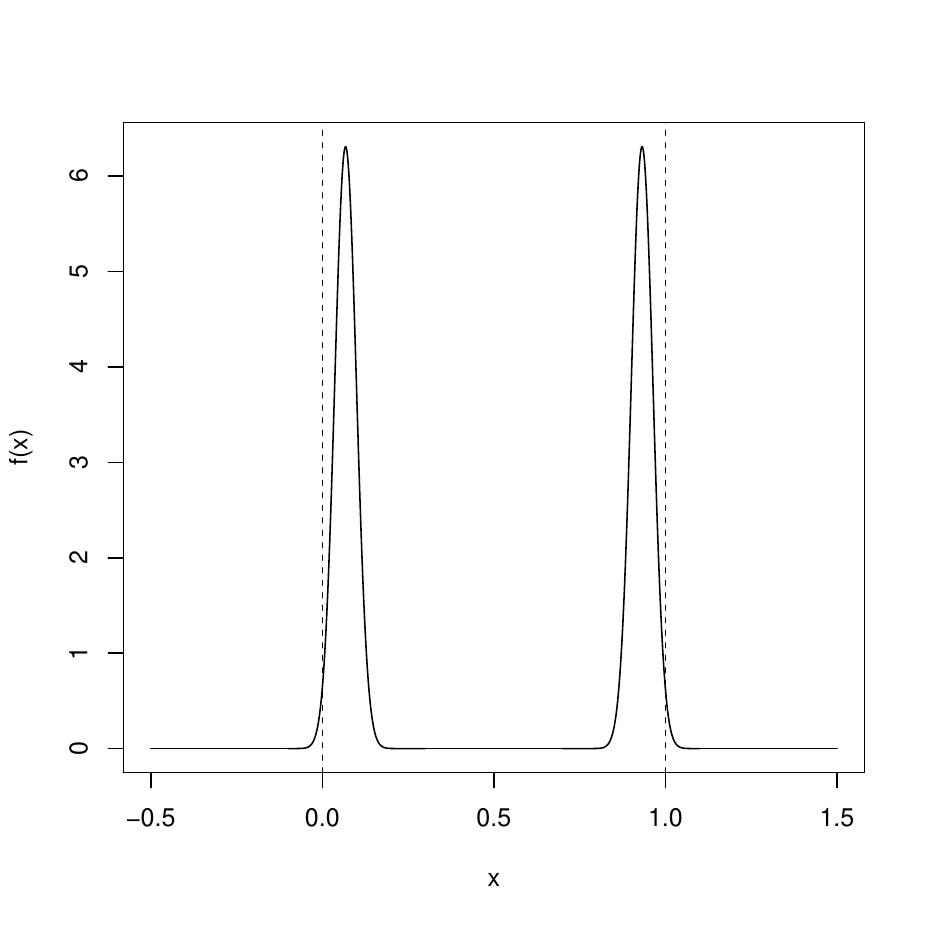}\\
M19
\includegraphics[width=.9\linewidth]{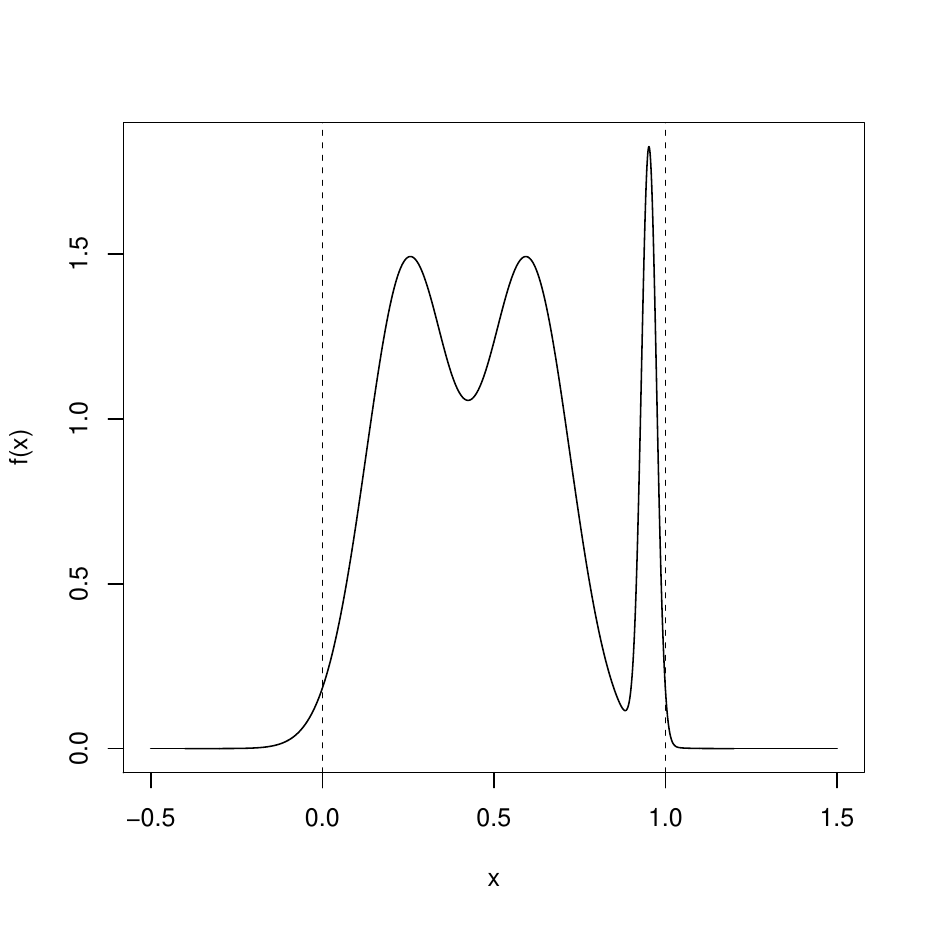}\\
M23
\includegraphics[width=.9\linewidth]{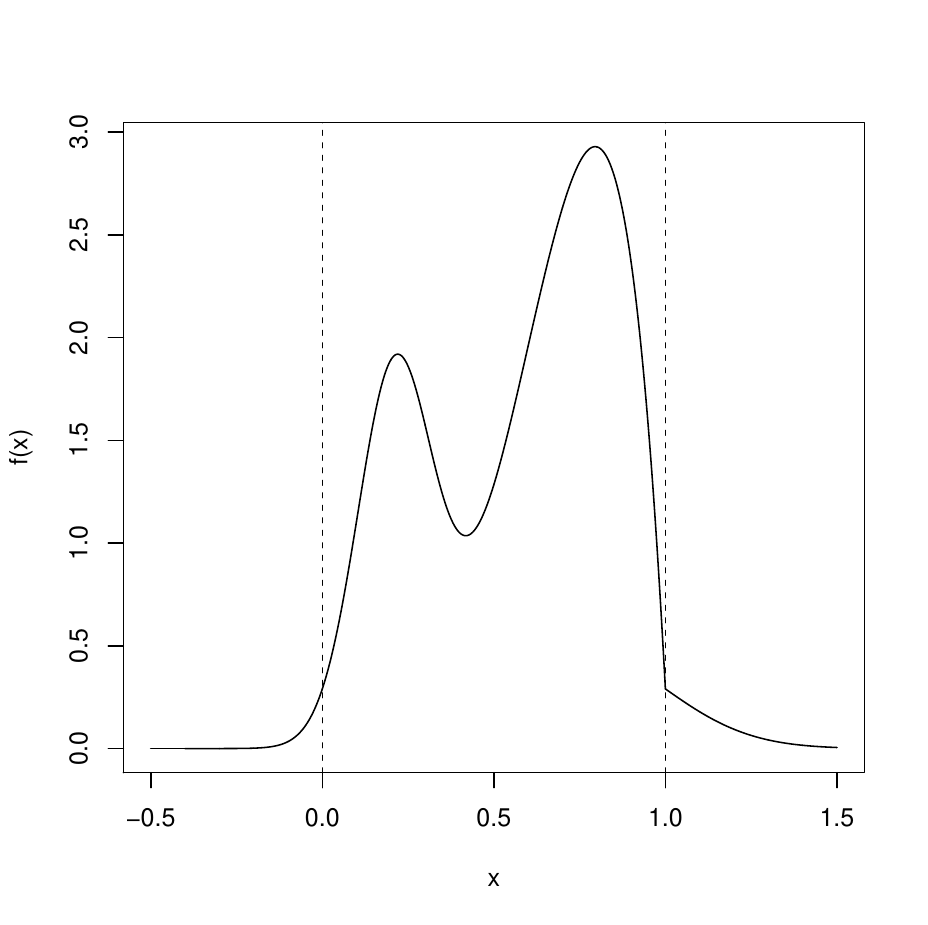}\\
M12
\includegraphics[width=.9\linewidth]{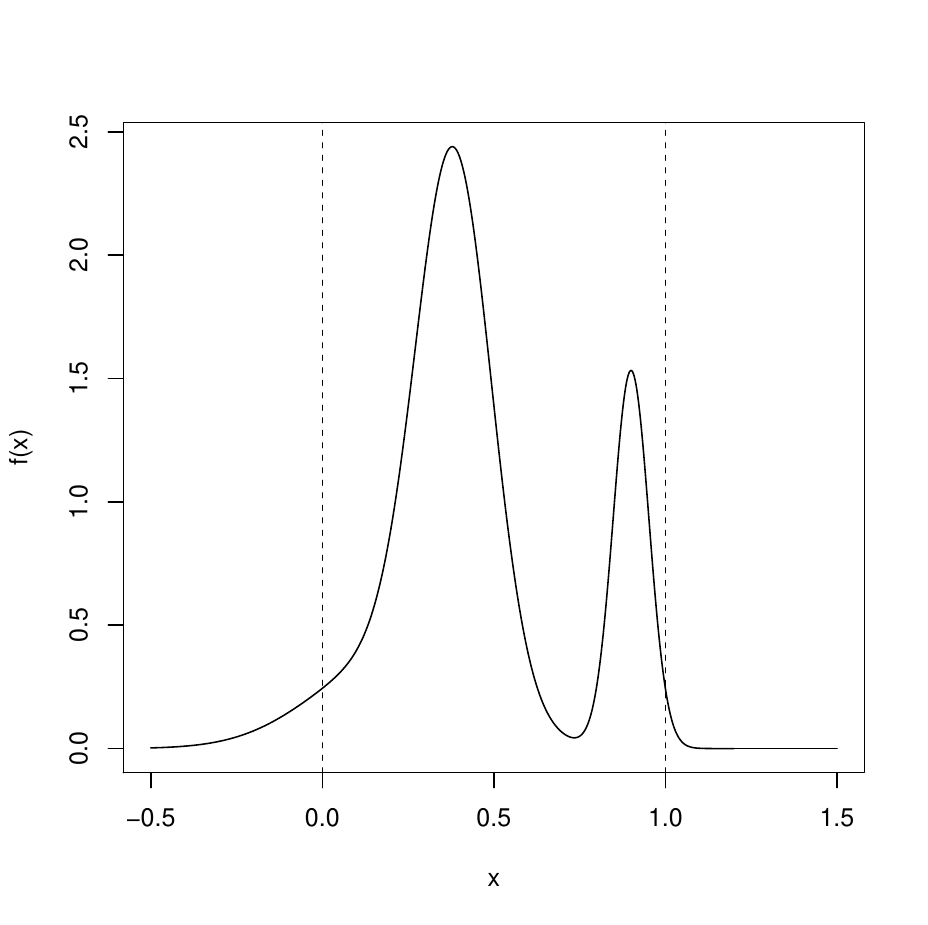}\\
M16
\includegraphics[width=.9\linewidth]{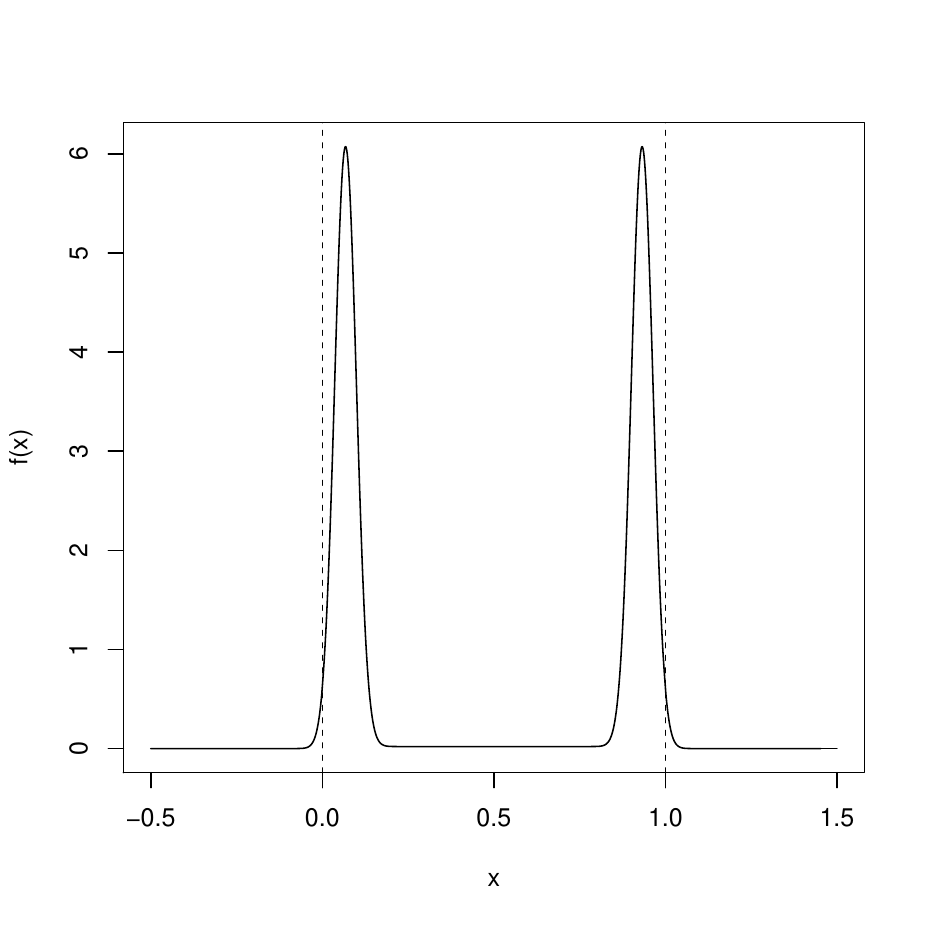}\\
M20
\includegraphics[width=.9\linewidth]{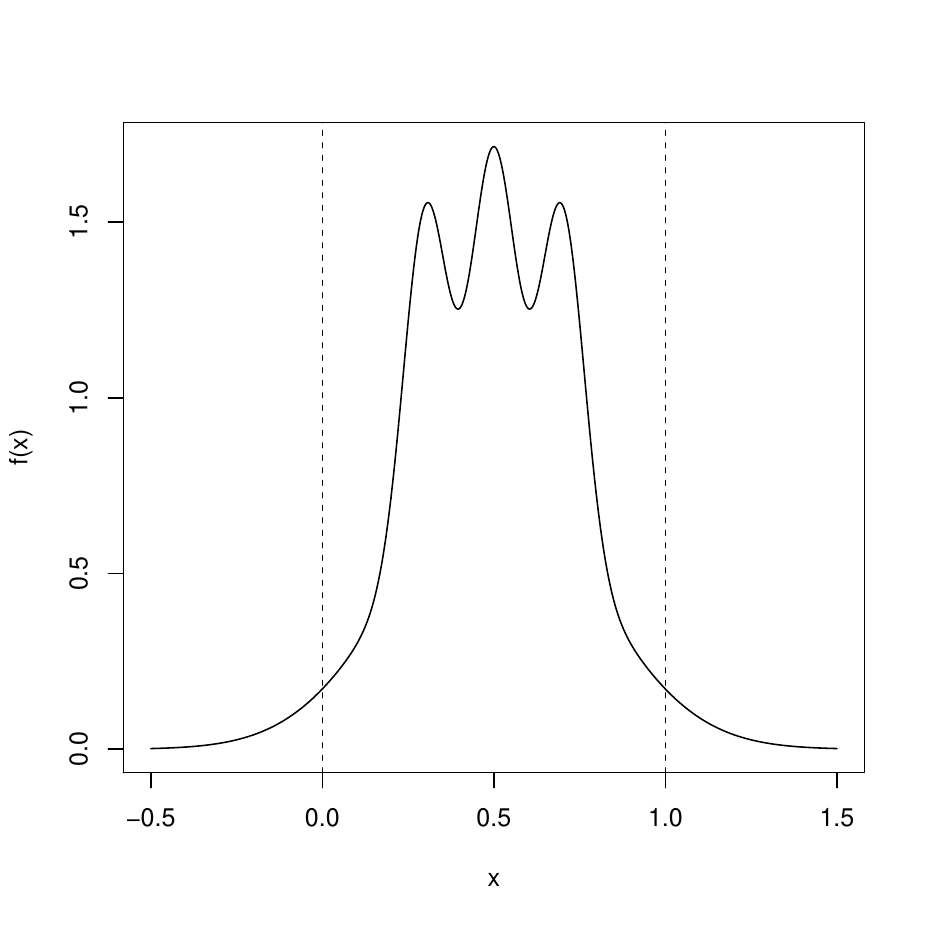}\\
M24
\includegraphics[width=.9\linewidth]{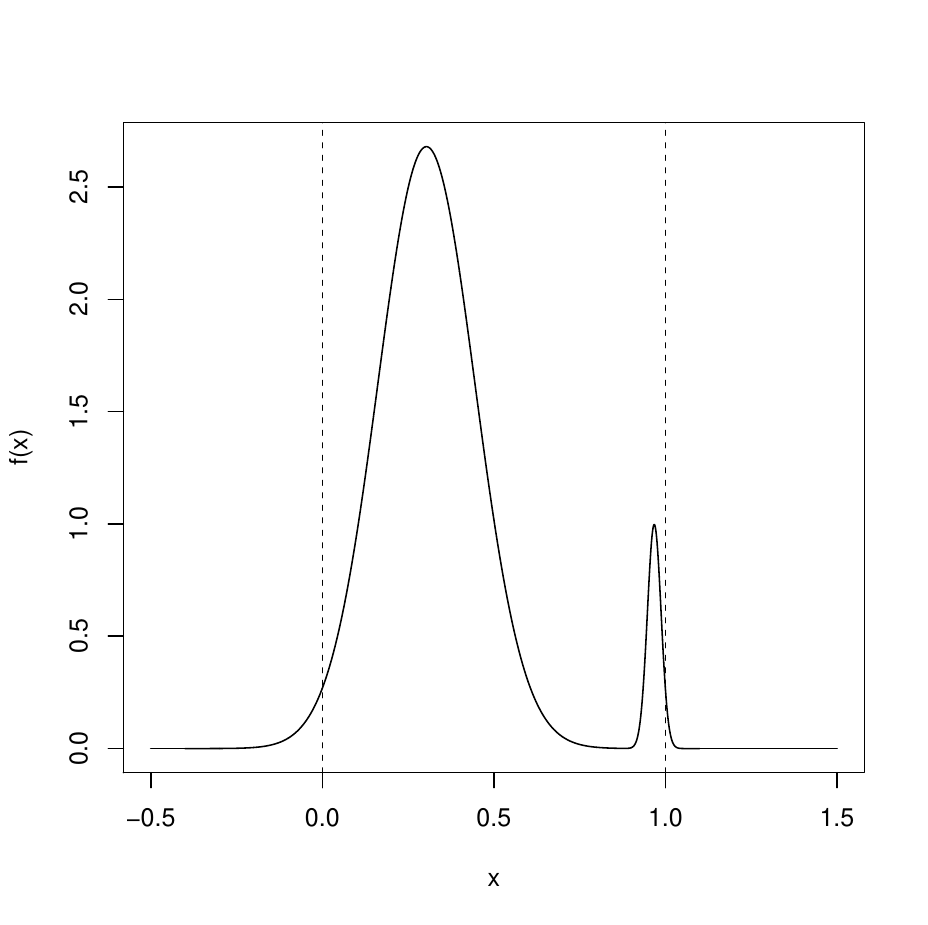}\\
M13
\includegraphics[width=.9\linewidth]{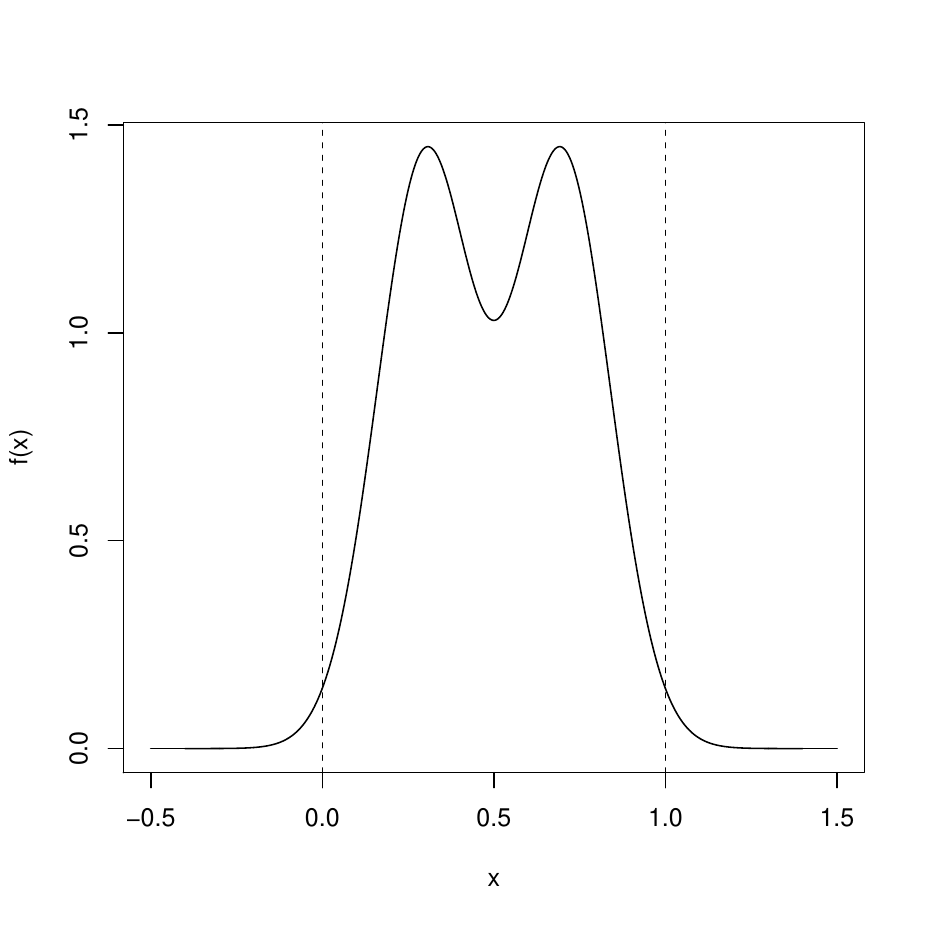}\\
M17
\includegraphics[width=.9\linewidth]{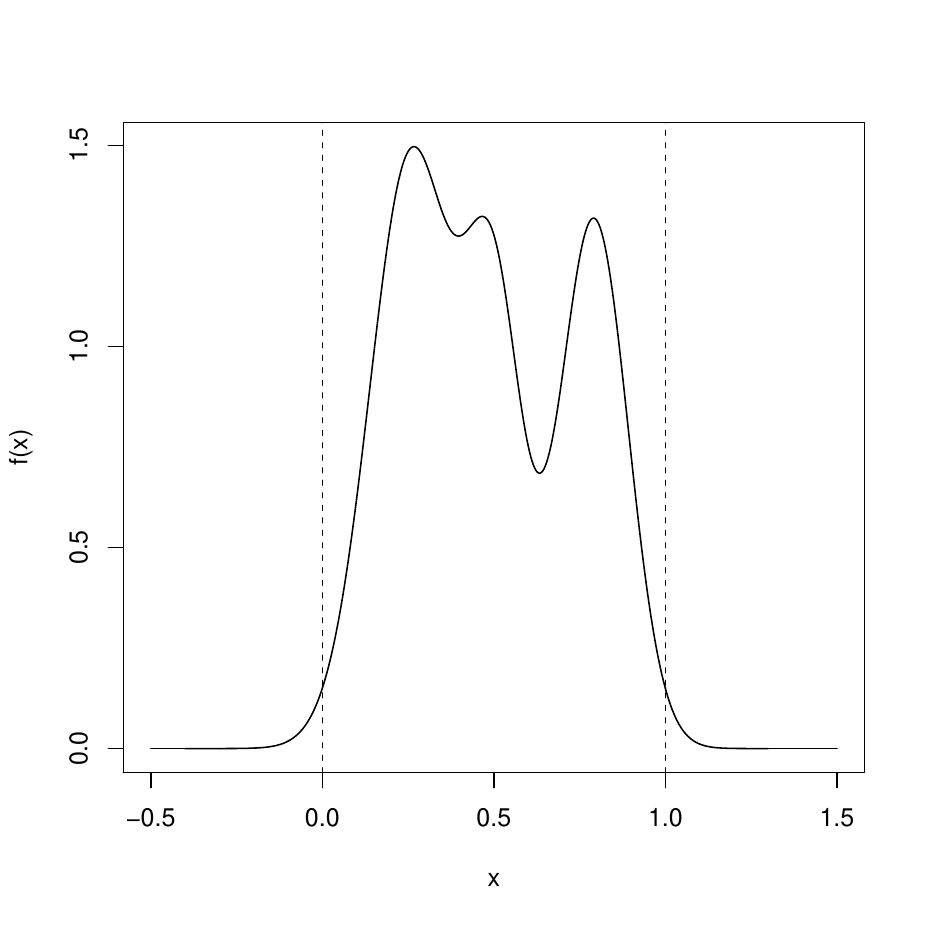}\\
M21
\includegraphics[width=.9\linewidth]{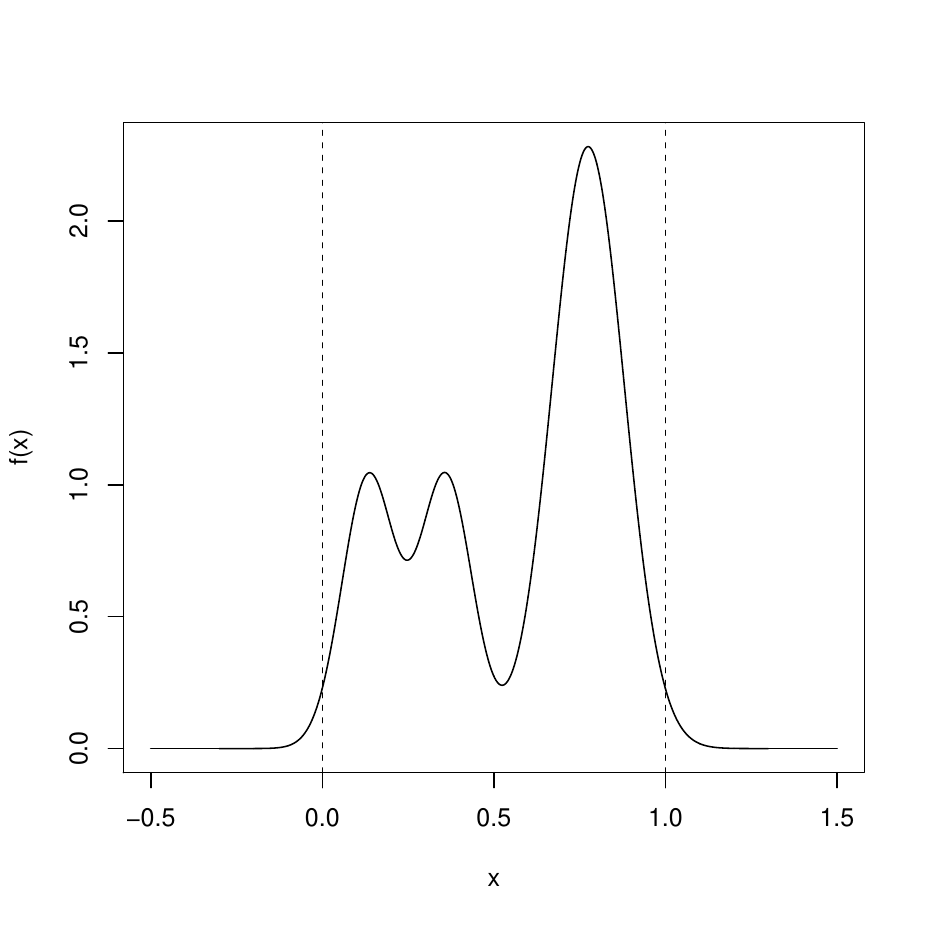}\\
M25
\includegraphics[width=.9\linewidth]{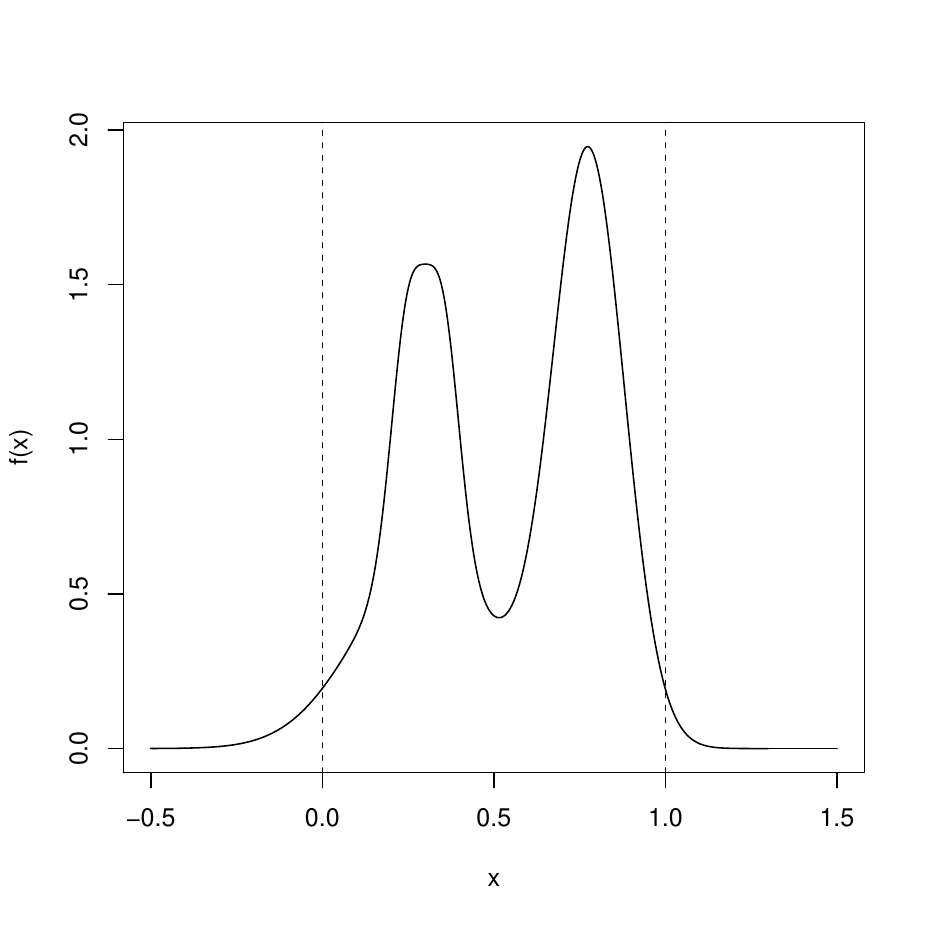}\\
M14
\includegraphics[width=.9\linewidth]{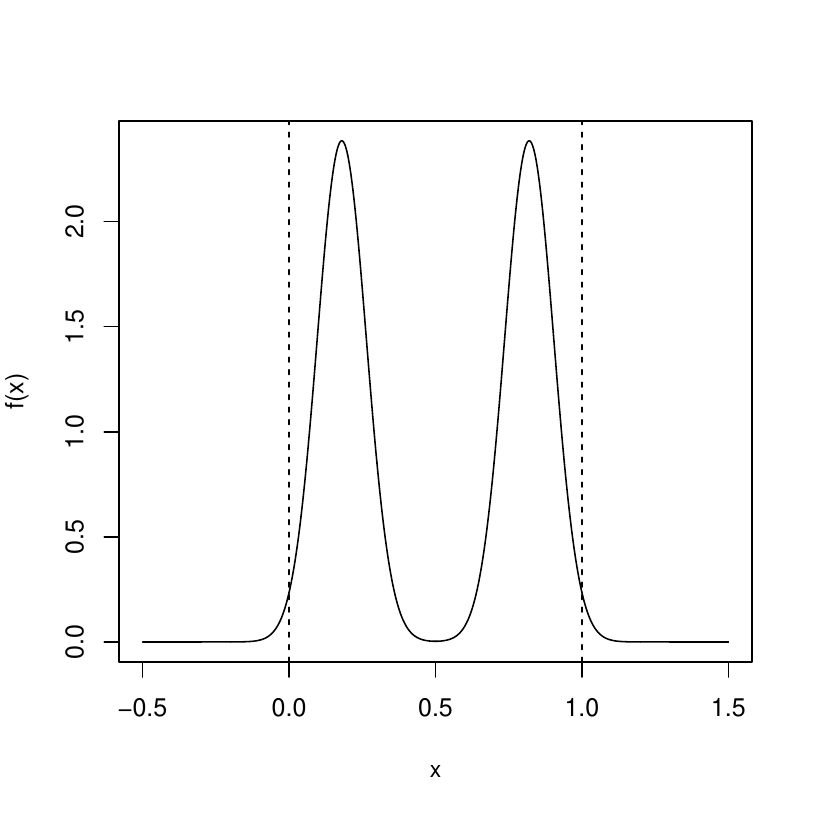}\\
M18
\includegraphics[width=.9\linewidth]{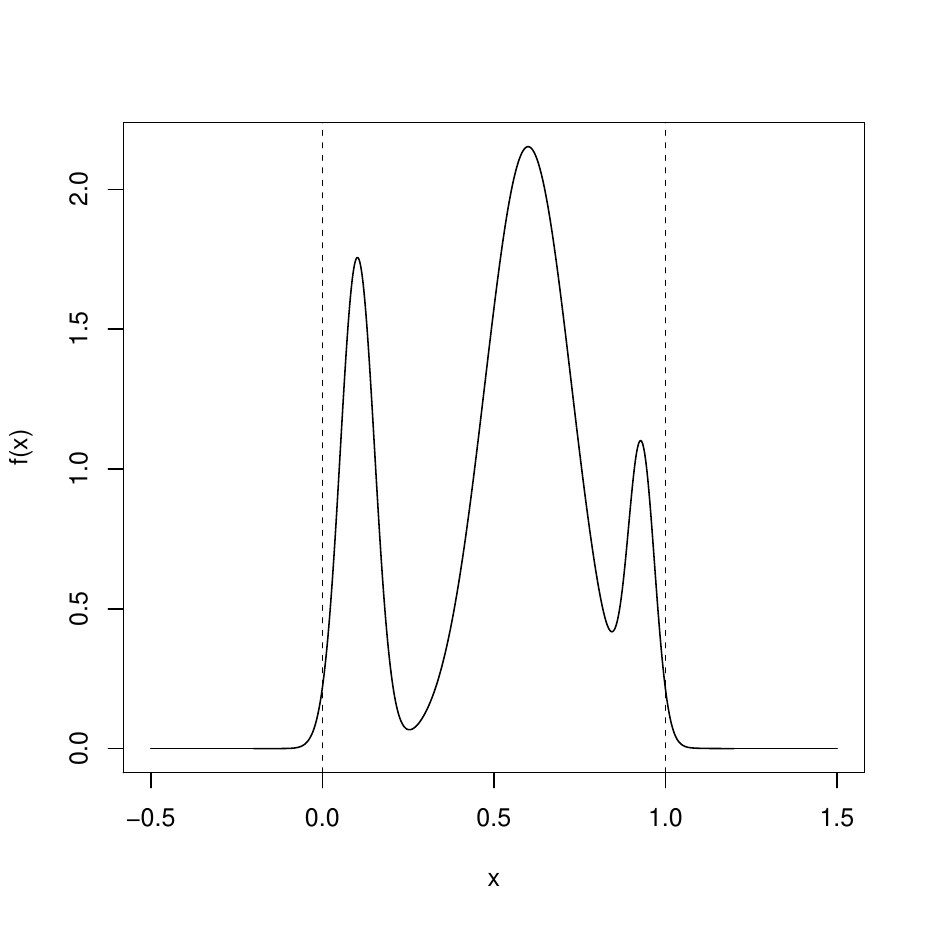}\\
M22
\includegraphics[width=.9\linewidth]{white_image}\\
\phantom{M28}
\end{multicols}
\caption{Density functions. M11--M20: bimodal models. M21--M25: trimodal models.}
\label{figs2}
\end{figure}

The specific formulas of those models considered in the simulation study carried out in Section \ref{simulation} are given here with the notation $\sum_{i=1}^l p_i \cdot \psi_i$, where each $\psi_i$ represents one of the component of the mixture and $p_i$ are the weights of these different components, with $i=1,\ldots, l$, satisfying $\sum_{i=1}^l p_i =1$. The unimodal density functions used as $\psi_i$ are the following models, as defined in \citet{Johnson95}: Beta$(\theta_i,\phi_i)$, Gamma$(\alpha_i,\beta_i)$, $N(\mu_i,\sigma_i^2)$ and Weibull$(\delta_i,c_i)$. All the models were created in such a way that $f(0) \approx f(1) \approx 0.1 \max_{x\in(0,1)} f(x)$. The unimodal probability density functions are represented in Figure \ref{figs1}, the bimodal and trimodal models appear in Figure \ref{figs2}.

\textit{Unimodal models:}

\begin{itemize}
\item M1: $0.44 \cdot N(0.372,0.03) + 0.44 \cdot N(0.67,0.022) + 0.12 \cdot N(0.5,0.2)$.
\item M2: $0.9 \cdot N(0.5,0.05) + 0.05 \cdot N(0.197,0.01) + 0.05 \cdot N(0.803,0.01)$.
\item M3: $0.6 \cdot N(0.62,0.04) + 0.2 \cdot N(0.218,0.1) + 0.2 \cdot N(0.5,0.00795)$.
\item M4: $N(0.5,0.05428)$.
\item M5: $0.9 \cdot N(0.5,0.0485) + 0.1 \cdot N(0.5,0.47)$.
\item M6: $0.6 \cdot N(0.5,0.0502) + 0.2 \cdot N(0.3,0.02) + 0.2 \cdot N(0.7,0.02)$.
\item M7: $0.5 \cdot \mbox{Beta}(10,3) + 0.5 \cdot N(0.5,0.137)$.
\item M8: $0.6 \cdot N(0.4985,0.0793) + 0.4 \cdot \mbox{Weibull}(3,0.5)$.
\item M9: $0.5 \cdot N(0.5,0.3) + 0.45 \cdot N(0.5,0.045) + 0.05 \cdot N(0.5,0.000135)$.
\item M10: $0.6 \cdot N(0.307,0.0518) + 0.4 \cdot \mbox{Gamma}(4,8)$. 
\item M26: $0.58 \cdot N(0.61,0.035) + 0.2 \cdot N(0.232,0.04) + 0.2 \cdot N(0.5,0.00795)+ 0.01 \cdot N(0.15,0.0028) + 0.01 \cdot N(0.98,0.0028)$.
\end{itemize}

\textit{Bimodal models:}

\begin{itemize}
\item M11: $0.75 \cdot N(0.458,0.0546) + 0.25 \cdot N(0.85,0.0041)$.
\item M12: $0.5 \cdot N(0.211,0.012)  + 0.3 \cdot N(0.75,0.062) + 0.2 \cdot \mbox{Beta}(5,2)$.
\item M13: $0.95 \cdot N(0.3035,0.02) + 0.05 \cdot N(0.96757,0.0004)$.
\item M14: $0.5 \cdot N(0.776,0.0109) + 0.3 \cdot N(0.3,0.04) + 0.1 \cdot N(0.25,0.0025) + 0.1 \cdot N(0.35,0.0025)$.
\item M15: $0.3 \cdot N(0.13,0.1)  + 0.3 \cdot N(0.81,0.1) + 0.2 \cdot \mbox{Gamma}(3,9)+ 0.2 \cdot \mbox{Beta}(7,2)$.
\item M16: $0.6 \cdot N(0.384,0.01202) + 0.2 \cdot N(0.2,0.05) + 0.2 \cdot N(0.9,0.00272)$.
\item M17: $0.5 \cdot N(0.3,0.0197) + 0.5 \cdot N(0.7,0.0197)$.
\item M18: $0.5 \cdot N(0.18,0.007) + 0.5 \cdot N(0.82,0.007)$.
\item M19: $0.5 \cdot N(0.06787,0.001) + 0.5 \cdot N(0.93213,0.001)$.
\item M20: $0.48 \cdot N(0.06777,0.001) + 0.48 \cdot N(0.93223,0.001)+ 0.02 \cdot \mbox{Beta}(1.1,2.37558)+ 0.02 \cdot \mbox{Beta}(2.37558,1.1)$.
\end{itemize}

\textit{Trimodal models:}

\begin{itemize}
\item M21: $0.45 \cdot N(0.26,0.01476) + 0.33 \cdot N(0.79145,0.01) + 0.22 \cdot N(0.5,0.007)$.
\item M22: $0.68 \cdot N(0.6,0.01588) + 0.22 \cdot N(0.10245,0.0025) + 0.1 \cdot N(0.93,0.0015)$.
\item M23: $0.45 \cdot N(0.25,0.015) + 0.45 \cdot N(0.6,0.015) + 0.1 \cdot N(0.95222,0.00049)$.
\item M24: $0.55 \cdot N(0.5,0.08425) + 0.15 \cdot N(0.3,0.004) + 0.15 \cdot N(0.5,0.004) + 0.15 \cdot N(0.7,0.004)$.
\item M25: $0.6 \cdot N(0.7749,0.011) + 0.2 \cdot N(0.1345,0.006) + 0.2 \cdot N(0.36,0.006)$.
\end{itemize}

\section{Technical proofs}\label{proofh1}

In this section the proofs of Theorem~\ref{th1} and Proposition \ref{th2} are provided. The first part for proving the asymptotic correct behaviour of our proposal can be derived by following \cite{ChengHall98}. Under the regularity conditions (RC1)--(RC4), \citet[][page 589]{ChengHall98} indicated that the distribution of the test statistic (\ref{msstatistic}) is independent from the underlying density $f$, except for the values in the modes and antimodes of $d_i={|f''(x_i)|}/{f^3(x_i)} \mbox{ with } i=1,\ldots,(2j-1).$ Also, assuming that $f$ has $k$ modes, they showed that the distribution of $\Delta_{n,k+1}$ can be approximated by $\Delta^*_{n,k+1}$, just with bootstrap values obtained from samples coming from a calibration distribution with $k$ modes. Its associated calibration density must satisfy the regularity conditions (RC1)--(RC4) and also that the estimated values $\widehat{d_i}$ converge in probability to $d_i$, as $n \rightarrow \infty$, for $i=1,\ldots,(2k-1)$.

As showed along the text, the calibration function $g$ defined in (\ref{gfunc}) is constructed to guarantee that it satisfies the regularity conditions (RC1)--(RC4) and that has $k$ modes. Then, the key point for obtaining the asymptotic correct behaviour is proving that the estimated values $\widehat{d_i}$, defined in (\ref{distar}), satisfy $\widehat{d_i}\overset{P}\rightarrow d_i$, for $i=1,\ldots,(2k-1)$. The application of the continuous mapping theorem leads that for proving this last convergence is enough with obtaining both $\widehat{f}_{h_k}(\widehat{x_i})\overset{P}\rightarrow f(x_i)$ and $\widehat{f}''_{h_{\tiny{\mbox{PI}}}}(\widehat{x_i}) \overset{P}\rightarrow f''(x_i)$, as the kernel density estimation always satisfies $\widehat{f}_{h}(x)>0$, for any $h>0$ and $x\in \mathbb{R}$ and $f(x_i)\neq 0$ by condition (RC3). 

For proving these last two convergences, let first introduce the following result. Under the regularity conditions, if $K$ satisfies some conditions \citep[see Corollary 1 of][]{Einmahl05} and, in particular, when using the Gaussian kernel, if $h_k$ verifies condition (CBC), then Remark 7 of \cite{Einmahl05} leads that  

\begin{equation}\label{romano1}
\underset{t}{\sup}|\hat{f}_{h_{k}}(t)-f(t)| \rightarrow 0  \quad a.s.
\end{equation}

Under the previous conditions as a consequence of the convergence in (\ref{romano1}) together with the continuous mapping theorem, the following result

\begin{equation}\label{romano0}
\hat{f}_{h_{k}} (\widehat{x_i}) \rightarrow f (x_i) \quad a.s.
\end{equation}
holds if $\widehat{x_i}\overset{a.s.}\rightarrow x_i$. Under the assumption that $\widehat{x_i}\overset{a.s.}\rightarrow x_i$, if a plug--in bandwidth is employed, a similar result to that one showed in (\ref{romano0}) can be derived for the $l$th derivative if $f$ is a bounded density with a $l$th continuous derivative in a neighbourhood of $x_i$ \citep[see Proposition 2.1 of][]{Romano88}. In particular, under (RC3) and (RC4) when employing $h_{\tiny{\mbox{PI}}}$ the following result is obtained if $\widehat{x_i}\overset{a.s.}\rightarrow x_i$,  

\begin{equation}\label{romano2}
\widehat{f}''_{h_{\tiny{\mbox{PI}}}}(\widehat{x_i}) \rightarrow f''(x_i) \quad a.s.
\end{equation}

Related with Remark \ref{remcbw5}, when replacing $h_{\tiny{\mbox{PI}}}$ by $h_k$ in (\ref{romano2}), results in \cite{Romano88} suggest that for obtaining this last convergence, the condition $n a_n/ \log n \rightarrow \infty$ in (CBC) should be replaced by $n a_n^5/ \log n \rightarrow \infty$. This last assumption seems that it is not fulfilled by the critical bandwidth (see Remark \ref{MHY}).

Finally, for proving the convergence $\widehat{x_i}\overset{a.s.}\rightarrow x_i$, let first denote as $x_1<\ldots<x_{2j-1}$ the ordered modes and antimodes of $f$, $x_0$ and $x_{2j}$ two values satisfying $x_0<x_1$ and $x_{2j}>x_{2j-1}$ and use conditions (RC1) and (RC3); then, if $x_i$ is a mode, defining as $\varepsilon_{i,1}=\min\{ (x_{i+1}-x_i)/2,(x_i-x_{i-1})/2\}$, for any $\varepsilon_{i,2}\in (0,\varepsilon_{i,1})$,

\begin{equation}\label{peakf}
 \underset{\{t: \varepsilon_{i,2}\leq |t-x_i|<\varepsilon_{i,1}\}}{\sup} f(t)<f(x_i),
\end{equation}
Now, combining (\ref{romano1}) and (\ref{peakf}), eventually, with probability one, 
\begin{equation}\label{peakfh}
 \underset{\{t: \varepsilon_{i,2}\leq |t-x_i|<\varepsilon_{i,1}\}}{\sup} \hat{f}_{h_{k}}(t)<f(x_i) 
\end{equation} 
The combination of results (\ref{romano1}) and (\ref{peakfh}) yields
\begin{equation}\label{ressm1}
\underset{\{t: \varepsilon_{i,2}\leq|t-x_i|<\varepsilon_{i,1}\}}{\sup} \hat{f}_{h_{k}}(t) < \underset{\{t: |t-x_i|<\varepsilon_{i,1}\}}{\sup}\hat{f}_{h_{k}}(t) 
\end{equation} 
Since the result (\ref{ressm1}) is true for any $\varepsilon_{i,2}\in (0,\varepsilon_{i,1})$, necessarily, $\hat{f}_{h_{k}}$ has a mode, namely $\hat{x}_{i}$, satisfying
$$
|\hat{x}_{i} - x_i|\rightarrow 0 \quad a.s.
$$ 
Similar arguments can be employed if $x_i$ is an antimode. Now, since $\hat{f}_{h_{k}}$ has $j$ modes and $(j-1)$ antimodes, for all the modes and antimodes of $\hat{f}_{h_{k}}$, $\hat{x}_{i}\overset{a.s.}\rightarrow x_i$, with $i=1,\ldots,(2j-1)$. 

\begin{remark}\label{MHY}
\cite{Mammenetal92} proof that if $f$ has a bounded support $[a,b]$ and is twice continuously differentiable on $(a,b)$, with $f'(a+)>0$ and $f'(b-)<0$, together with the regularity conditions (RC1) and (RC3), then $n^{1/5}h_j$ converges in distribution to one random variable that just depends on the values $c_i$, with $i=1,\ldots,(2j-1)$ (see Section \ref{background:critical}). Also, according to \cite{HallYork01}, this convergence in distribution can be derived when employing their critical bandwidth, with the interval $I=[a,b]$, if $f''$ is bounded and continuous in an open interval containing $I$, $f'(a+)>0$, $f'(b-)<0$ and $f$ does not have modes or antimodes outside $(a,b)$.
\end{remark}

\section{New proposal when the support is known}\label{approachA3}

When the modes and antimodes lie in a known closed interval $I$, an alternative approach for the new proposal can be used in order to get better results in practice. This new proposal consists in replacing the critical bandwidth of \citet{Silverman81} for the one of \citet{HallYork01} in the definition of the calibration function $g$. If the number of modes in the entire support is equal to $k$ when testing $H_0:j=k$ (with $(k-1)$ antimodes) in $[a,b]$, then no more changes are needed. If modes appear outside $[a,b]$, then the link function (\ref{lfunc}) can be used in order to preserve the required regularity conditions. Denoting as $a<\widehat{x}_{1}<\ldots<\widehat{x}_{2k-1}<b$, being $\widehat{x}_{1}$ and $\widehat{x}_{2k-1}$ modes, the points $\widehat{x}_{0}$ and $\widehat{x}_{2k}$, needed to obtain the values in (\ref{variabJ}), will be redefined to remove the modes outside $[a,b]$. If there are modes lower than $\widehat{x}_{1}$, then $\widehat{x}_{0}=\min\{x: x\geq a \mbox{ and }\widehat{f}'_{h_{\tiny{\mbox{HY}},k}}(x)>0\}$ and if there are modes greater than $\widehat{x}_{2k-1}$, then $\widehat{x}_{2k}=\max\{x: x\leq b \mbox{ and }\widehat{f}'_{h_{\tiny{\mbox{HY}},k}}(x)<0\}$. Once this change is done, two extra values, $\mathfrak{a}<\widehat{x}_{0}$ and $\mathfrak{b}>\widehat{x}_{2k}$, are needed in order to use the link function. The steps to obtain these two values will be defined later and, from them, the calibration function in (\ref{gfunc}) can be modified in its tails to define $g(x;h_{\tiny{\mbox{HY}},k},h_{\tiny{\mbox{PI}}},\boldsymbol{\varsigma},\mathfrak{a},\mathfrak{b})$ as follows
{\small
\begin{equation*}
  \begin{cases}
	0 &\mbox{ if } x\leq\mathfrak{a} \mbox{ and }  \widehat{f}_{h_{\tiny{\mbox{HY}},k}}  \mbox{has modes lower than } a,\\
	l(x;\mathfrak{a},\widehat{x}_{0},0,\widehat{f}_{h_{\tiny{\mbox{HY}},k}}(\widehat{x}_{0}),0,\widehat{f}'_{h_{\tiny{\mbox{HY}},k}}(\widehat{x}_{0})) &\mbox{ if } x\in (\mathfrak{a},\widehat{x}_{0}) \mbox{ and }  \widehat{f}_{h_{\tiny{\mbox{HY}},k}}  \mbox{has modes lower than } a,\\
	J(x;\widehat{x_i},h_{\tiny{\mbox{HY}},k},h_{\tiny{\mbox{PI}}},\varsigma_i) &\mbox{ if } x\in (\mathfrak{r}_i,\mathfrak{s}_i) \mbox{ for some } i \in \{1,\ldots, (2k-1)\},  \\
	L(x;\zeta_p,h_{\tiny{\mbox{HY}},k}) &\mbox{ if } x\in (z_{(2p-1)},z_{(2p)}) \mbox{ for some } p \in \{1,\ldots, t\},\\
	&\mbox{ and }  \zeta_p \notin (\mathfrak{r}_i,\mathfrak{s}_i) \mbox{ for any } i \in \{1,\ldots, (2k-1)\}, \\
	l(x;\widehat{x}_{2k},\mathfrak{b},\widehat{f}_{h_{\tiny{\mbox{HY}},k}}(\widehat{x}_{2k}),0,\widehat{f}'_{h_{\tiny{\mbox{HY}},k}}(\widehat{x}_{2k}),0) &\mbox{ if } x\in (\widehat{x}_{2k},\mathfrak{b})\mbox{ and }  \widehat{f}_{h_{\tiny{\mbox{HY}},k}}  \mbox{has modes greater than } b,\\
	0 &\mbox{ if } x\geq\mathfrak{b} \mbox{ and }  \widehat{f}_{h_{\tiny{\mbox{HY}},k}}  \mbox{has modes greater than } b,\\
	\widehat{f}_{h_{\tiny{\mbox{HY}},k}}(x) &\mbox{ otherwise,}\\
	\end{cases}
\end{equation*}
} 
the functions $J$ and $L$ are defined as in Section \ref{background:new}, replacing the kernel density estimator $\widehat{f}_{h_k}$ by $\widehat{f}_{h_{\tiny{\mbox{HY}},k}}$ and changing the values of $\widehat{x}_{0}$ and $\widehat{x}_{2k}$ as it was pointed out. The neighborhood in which the $J$ functions are defined is chosen by the same method as in the approach described in the main text. To guarantee that the calibration function is a density, it is also necessary to select correctly the values of $\mathfrak{a}$ (if $\widehat{f}_{h_{\tiny{\mbox{HY}},k}}$ has modes lower than $a$) and $\mathfrak{b}$ (if it has modes greater than $b$) to obtain an integral equal to one. An option is to employ $\mathfrak{a}$ and $\mathfrak{b}$ satisfying 
\begin{eqnarray}
\int_{-\infty}^{\widehat{x}_{0}}  g(x;h_{\tiny{\mbox{HY}},k},h_{\tiny{\mbox{PI}}},\boldsymbol{\varsigma},\mathfrak{a},\mathfrak{b}) dx +& \int_{\widehat{x}_{2k}}^{\infty} g(x;h_{\tiny{\mbox{HY}},k},h_{\tiny{\mbox{PI}}},\boldsymbol{\varsigma},\mathfrak{a},\mathfrak{b}) dx =  \nonumber\\
 \int_{-\infty}^{\widehat{x}_{0}}  \widehat{f}_{h_k} (x) dx + \int_{\widehat{x}_{2k}}^{\infty} \widehat{f}_{h_k} (x) dx.& \label{tailsconst}
\end{eqnarray}
It may happen that the equality (\ref{tailsconst}) is not satisfied for any pair $(\mathfrak{a},\mathfrak{b})$, being $\mathfrak{a}\in (-\infty,\widehat{x}_{0})$ and $\mathfrak{b}\in (\widehat{x}_{2k},\infty)$. In this case, the calibration function can be divided by the normalizing constant to correct the value of the integral. Another alternative can be to take other values of $\widehat{x}_{0}<\widehat{x}_{1}$ and $\widehat{x}_{2k}>\widehat{x}_{2k-1}$, such as $\widehat{f}'_{h_{\tiny{\mbox{HY}},k}}(x)>0$, for all $x\in[\widehat{x}_{0},\widehat{x}_{1})$, and $\widehat{f}'_{h_{\tiny{\mbox{HY}},k}}(x)<0$, for all $x\in(\widehat{x}_{2k-1},\widehat{x}_{2k}]$.

The approach considered in the simulation study (when the support is known) is, in the tails, try to find the value of $\mathfrak{a}$ in the interval $[\widehat{x}_{0}-b+a,\widehat{x}_{0})$ and the value of $\mathfrak{b}$ in $(\widehat{x}_{2k},\widehat{x}_{2k}+b-a]$. If for all the possible values of $\mathfrak{a}$ and $\mathfrak{b}$ the integral
\begin{equation*}
q_2=  \int_{-\infty}^{\infty}  g(x;h_{\tiny{\mbox{HY}},k},h_{\tiny{\mbox{PI}}},\boldsymbol{\varsigma},\mathfrak{a},\mathfrak{b}) dx,
\end{equation*}
is not equal to 1, then the solution is take $\mathfrak{a}$ and $\mathfrak{b}$ in such a way that $q_2$ is as close as possible to one and, then, employ the quotient $g(\cdot;h_{\tiny{\mbox{HY}},k},h_{\tiny{\mbox{PI}}},\boldsymbol{\varsigma},\mathfrak{a},\mathfrak{b})/q_2$ as the calibration function.

\section{Testing $k$--modality when the true density has less than $k$ modes}\label{testbimuni}

As it has already been mentioned, the methods proposed by \citet{Silverman81} and \citet{FisMar01} can be extended from unimodality to test a general null hypothesis as $H_0:j\leq k$. Nevertheless, the proposal presented in this work just allows to test $H_0:j=k$ vs. $H_0:j>k$.  The reason why the $k$--modal test should not be used when the true underlying density has less than $k$ modes is that the test statistic in the bootstrap resamples converge in distribution to a random variable, depending only on the values $\widehat{d_i}$ with $i=1,\ldots,(2k-1)$ \citep[see][]{ChengHall98}. When $j<k$, in the calibration function $g$, there exist $(2k-2j)$ turning points that they will not converge to any fixed value depending on the real density function. As the (asymptotic) distribution of the test statistic in the bootstrap resamples depends also on this $(2k-2j)$ values, one would expect that the sample distribution of the test statistic will not be correctly approximated with the bootstrap resamples. 

Testing $H_0:j=k$ instead of $H_0:j \leq k$ is not in general an important limitation for practical purposes. As it is done in the stamp example in Section \ref{data}, the usual procedure is to perform a stepwise algorithm starting with one mode and, if the null hypothesis is rejected, increasing the number of modes in the null hypothesis by one until there is no evidences for rejection. Despite this note of caution, it can be seen that, generally, testing $H_0:j=k$, when $j<k$, reports also good calibration results.

In order to show the accuracy in practice when the bimodality test is employed in unimodal cases, Table \ref{estsim8} reports the percentages of rejections for significance levels $\alpha=0.01$, $\alpha=0.05$ and $\alpha=0.10$ for testing bimodality employing the proposals by \citet{Silverman81} (SI), \citet{FisMar01} (FM) and the new proposal (NP) presented in this paper. With this goal, samples of size $n=50$, $n=200$ and $n=1000$ were drawn from 10 unimodal distributions (models M1--M10). Again, for each choice of sampling distribution and sample size, 500 realizations were generated. Conditionally on each of those samples, for testing purposes, 500 resamples of size $n$ were drawn from the population. 

The conclusions from the results reported in Table \ref{estsim8} are quite similar to those given previously in Section \ref{simulation} when $H_0:j=1$ was tested. First, although SI still reports a percentage of rejections below the significance level, it is less conservative than for the initial results (testing $H_0:j=1$). For all the available models and all sample sizes and significance levels, the percentage of rejections employing the bimodality test is greater or equal than the one obtained when the unimodality test was applied but lower than the significance level. Regarding FM, again a systematic behaviour cannot be concluded, being the results similar to those ones reported in Section \ref{simulation} when unimodality was tested on the same models. Finally, the results for the new proposal seem to be again quite satisfactory, with a slightly conservative performance in some models, such as M2, M3 ($n=1000$ and $\alpha=0.10$), M4, M5 ($n=200$), M7, M9 and M10 ($n=50$). Observing these results, it seems that, in practice, NP can be used for testing $H_0:j \leq k$, but it should be kept in mind that a correct calibration is not guaranteed. An example of poor behaviour can be observed for (unimodal) model M26. Analysing the results reported in Table \ref{estsim9} for NP, it can be seen that, for $n=1000$, when testing $H_0:j=1$ the percentage of rejections is close to the significance level, whereas when testing $H_0:j=2$, the percentage of rejections is bellow $\alpha$, even employing the correction provided when the support is known.

\setlength{\tabcolsep}{2.2pt}
\setlength\extrarowheight{0.3pt}

\begin{table}
\centering
\scalebox{0.48}{
\begin{tabular}{|c |c| c|c c c |c |c| c|c c c |}
\hline
 & & $\alpha$ & 0.01 & 0.05& 0.10& & & $\alpha$ & 0.01 & 0.05& 0.10\\ 
  \hline 
 M1  &  \multirow{3}{*}{SI}  &  $n=50$  &  0(0) & 0(0) & 0.008(0.008)  &  M6  &  \multirow{3}{*}{SI}  &  $n=50$  &  0(0) & 0(0) & 0.008(0.008)  \\ 
   \multirow{8}{*}{\includegraphics[width=23mm]{120a2.eps}}  &  &  $n=200$  &  0(0) & 0(0) & 0.022(0.013)  &  \multirow{8}{*}{\includegraphics[width=23mm]{120f.eps}}  &  &  $n=200$  &  0(0) & 0.004(0.006) & 0.020(0.012)  \\ 
   &  &  $n=1000$  &  0(0) & 0(0) & 0.002(0.004)  &  &  &  $n=1000$  &  0(0) & 0.012(0.010) & 0.038(0.017)  \\ 
\cline{2-6} \cline{8-12} &  \multirow{3}{*}{FM}  &  $n=50$  &  0.004(0.006)  &  0.040(0.017)  &  0.120(0.028)  &  &  \multirow{3}{*}{FM}  &  $n=50$  &  0(0)  &  0.012(0.010)  &  0.036(0.016)  \\ 
   &  &  $n=200$  &  0.008(0.008)  &  0.094(0.026)  &  0.180(0.034)  &  &  &  $n=200$  &  0.010(0.009)  &  0.042(0.018)  &  0.072(0.023)  \\ 
   &  &  $n=1000$  &  0.010(0.009)  &  0.052(0.020)  &  0.138(0.030)  &  &  &  $n=1000$  &  0.002(0.004)  &  0.030(0.015)  &  0.082(0.024)  \\ 
\cline{2-6} \cline{8-12} &  \multirow{3}{*}{NP}  &  $n=50$  &  0.014(0.010)  &  0.054(0.020)  &  0.120(0.028)   &  &  \multirow{3}{*}{NP}  &  $n=50$  &  0.010(0.009)  &  0.048(0.019)  &  0.110(0.027)   \\ 
   &  &  $n=200$  &  0.010(0.009)  &  0.042(0.018)  &  0.096(0.026)   &  &  &  $n=200$  &  0.008(0.008)  &  0.058(0.020)  &  0.108(0.027)   \\ 
   &  &  $n=1000$  &  0.018(0.011)  &  0.052(0.020)  &  0.092(0.025)   &  &  &  $n=1000$  &  0.014(0.010)  &  0.052(0.019)  &  0.098(0.026)   \\ 
  \hline   M2  &  \multirow{3}{*}{SI}  &  $n=50$  &  0(0) & 0(0) & 0.004(0.006)  &  M7  &  \multirow{3}{*}{SI}  &  $n=50$  &  0(0) & 0(0) & 0.006(0.007)  \\ 
   \multirow{8}{*}{\includegraphics[width=23mm]{120b.eps}}  &  &  $n=200$  &  0(0) & 0(0) & 0.014(0.010)  &  \multirow{8}{*}{\includegraphics[width=23mm]{120g.eps}}  &  &  $n=200$  &  0(0) & 0.002(0.004) & 0.022(0.013)  \\ 
   &  &  $n=1000$  &  0(0) & 0(0) & 0.014(0.010)  &  &  &  $n=1000$  &  0(0) & 0(0) & 0.016(0.011)  \\ 
\cline{2-6} \cline{8-12} &  \multirow{3}{*}{FM}  &  $n=50$  &  0.002(0.004)  &  0.020(0.012)  &  0.046(0.018)  &  &  \multirow{3}{*}{FM}  &  $n=50$  &  0.070(0.022)  &  0.186(0.034)  &  0.338(0.041)  \\ 
   &  &  $n=200$  &  0(0)  &  0.014(0.010)  &  0.060(0.021)  &  &  &  $n=200$  &  0.064(0.021)  &  0.188(0.034)  &  0.320(0.041)  \\ 
   &  &  $n=1000$  &  0(0)  &  0.016(0.011)  &  0.038(0.017)  &  &  &  $n=1000$  &  0.038(0.017)  &  0.160(0.032)  &  0.262(0.039)  \\ 
\cline{2-6} \cline{8-12} &  \multirow{3}{*}{NP}  &  $n=50$  &  0.006(0.007)  &  0.060(0.021)  &  0.126(0.029)   &  &  \multirow{3}{*}{NP}  &  $n=50$  &  0.002(0.004)  &  0.016(0.011)  &  0.042(0.018)   \\ 
   &  &  $n=200$  &  0.008(0.008)  &  0.032(0.015)  &  0.088(0.025)   &  &  &  $n=200$  &  0.004(0.006)  &  0.034(0.016)  &  0.078(0.024)   \\ 
   &  &  $n=1000$  &  0.010(0.009)  &  0.042(0.018)  &  0.072(0.023)   &  &  &  $n=1000$  &  0.008(0.008)  &  0.042(0.018)  &  0.092(0.025)   \\ 
  \hline   M3  &  \multirow{3}{*}{SI}  &  $n=50$  &  0(0)  &  0(0)  &  0(0)  &  M8  &  \multirow{3}{*}{SI}  &  $n=50$  &  0(0) & 0(0) & 0.002(0.004)  \\ 
   \multirow{8}{*}{\includegraphics[width=23mm]{120c.eps}}  &  &  $n=200$  &  0(0) & 0.002(0.004) & 0.008(0.008)  &  \multirow{8}{*}{\includegraphics[width=23mm]{120i2.eps}}  &  &  $n=200$  &  0(0) & 0.002(0.004) & 0.004(0.006)  \\ 
   &  &  $n=1000$  &  0(0) & 0(0) & 0.012(0.010) &  &  &  $n=1000$  &  0(0) & 0(0) & 0.014(0.010)  \\ 
\cline{2-6} \cline{8-12} &  \multirow{3}{*}{FM}  &  $n=50$  &  0.010(0.009)  &  0.054(0.020)  &  0.146(0.031)  &  &  \multirow{3}{*}{FM}  &  $n=50$  &  0.002(0.004)  &  0.024(0.013)  &  0.050(0.019)  \\ 
   &  &  $n=200$  &  0.008(0.008)  &  0.064(0.021)  &  0.150(0.031)  &  &  &  $n=200$  &  0(0)  &  0.024(0.013)  &  0.064(0.021)  \\ 
   &  &  $n=1000$  &  0.002(0.004)  &  0.042(0.018)  &  0.110(0.027)  &  &  &  $n=1000$  &  0(0)  &  0.020(0.012)  &  0.058(0.020)  \\ 
\cline{2-6} \cline{8-12} &  \multirow{3}{*}{NP}  &  $n=50$  &  0.004(0.006)  &  0.032(0.015)  &  0.064(0.021)   &  &  \multirow{3}{*}{NP}  &  $n=50$  &  0.008(0.008)  &  0.034(0.016)  &  0.084(0.024)   \\ 
   &  &  $n=200$  &  0.004(0.006)  &  0.028(0.014)  &  0.066(0.022)   &  &  &  $n=200$  &  0.006(0.007)  &  0.040(0.017)  &  0.076(0.023)   \\ 
   &  &  $n=1000$  &  0.006(0.007)  &  0.034(0.016)  &  0.062(0.021)   &  &  &  $n=1000$  &  0.008(0.008)  &  0.044(0.018)  &  0.102(0.027)   \\ 
  \hline   M4  &  \multirow{3}{*}{SI}  &  $n=50$  &  0(0) & 0(0) & 0.006(0.007)  &  M9  &  \multirow{3}{*}{SI}  &  $n=50$  &  0(0) & 0(0) & 0.010(0.009) \\ 
   \multirow{8}{*}{\includegraphics[width=23mm]{120d.eps}}  &  &  $n=200$  &  0(0) & 0(0) & 0.008(0.008)  &  \multirow{8}{*}{\includegraphics[width=23mm]{120h.eps}}  &  &  $n=200$  &  0(0) & 0(0) & 0.008(0.008)  \\ 
   &  &  $n=1000$  &  0(0) & 0.002(0.004) & 0.008(0.008)  &  &  &  $n=1000$  &  0(0) & 0(0) & 0.018(0.012)  \\ 
\cline{2-6} \cline{8-12} &  \multirow{3}{*}{FM}  &  $n=50$  &  0.004(0.006)  &  0.020(0.012)  &  0.052(0.019)  &  &  \multirow{3}{*}{FM}  &  $n=50$  &  0.014(0.010)  &  0.064(0.021)  &  0.132(0.030)  \\ 
   &  &  $n=200$  &  0.002(0.004)  &  0.008(0.008)  &  0.026(0.014)  &  &  &  $n=200$  &  0.036(0.016)  &  0.222(0.036)  &  0.376(0.042)  \\ 
   &  &  $n=1000$  &  0(0)  &  0.014(0.010)  &  0.048(0.019)  &  &  &  $n=1000$  &  0.032(0.015)  &  0.188(0.034)  &  0.334(0.041)  \\ 
\cline{2-6} \cline{8-12} &  \multirow{3}{*}{NP}  &  $n=50$  &  0.010(0.009)  &  0.066(0.022)  &  0.112(0.028)   &  &  \multirow{3}{*}{NP}  &  $n=50$  &  0.004(0.006)  &  0.024(0.013)  &  0.066(0.022)   \\ 
   &  &  $n=200$  &  0.012(0.010)  &  0.028(0.014)  &  0.050(0.019)   &  &  &  $n=200$  &  0.010(0.009)  &  0.044(0.018)  &  0.092(0.025)   \\ 
   &  &  $n=1000$  &  0.006(0.007)  &  0.046(0.018)  &  0.102(0.027)   &  &  &  $n=1000$  &  0.010(0.009)  &  0.048(0.019)  &  0.110(0.027)   \\ 
 \hline    M5  &  \multirow{3}{*}{SI}  &  $n=50$  &  0(0) & 0(0) & 0.016(0.011)  &  M10  &  \multirow{3}{*}{SI}  &  $n=50$  &  0(0)  &  0(0)  &  0(0)  \\ 
   \multirow{8}{*}{\includegraphics[width=23mm]{120e.eps}}  &  &  $n=200$  &  0(0) & 0.002(0.004) & 0.020(0.012)  &  \multirow{8}{*}{\includegraphics[width=23mm]{120j.eps}}  &  &  $n=200$  &  0(0) & 0.002(0.004) & 0.018(0.012)  \\ 
   &  &  $n=1000$  &  0(0) & 0(0) & 0.002(0.004)  &  &  &  $n=1000$  &  0(0) & 0(0) & 0.020(0.012)  \\ 
\cline{2-6} \cline{8-12} &  \multirow{3}{*}{FM}  &  $n=50$  &  0.056(0.020)  &  0.168(0.033)  &  0.244(0.038)  &  &  \multirow{3}{*}{FM}  &  $n=50$  &  0(0)  &  0.026(0.014)  &  0.066(0.022)  \\ 
   &  &  $n=200$  &  0.130(0.029)  &  0.298(0.040)  &  0.414(0.043)  &  &  &  $n=200$  &  0.004(0.006)  &  0.034(0.016)  &  0.082(0.024)  \\ 
   &  &  $n=1000$  &  0.072(0.023)  &  0.228(0.037)  &  0.364(0.042)  &  &  &  $n=1000$  &  0.006(0.007)  &  0.048(0.019)  &  0.096(0.026)  \\ 
\cline{2-6} \cline{8-12} &  \multirow{3}{*}{NP}  &  $n=50$  &  0.008(0.008)  &  0.044(0.018)  &  0.094(0.026)   &  &  \multirow{3}{*}{NP}  &  $n=50$  &  0.002(0.004)  &  0.026(0.014)  &  0.072(0.023)   \\ 
   &  &  $n=200$  &  0.002(0.004)  &  0.026(0.014)  &  0.084(0.024)   &  &  &  $n=200$  &  0.008(0.008)  &  0.042(0.018)  &  0.098(0.026)   \\ 
   &  &  $n=1000$  &  0.010(0.009)  &  0.046(0.018)  &  0.094(0.026)   &  &  &  $n=1000$  &  0.002(0.004)  &  0.042(0.018)  &  0.088(0.025)    \\ 
   \hline
		\end{tabular}
}
\caption{Percentages of rejections for testing $H_0:j = 2$, with $500$ simulations ($1.96$ times their estimated standard deviation in parenthesis) and $B = 500$ bootstrap samples.}
\label{estsim8}
\end{table}
	
\begin{table}
\centering
\scalebox{0.75}{
\begin{tabular}{|c|c|c|c c c|}
  \hline
 & & $\alpha$ & 0.01 & 0.05& 0.10 \\
  \hline
	M26 & \multirow{3}{*}{NP (unknown support)} &  $n=50$  &  0.008(0.008)  &  0.032(0.015)  &  0.066(0.022)  \\ 
\multirow{8}{*}{\includegraphics[width=23mm]{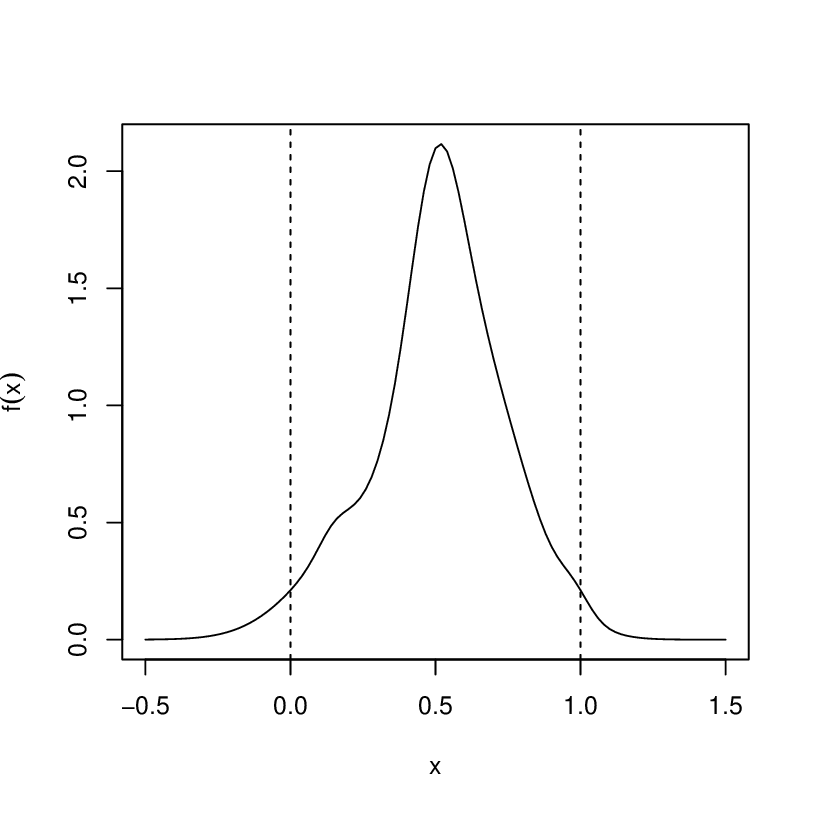}}   &  &  $n=200$  &  0(0)  &  0.022(0.013)  &  0.068(0.022)  \\ 
   &  &  $n=1000$  &  0.004(0.006)  &  0.040(0.017)  &  0.080(0.024)  \\ 
 \cline{2-6}   & \multirow{3}{*}{NP (unknown support)}   &  $n=50$  &  0.004(0.006)  &  0.028(0.014)  &  0.048(0.019)  \\ 
   &  &  $n=200$  &  0.006(0.007)  &  0.030(0.015)  &  0.074(0.023)  \\ 
   &  &  $n=1000$  &  0(0)  &  0.020(0.012)  &  0.054(0.020)  \\ 
  \cline{2-6}  & \multirow{3}{*}{NP (known support)}   &  $n=50$  &  0.002(0.004)  &  0.022(0.013)  &  0.076(0.023)   \\ 
   &  &  $n=200$  &  0.004(0.006)  &  0.034(0.016)  &  0.062(0.021)   \\ 
   &  &  $n=1000$  &  0(0)  &  0.026(0.014)  &  0.052(0.019)   \\ 
  \hline
	\end{tabular}   
}
\caption{Percentages of rejections for testing $H_0:j = 1$ (first column) and $H_0:j = 2$ (second and third column), with $500$ simulations ($1.96$ times their estimated standard deviation in parenthesis) and $B = 500$ bootstrap samples.}
\label{estsim9}
\end{table}

\section{Numerical approximations}\label{numericas}

Details of the numerical approaches used in this paper can be found bellow. All the functions were implemented in the statistical software \citet{R14}.

\textbf{Practical computation of the critical bandwidth}

To obtain both the critical bandwidth of \citet{Silverman81} and that one of \citet{HallYork01}, a binary search procedure was used. In each step of the algorithm, denoting $h_a$ as the bandwidth in such a way than $\widehat{f}_{h_a}$ has at most $k$ modes and $h_b$ the bandwidth for which the kernel density estimation has more than $k$ modes (both in the interval $I$ if the critical bandwidth of \citet{HallYork01} is being calculated). Then the dichotomy algorithm is stopped when $(h_b-h_a)<(h_{a_0}/2^{10})$, where $h_{a_0}$ is the initial value of $h_a$. The last calculated value of $h_a$ is the one employed as the critical bandwidth.

\textbf{Practical computation of the excess mass}

To obtain the excess mass defined by \citet{MulSaw91} when the null hypothesis of unimodality is being tested, the following result is used: the value of the excess mass statistic is exactly twice the value of the \textit{dip} statistic introduced by \citet{Hartigan85}. The value of the \textit{dip} was obtained using the \texttt{diptest} package implemented by \citet{Maechler13}.

For the general case, the following algorithm will be employed to obtain the excess mass statistic when the null hypothesis $H_0:j= k$, with $k>1$, is tested. First, assume that $d_{k}(p)$ is the minimum distance of the union of $k$ intervals containing $p$ data points. To get the possible values of $\lambda$ corresponding to $E_{n,k}(\mathbb{P}_n,\lambda)$, and candidates to minimize the difference 

$$
D_{n,k+1}(\lambda)=\{E_{n,k+1}(\mathbb{P}_n,\lambda) - E_{n,k}(\mathbb{P}_n,\lambda)\},
$$
the first step begins in $q_k(1)=n$ (where the number in parenthesis is the current iteration). Then, it search from $q_k(1)-1$ until $(k+1)$ the integer $q_k(2)$ minimizing the following expression

$$
\lambda_{k}(1)=\underset{q_k(2)}\min \frac{q_k(1)-q_k(2)}{n(d_{k}(q_k(1))-d_{k}(q_k(2)))},
$$
the value of $\lambda_{k}(1)$ is one of the possible values of $\lambda$ minimizing $D_{n,k+1}(\lambda)$. Then, if $q_k(2)=(k+1)$ is the value minimizing the previous expression, the algorithm will be stopped, otherwise it is continued until $q_k(t_{k})=(k+1)$ (where $t_k$ is the realised number of iterations). After obtaining the vector $\boldsymbol{\lambda}_{k}$ of possible values of $\lambda$ minimizing $D_{n,k+1}(\lambda)$, the algorithm is repeated for $(k+1)$ to get the vector $\boldsymbol{\lambda}_{k+1}$. Then the excess mass statistic is easy to obtain using that 
$$
\Delta_{n,k+1}=\underset{\boldsymbol{\lambda}_{k} \cup \boldsymbol{\lambda}_{k+1}}{\min} D_{n,k+1}(\lambda).
$$

To get the values of $d_{k}(p)$, one can obtain the exact result of the excess mass employing the algorithm provided by \citet{MulSaw91}. A similar algorithm was implemented and employed to get the exact results in Section \ref{data}. The problem of employing this algorithm is the high computing cost for an extensive study. For this reason an approximation was employed in Section \ref{simulation} to get the excess mass statistic, $\Delta_{n,3}$. The new algorithm consist in, first, calculate $d_{1}(p)$ to obtain the exact values of $\boldsymbol{\lambda}_{1}$ and secondly create a grid of $l$ possible values between each $\lambda_{1}(j)$ and $\lambda_{1}(j+1)$, with $j$ and entire value between 1 and $(t_k-1)$. Finally, to get the test statistic, that value of $\lambda$ belonging the entire grid and minimising $D_{n,3}(\lambda)$ is chosen. The employed size $l$ was: $l=100$ when $n=50$, $l=40$ for $n=100$, $l=20$ if $n=200$ and $l=5$ when $n=1000$. This selection of points represents a balance between the accuracy and the computation time, as, in general, if $n$ is large then the length of the vector $t_k$ is also large.

\section{Further details on real data analysis}
\label{moredata}

As explained in the Introduction, the value of stamps depends on its scarcity, and thickness is determinant in this sense. However, in general, the designation of thick, medium or thin stamps is relative and can only refer to a particular stamp issue. Otherwise, making uniform categories for all stamp issues may lead to inaccurate classifications. In addition, there is not such a differentiation between groups available in stamps catalogs, leaving this classification to a personal subjective judgment. The importance of establishing an objective criterion specially appears in stamp issues printed on a mixture of paper types, with possible differences in their thickness.

A stamp issue where the problem of determining the number of different groups of stamps appears is in the 1872 Hidalgo issue. First, for this particular issue, and in general in the Mexican ones, it is known that the handmade paper presents a high variability in the thickness of the paper. Second, since of scarcity of ordinary white wove paper, other types of paper were used to produce the Hidalgo issue. A small quantity of ``vertically laid'' paper, a fiscal type of white wove paper denominated \textit{Papel Sellado} (some of them were watermarked vertically), other type of white wove, the \textit{La Croix--Freres} of France (some of them with a watermark of \textit{LA+-F}) and also another unwatermarked white wove paper might also have been used. It is estimated than the watermark of \textit{Papel Sellado} can appear in between 6 and 18 stamps in each sheet of 100. For the \textit{La Croix--Freres} watermark, it is estimated that the symbol appears in between 4 and 10 stamps if the sheet was watermarked, and some authors suggested that this watermark appears only once of every 4 sheets. In order to get more information about this particular problem and to obtain some further references, see \citet{IzenSom88}.

This particular example has been explored in several references in the literature for determining the number of groups. From a non--parametric point of view, some examples of its utilization for mode testing can be shown in \citet[][Ch. 16]{Efron94}, \citet{IzenSom88} or in \citet{FisMar01}. Also it was analysed using non--parametric exploratory tools in \citet{Wilson83}, \citet{MinSco93} and in \citet{ChMar99}. Some parametric studies of the 1872 Hidalgo issue can be found in \citet{Basford97} or in \citet[][Ch. 6]{McLachlan00}.

Taking a subsample of 437 stamps on white wove, \citet{Wilson83} made a histogram and the conclusion was that only two kinds of paper were used, the \textit{Papel Sellado} and the \textit{La Croix--Freres}, and that there was not a third kind of paper. \citet{IzenSom88} revisited the example considering a more complete collection, with 485 stamps. A histogram with the same parameters as those used by \citet{Wilson83} (same starting point and bin width) is shown in Figure \ref{fig1} (top--left panel), revealing the same features as those noticed in the original reference. Two groups are also shown by a kernel density estimator, shown in the same plot, considering a gaussian kernel and a rule of thumb bandwidth \citep[see][Ch. 3.2]{wandjones}. However, both approximations (histogram and kernel density estimator) depend heavily on the bin width and bandwidth, respectively. Specifically, the use of an automatic rule for selecting the bandwidth value (focused on the global estimation of the entire density function) does not guarantee an appropriate recovery of the modes. In fact, using another automatic rule as the plug--in bandwidth (Figure \ref{fig1}, bottom--left panel), nine modes are observed. A histogram with a smaller bin width is also included in this plot, exhibiting apparently more modes than the initial one.

Given that the exploratory tools did not provide a formal way of determining if there are more than two groups, \textit{Papel Sellado} and \textit{La Croix--Freres}, \citet{IzenSom88} employed the multimodality test of \citet{Silverman81}. Note that for this purpose, just FM, SI and the new proposal NP can be used, and the first two proposals present a poor calibration, as shown in the simulation study. Results from \citet{IzenSom88}, applying SI with $B=100$, are shown in Table \ref{sellos1} (note that with $B=500$, different p--values are obtained). For $\alpha=0.05$, the conclusions are the same, except in the crucial case of testing $H_0:j\leq 2$, where for $B=500$, there are no evidences to reject the null hypothesis. These differences may be caused by the approximations implemented by \citet{IzenSom88} to obtain the critical bandwidth. Both \citet[][Ch. 16]{Efron94} (using $B=500$ bootstrap replicates) and \citet{Salgado98} (employing $B=600$) obtained similar results to ours. Hence, the null hypothesis must not be rejected when the hypothesis is that the distribution has at most two modes, but it has to be rejected when $H_0$ is that the distribution has at most six modes. This strange behaviour also happens in \citet{IzenSom88} analysis, when testing $H_0:j\leq 3$ and $H_0:j\leq 6$.

\begin{table}
\begin{center}
\scalebox{0.8}{
\begin{tabular}{ |c |c  | c c c c c c c c c| }
\hline
& $k$ & 1 & 2 & 3 & 4 & 5 & 6 & 7 & 8 & 9 \\ \hline
SI & $B=100$& 0 & 0.04 & 0.06 & 0.01 & 0 & 0 & 0.44 & 0.31 & 0.82 \\ 
& $B=500$ & 0.018 & 0.394 & 0.090 & 0.008 & 0.002 & 0.002 & 0.488 & 0.346 & 0.614 \\  \hline
FM &  & 0 & 0.04 & 0 & 0 & 0 & 0 & 0.06 & 0.01 & 0.06 \\ \hline
NP& & 0 & 0.022 & 0.004 & 0.506 & 0.574 & 0.566 & 0.376 & 0.886 & 0.808 \\ \hline
\end{tabular}
}
\caption{P--values obtained using different proposals for testing $k$--modality, with $k$ between 1 and 9. Methods: SI, FM and NP. For SI method, $B=100$ \citep[first row;][]{IzenSom88} and $B=500$.}
\label{sellos1}
\end{center}
\end{table}

\citet{IzenSom88} suggested non--rejecting the null hypothesis the first time that the p--value is higher than $0.4$. The consideration of a \emph{flexible} rule for rejecting the null hypothesis is justified by the fluctuations in the p--values of SI and, as \citet{IzenSom88} mentioned, by the ``conservative'' nature of this test. Under this premise, the result when applying SI would be that the null hypothesis is rejected until it is tested $H_0:j\leq 7$. Hence, \citet{IzenSom88} conclude that the number of groups in the 1872 Hidalgo Issue is seven. 

As shown in Section \ref{simulation}, SI does not present a good calibration and sometimes it can be also anticonservative. It is not surprising that SI behaves differently when testing $H_0:j\leq k$ for $k=2,3$, with respect to the rest of cases until $k=7$. Since NP has a good calibration behaviour, even with ``small'' sample sizes, this method is going to be used, first for testing the important case $H_0:j= 2$ vs. $H_a:j> 2$ and then to figure out how many groups are there in the 1872 Hidalgo Issue.

The computation of the excess mass statistic requires a non--discrete sample and the original data (denoted as $\mathcal{X}$) contained repeated values, the artificial sample $\mathcal{Y}=\mathcal{X}+\mathcal{E}$ will be employed for testing the number of modes, where $\mathcal{E}$ is a sample of size 485 from the $U(-5\cdot10^{-4},5\cdot 10^{-4})$ distribution. This modification of the data was also considered by \cite {FisMar01}. The p--values obtained in their studio (using $B=200$ bootstrap replicates) are shown in Table \ref{sellos1}: it is not clear which conclusion has to be made. They mentioned that their results are consistent with the previous studies, detecting 7 modes. But it should be noticed that, as shown in the simulation, FM does not present a good calibration behaviour. 

Finally, the p--values obtained with NP are also shown in the Table \ref{sellos1}, with $B=500$. Similar results can be obtained employing the interval $I=[0.04,0.15]$ in NP with known support, as \citet{IzenSom88} notice that the thickness of the stamps is always in this interval $I$. Employing a significance level $\alpha=0.05$ for testing $H_0:j= 2$, it can be observed that the null hypothesis is rejected. It can be seen that the null hypothesis is rejected until $k=4$, and then there is no evidences to reject $H_0$ employing greater values of $k$. Then, applying our new procedure, the conclusion is that the number of groups in the 1872 Hidalgo Issue is four.

In order to compare the results obtained by \citet{IzenSom88} and the ones derived applying the new proposal, two kernel density estimators, with gaussian kernel and critical bandwidths $h_4$ and $h_7$ are depicted in Figure \ref{fig1} (bottom--right panel). \citet{IzenSom88} conclude that seven modes were present, and argued that the stamps could be divided in, first, three groups (pelure paper with  mode at 0.072 mm, related with the forged stamps; the medium paper in the point 0.080 mm; and the thick paper at 0.090 mm). Given the efforts made in the new issue in 1872 to avoid forged stamps, it seems quite reasonable to assume that the group associated with the pelure paper had disappeared in this new issue. In that case, the asymmetry in the first mode using $h_4$ can be attributed to the modifications in the paper made by the manufacturers. Also, this first and asymmetric group, justifies the application of non--parametric techniques to determine the number of groups. It can be seen, in the Section 7 of \citet{IzenSom88} and in other references using mixtures of gaussian densities to model this data \citep[see, for example,][Ch. 6]{McLachlan00}, that these parametric techniques have problems capturing this asymmetry, and they always determine that there are two modes in this first part of the density, one near the point 0.07 mm and another one near 0.08 mm. For the two modes near the points 0.10 and 0.11 mm, both corresponding to stamps produced in 1872. As \citet{IzenSom88} noticed, it seems that the stamps of 1872 were printed on two different paper types, one with the same characteristics as the unwatermarked white wove paper used in the 1868 issue, and a second much thicker paper that disappeared completely by the end of 1872. Using this explanation, it seems quite reasonable to think that the two final modes using $h_4$, corresponds with the medium paper and the thick paper in this second block of stamps produced in 1872. Finally, for the two minor modes appearing near 0.12 and 0.13 mm, when $h_7$ is used, \citet{IzenSom88} do not find an explanation and they mention that probably they could be artefacts of the estimation procedure. This seems to confirm the conclusions obtained with our new procedure. The reason of determining more groups than the four obtained with our proposal, seems to be quite similar to that of the model M20 in our simulation study. This possible explanation is that the spurious data in the right tail of the last mode are causing the rejection of $H_0$, when SI is used.

\bibliographystyle{chicago}
\bibliography{mode_testing_biblio}

\begin{thebibliography}{}

\bibitem[\protect\citeauthoryear{Basford, McLachlan, and York}{Basford
  et~al.}{1997}]{Basford97}
Basford, K.~E., G.~J. McLachlan, and M.~G. York (1997).
\newblock Modelling the distribution of stamp paper thickness via finite normal
  mixtures: The 1872 {H}idalgo stamp issue of {M}exico revisited.
\newblock {\em Journal of Applied Statistics\/}~{\em 24}, 169--180.

\bibitem[\protect\citeauthoryear{Chaudhuri and Marron}{Chaudhuri and
  Marron}{1999}]{ChMar99}
Chaudhuri, P. and J.~S. Marron (1999).
\newblock {SiZer} for exploration of structures in curves.
\newblock {\em Journal of the American Statistical Association\/}~{\em 94},
  807--823.

\bibitem[\protect\citeauthoryear{Cheng and Hall}{Cheng and
  Hall}{1998}]{ChengHall98}
Cheng, M.~Y. and P.~Hall (1998).
\newblock Calibrating the excess mass and dip tests of modality.
\newblock {\em Journal of the Royal Statistical Society. Series B\/}~{\em 60},
  579--589.

\bibitem[\protect\citeauthoryear{Efron and Tibshirani}{Efron and
  Tibshirani}{1994}]{Efron94}
Efron, B. and R.~J. Tibshirani (1994).
\newblock {\em An Introduction to the Bootstrap}.
\newblock United States of America: Chapman and Hall.

\bibitem[\protect\citeauthoryear{Einmahl, Mason, et~al.}{Einmahl
  et~al.}{2005}]{Einmahl05}
Einmahl, U., D.~M. Mason, et~al. (2005).
\newblock Uniform in bandwidth consistency of kernel-type function estimators.
\newblock {\em The Annals of Statistics\/}~{\em 33}, 1380--1403.

\bibitem[\protect\citeauthoryear{Fisher and Marron}{Fisher and
  Marron}{2001}]{FisMar01}
Fisher, N.~I. and J.~S. Marron (2001).
\newblock Mode testing via the excess mass estimate.
\newblock {\em Biometrika\/}~{\em 88}, 419--517.

\bibitem[\protect\citeauthoryear{Good and Gaskins}{Good and
  Gaskins}{1980}]{GoodGask80}
Good, I.~J. and R.~A. Gaskins (1980).
\newblock Density estimation and bump-hunting by the penalized likelihood
  method exemplified by scattering and meteorite data.
\newblock {\em Journal of the American Statistical Association\/}~{\em 75},
  42--56.

\bibitem[\protect\citeauthoryear{Hall and York}{Hall and
  York}{2001}]{HallYork01}
Hall, P. and M.~York (2001).
\newblock On the calibration of {S}ilverman's test for multimodality.
\newblock {\em Statistica Sinica\/}~{\em 11}, 515--536.

\bibitem[\protect\citeauthoryear{Hartigan and Hartigan}{Hartigan and
  Hartigan}{1985}]{Hartigan85}
Hartigan, J.~A. and P.~M. Hartigan (1985).
\newblock The dip test of unimodality.
\newblock {\em The Annals of Statistics\/}~{\em 13}, 70--84.

\bibitem[\protect\citeauthoryear{Izenman and Sommer}{Izenman and
  Sommer}{1988}]{IzenSom88}
Izenman, A.~J. and C.~J. Sommer (1988).
\newblock Philatelic mixtures and multimodal densities.
\newblock {\em Journal of the American Statistical Association\/}~{\em 83},
  941--953.

\bibitem[\protect\citeauthoryear{Johnson, Kotz, and Balakrishnan}{Johnson
  et~al.}{1995}]{Johnson95}
Johnson, N.~L., S.~Kotz, and N.~Balakrishnan (1995).
\newblock {\em Continuous Univariate Distributions}, Volume 1 and 2.
\newblock New York: Wiley Series in Probability and Statistics.

\bibitem[\protect\citeauthoryear{Maechler}{Maechler}{2015}]{Maechler13}
Maechler, M. (2015).
\newblock {\em {diptest}: {H}artigan's Dip Test Statistic for Unimodality -
  Corrected}.
\newblock R package version 0.75-7.

\bibitem[\protect\citeauthoryear{Mammen, Marron, and Fisher}{Mammen
  et~al.}{1992}]{Mammenetal92}
Mammen, E., J.~S. Marron, and N.~I. Fisher (1992).
\newblock Some asymptotics for multimodality tests based on kernel density
  estimates.
\newblock {\em Probability Theory and Related Fields\/}~{\em 91}, 115--132.

\bibitem[\protect\citeauthoryear{Marron and Schmitz}{Marron and
  Schmitz}{1992}]{MarSch92}
Marron, J.~S. and H.~P. Schmitz (1992).
\newblock Simultaneous density estimation of several income distributions.
\newblock {\em Econometric Theory\/}~{\em 8}, 476--488.

\bibitem[\protect\citeauthoryear{McLachlan and Peel}{McLachlan and
  Peel}{2000}]{McLachlan00}
McLachlan, G. and D.~Peel (2000).
\newblock {\em Finite Mixture Models}.
\newblock United States of America: John Wiley \& Sons.

\bibitem[\protect\citeauthoryear{Minnotte, Marchette, and Wegman}{Minnotte
  et~al.}{1998}]{Minetal98}
Minnotte, M.~C., D.~J. Marchette, and E.~J. Wegman (1998).
\newblock The bumpy road to the mode forest.
\newblock {\em Journal of Computational and Graphical Statistics\/}~{\em 7},
  239--251.

\bibitem[\protect\citeauthoryear{Minnotte and Scott}{Minnotte and
  Scott}{1993}]{MinSco93}
Minnotte, M.~C. and D.~W. Scott (1993).
\newblock The mode tree: A tool for visualization of nonparametric density
  features.
\newblock {\em Journal of Computational and Graphical Statistics\/}~{\em 2},
  51--68.

\bibitem[\protect\citeauthoryear{Mitchell, Sundberg, and Reynolds}{Mitchell
  et~al.}{2007}]{Mitchell07}
Mitchell, J.~F., K.~A. Sundberg, and J.~H. Reynolds (2007).
\newblock Differential attention-dependent response modulation across cell
  classes in macaque visual area {V4}.
\newblock {\em Neuron\/}~{\em 55}, 131--141.

\bibitem[\protect\citeauthoryear{M\"uller and Sawitzki}{M\"uller and
  Sawitzki}{1991}]{MulSaw91}
M\"uller, D.~W. and G.~Sawitzki (1991).
\newblock Excess mass estimates and tests for multimodality.
\newblock {\em Journal of the American Statistical Association\/}~{\em 86},
  738--746.

\bibitem[\protect\citeauthoryear{Olden, Hogan, and Zanden}{Olden
  et~al.}{2007}]{Oldenetal07}
Olden, J.~D., Z.~S. Hogan, and M.~Zanden (2007).
\newblock Small fish, big fish, red fish, blue fish: size-biased extinction
  risk of the world's freshwater and marine fishes.
\newblock {\em Global Ecology and Biogeography\/}~{\em 16}, 694--701.

\bibitem[\protect\citeauthoryear{{R Core Team}}{{R Core Team}}{2016}]{R14}
{R Core Team} (2016).
\newblock {\em R: A Language and Environment for Statistical Computing}.
\newblock Vienna, Austria: R Foundation for Statistical Computing.

\bibitem[\protect\citeauthoryear{Roeder}{Roeder}{1990}]{Roeder90}
Roeder, K. (1990).
\newblock Density estimation with confidence sets exemplified by superclusters
  and voids in the galaxies.
\newblock {\em Journal of the American Statistical Association\/}~{\em 85},
  617--624.

\bibitem[\protect\citeauthoryear{Romano}{Romano}{1988}]{Romano88}
Romano, J.~P. (1988).
\newblock On weak convergence and optimality of kernel density estimates of the
  mode.
\newblock {\em The Annals of Statistics\/}~{\em 16}, 629--647.

\bibitem[\protect\citeauthoryear{Salgado-Ugarte, Shimizu, and
  Taniuchi}{Salgado-Ugarte et~al.}{1998}]{Salgado98}
Salgado-Ugarte, I.~H., M.~Shimizu, and T.~Taniuchi (1998).
\newblock Nonparametric assessment of multimodality for univariate data.
\newblock {\em Stata Technical Bulletin\/}~{\em 7}, 27--35.

\bibitem[\protect\citeauthoryear{Silverman}{Silverman}{1981}]{Silverman81}
Silverman, B.~W. (1981).
\newblock Using kernel density estimates to investigate multimodality.
\newblock {\em Journal of the Royal Statistical Society. Series B\/}~{\em 43},
  97--99.

\bibitem[\protect\citeauthoryear{Wand and Jones}{Wand and
  Jones}{1995}]{wandjones}
Wand, M.~P. and M.~C. Jones (1995).
\newblock {\em Kernel Smoothing}.
\newblock Great Britain: Chapman and Hall.

\bibitem[\protect\citeauthoryear{Wilson}{Wilson}{1983}]{Wilson83}
Wilson, I.~G. (1983).
\newblock Add a new dimension to your philately.
\newblock {\em The American Philatelist\/}~{\em 97}, 342--349.

\end{thebibliography}

\end{document}